\documentclass[12pt]{livrev}          
\usepackage{epsfig, a4wide}
\begin{document}

\tableofcontents

 \def\etal{{\sl et al.}\,}
 \def\ie{{\sl i.e.,}\,}
 \def\eg{{\sl e.g.,}\,}
 \def\sun{\odot}
 \def\ms{M_{\sun}}

 \def\rund#1{\left( #1 \right)}

 \def\Ws{W^{\ast}}

 \def\that{\widehat \tau}
 \def\thata{\widehat \tau_a}
 \def\thats{\widehat \tau^{\ast}}
 \def\thatas{\widehat \tau_a^{\ast}}
 \def\thatbs{\widehat \tau_b^{\ast}}

 \def\vecr{{\mathbf r}}
 \def\vecv{{\mathbf v}}
 \def\vecj{{\mathbf j}}

 \def\tpiab{{\mathbf \Pi}_{ab}}
 \def\tomab{{\mathbf \Omega}_{ab}}

 \def\vecra{{\mathbf r}_a}
 \def\vecrb{{\mathbf r}_b}

 \def\vecS{{\mathbf S}}

 \def\Shat{\widehat {\mathbf S}}
 \def\Shata{\widehat {\mathbf S}_a}
 \def\Shats{\widehat {\mathbf S}^{\ast}}
 \def\Shatas{\widehat {\mathbf S}_a^{\ast}}
 \def\Shatbs{\widehat {\mathbf S}_b^{\ast}}

 \def\vecva{{\mathbf v}_a}
 \def\vecvb{{\mathbf v}_b}

 \def\vecvab{{\mathbf v}_{ab}}
 \def\vecrab{{\mathbf r}_{ab}}

 \def\ARAA#1{{\it Ann. Rev. Astron. Astrophys. {\bf#1}}}
 \def\ARFM#1{{\it Ann. Rev. Fluid Mech. {\bf#1}}}
 \def\AaA#1{{\it Astron. Astrophys. {\bf#1}}}
 \def\AaAS#1{{\it Astron. Astrophys. Suppl. {\bf#1}}}
 \def\AJ#1{{\it Astronom. J. {\bf#1}}}
 \def\ApJ#1{{\it Astrophys. J. {\bf#1}}}
 \def\ApJS#1{{\it Astrophys. J. Suppl. {\bf#1}}}
 \def\CPAM#1{{\it Comm. Pure Appl. Math. {\bf#1}}}
 \def\CPC#1{{\it Comput. Phys. Commun. {\bf#1}}}
 \def\CPR#1{{\it Comput. Phys. Rep. {\bf#1}}}
 \def\CQG#1{{\it Class. Quant. Grav. {\bf#1}}}
 \def\JCP#1{{\it J. Comp. Phys. {\bf#1}}}
 \def\Nat#1{{\it Nature {\bf#1}}}
 \def\MNRAS#1{{\it Mon. Not. Roy. Astronom. Soc. {\bf#1}}}
 \def\PTP#1{{\it Prog. Theor. Phys. {\bf#1}}}
 \def\SINUM#1{{\it SIAM J. Numer. Anal. {\bf#1}}}
 \def\SISC#1{{\it SIAM J. Sci. Comput. {\bf#1}}}

\title{NUMERICAL HYDRODYNAMICS IN SPECIAL RELATIVITY}
\author{J. M. Mart\'{\i} \\
        Departamento de Astronom\'{\i}a y Astrof\'{\i}sica \\
        Universidad de Valencia \\
        46100 Burjassot (Valencia), Spain\\
\and
        E. M\"uller \\
        Max-Planck-Institut f\"ur Astrophysik\\
        Karl-Schwarzschild-Str.\,1, 85748 Garching, Germany }
%
%\date{}
%
\maketitle
\begin{center}
\section*{ABSTRACT}
\end{center}
  This review is concerned with a discussion of numerical methods for
the solution of the equations of special relativistic hydrodynamics
(SRHD).  Particular emphasis is put on a comprehensive review of the
application of high--resolution shock--capturing methods in SRHD.
Results obtained with different numerical SRHD methods are compared,
and two astrophysical applications of SRHD flows are discussed.  An
evaluation of the various numerical methods is given and future
developments are analyzed.
\bigskip

\section{INTRODUCTION}
%        ############
\label{intro}

\subsection{Current fields of research}
%           --------------------------
\label{ss:current}

  Relativity is a necessary ingredient for describing astrophysical
phenomena involving compact objects. Among these phenomena are core
collapse supernovae, X-ray binaries, pulsars, coalescing neutron
stars, formation of black holes, micro--quasars, active galactic
nuclei, superluminal jets and gamma-ray bursts.  General relativistic
effects must be considered when strong gravitational fields are
encountered as, for example, in the case of coalescing neutron stars
or near black holes. The significant gravitational wave signal
produced by some of these phenomena can also only be understood in the
framework of general theory of relativity.  There are, however,
astrophysical phenomena which involve flows at relativistic speeds but
no strong gravitational fields and thus at least certain aspects of
these phenomena can be described within the framework of special
relativity.

  Another field of research, where special relativistic ``flows'' are
encountered, are present-day heavy-ion collision experiments taking
place in large particle accelerators. The heavy ions are accelerated
to ultra-relativistic velocities very close to the speed of light
($\sim 99.998\%$ \cite{SH97}) to study the equation of state for hot
dense nuclear matter.

\subsection{Overview of the numerical methods}
%           ---------------------------------
\label{ss:overview}

  The first attempt to solve the equations of relativistic
hydrodynamics (SRHD) was made by Wilson \cite{Wi72, Wi79} and
collaborators \cite{CW84, HS84} using an Eulerian explicit finite
difference code with monotonic transport. The code relies on
artificial viscosity techniques \cite{vR50, RM67} to handle shock
waves. It has been widely used to simulate flows encountered in
cosmology, axisymmetric relativistic stellar collapse, accretion onto
compact objects and, more recently, collisions of heavy ions. Almost
all the codes for numerical (both special and general) relativistic
hydrodynamics developed in the eighties \cite{Pi80, SP87, NM80, Na81,
NS82, Ev86} were based on Wilson's procedure. However, despite its
popularity it turned out to be unable to accurately describe extremely
relativistic flows (Lorentz factors larger than 2; see, e.g.,
\cite{CW84}).

  In the mid eighties, Norman \& Winkler \cite{NW86} proposed a
reformulation of the difference equations with artificial viscosity
consistent with the relativistic dynamics of non--perfect fluids. The
strong coupling introduced in the equations by the presence of the
viscous terms in the definition of relativistic momentum and total
energy densities required an implicit treatment of the difference
equations. Accurate results across strong relativistic shocks with
large Lorentz factors were obtained in combination with adaptive mesh
techniques. However, no multidimensional version of this code was
developed.

  Attempts to integrate the SRHD equations avoiding the use of
artificial viscosity were performed in the early nineties. Dubal
\cite{Du91} developed a 2D code for relativistic magneto-hydrodynamics
based on an explicit second--order Lax--Wendroff scheme incorporating
a flux--corrected transport (FCT) algorithm \cite{BB73}. Following a
completely different approach Mann \cite{Ma91} proposed a
multidimensional code for general relativistic hydrodynamics based on
smoothed particle hydrodynamics (SPH) techniques \cite{Mo92}, which he
applied to relativistic spherical collapse \cite{Ma93}.  When tested
against 1D relativistic shock tubes all these codes performed similar
to the code of Wilson. More recently, Dean \etal \cite{DB93} have
applied flux correcting algorithms for the SRHD equations in the
context of heavy ion collisions. Recent developments in relativistic
SPH methods \cite{CM97,SR99} are discussed in Section~(\ref{ss:SPH}).

   A major break--through in the simulation of ultra--relativistic flows
was accomplished when high--resolution shock--capturing (HRSC)
methods, specially designed to solve hyperbolic systems of
conservations laws, were applied to solve the SRHD equations
\cite{MI91, MM92, Eu93, EM95}. This review is intended to provide a
comprehensive discussion of different HRSC methods and of related
methods used in SRHD.  Numerical methods for special relativistic MHD
flows are not included, because they are beyond the scope of this
review. However, we may include such a discussion in a future update
of this article.

\subsection{Plan of the Review}
%           ------------------
\label{ss:plan}

  The review is organized as follows. Section~(\ref{s:rhd}) contains a
derivation of the equations of special relativistic (perfect) fluid
dynamics, as well as a discussion of their main properties.  In
Section~(\ref{s:hrsc}) the most recent developments in numerical
methods for SRHD are reviewed paying particular attention to
high--resolution shock--capturing methods. Other developments in
special relativistic numerical hydrodynamics are discussed in
Section~(\ref{s:other}).  Numerical results obtained with different
methods as well as analytical solutions for several test problems are
presented in Section~(\ref{s:tests}). Two astrophysical applications
of SRHD are discussed in Section~(\ref{s:appl}). An evaluation of the
various numerical methods is given in Section~(\ref{s:concl}) together
with an outlook for future developments. Finally, some additional
technical information is presented in Section~(\ref{s:additional}).

  The reader is assumed to have basic knowledge in classical
\cite{LL87, CF76} and relativistic fluid dynamics \cite{Ta78, An89},
as well as in finite difference/volume methods for partial
differential equations \cite{Po77, OB87}.  A discussion of modern
finite volume methods for hyperbolic systems of conservation laws can
be found, \eg in \cite{Le92, Le97, La98}. The theory of spectral
methods for fluid dynamics is developed in \cite{CH88}, and smoothed
particle hydrodynamics (SPH) is reviewed in \cite{Mo92}.

\section{SPECIAL RELATIVISTIC HYDRO-DYNAMICS}
%        ##################################
\label{s:rhd}

\subsection{Equations}
%           ---------
\label{ss:eqs}

Using the Einstein summation convention the equations describing the
motion of a relativistic fluid are given by the five conservation laws
\begin{equation}
(\rho u^\mu)_{; \mu} = 0 \, ,
\label{cont}
\end{equation}
\begin{equation}
T^{\mu \nu}_{\,\,\,\, ; \nu} = 0  \, ,
\label{euler}
\end{equation}
where $(\mu, \nu = 0, \ldots, 3)$, and where ${;\mu}$ denotes the
covariant derivative with respect to coordinate $x^\mu$. Furthermore,
$\rho$ is the proper rest--mass density of the fluid, $u^\mu$ its
four--velocity, and $T^{\mu \nu}$ is the stress--energy tensor, which
for a perfect fluid can be written as
\begin{equation}
T^{\mu \nu} = \rho h u^\mu u^\nu + p g^{\mu \nu} \, .
\label{tmunu}
\end{equation}
Here, $g^{\mu \nu}$ is the metric tensor, $p$ the fluid pressure, and
and $h$ the specific enthalpy of the fluid defined by
\begin{equation}
h = 1 + \varepsilon + p/\rho \, ,
\label{enthalpy}
\end{equation}
where $\varepsilon$ is the specific internal energy. Note that we use
natural units (\ie the speed of light $c=1$) throughout this review.

In Minkowski space--time and Cartesian coordinates $(t,x^1,x^2,x^3)$,
the conservation equations (\ref{cont}), (\ref{euler}) can be written
in vector form as
\begin{equation}
\frac{\partial {\bf u}}{\partial t} +
\frac{\partial {\bf F}^i({\bf u})}{\partial x^i}=0 \, ,
\label{21}
\end{equation}
where $i = 1,2,3$. The state vector ${\bf u}$ is defined by 
\begin{equation}
{\bf u} = (D, S^1, S^2, S^3, \tau)^{\rm T}
\label{22}
\end{equation}
and the flux vectors ${\bf F}^i$ are given by 
\begin{equation}
{\bf F}^i = (Dv^i,\, S^1v^i+p \delta^{1i},\, S^2v^i+p \delta^{2i},\,
                     S^3v^i+p \delta^{3i},\, S^i-Dv^i)^{\rm T} \, .
\label{23}
\end{equation} 
The five conserved quantities $D$, $S^1, S^2, S^3$ and $\tau$ are the
rest--mass density, the three components of the momentum density, and
the energy density (measured relative to the rest mass energy
density), respectively. They are all measured in the laboratory frame,
and are related to quantities in the local rest frame of the fluid
(primitive variables) through
\begin{equation}
D=\rho W \, ,
\label{4}
\end{equation}
\begin{equation}
S^i=\rho h W^2 v^i \quad i = 1, 2, 3  \, ,
\label{5}
\end{equation}
\begin{equation}
\tau=\rho h W^2 - p - D \, ,
\label{6}
\end{equation}
where $v^i$ are the components of the three--velocity of the fluid
\begin{equation}
v^i = u^i/u^0
\label{v}
\end{equation}
and $W$ is the Lorentz factor  
\begin{equation}
W=u^0=\frac{1}{\sqrt{1-v^iv_i}} \, .
\label{lorentz}
\end{equation}
The system of equations (\ref{21}) with definitions (\ref{22}) --
(\ref{lorentz}) is closed by means of an equation of state (EOS),
which we shall assume to be given in the form
\begin{equation}
p=p\,(\rho,\varepsilon) \, . 
\label{12}
\end{equation}

In the non-relativistic limit (\ie $v \ll 1$, $h \rightarrow 1$) $D$,
$S^i$ and $\tau$ approach their Newtonian counterparts $\rho$, $\rho
v^i$ and $\rho E = \rho \varepsilon + \rho v^2/2$, and equations
(\ref{21}) reduce to the classical ones. In the relativistic case the
equations of system (\ref{21}) are strongly coupled via the Lorentz
factor and the specific enthalpy, which gives rise to numerical
complications (see Section\,\ref{ss:esrp}).

In classical numerical hydrodynamics it is very easy to obtain $v^i$
from the conserved quantities (\ie $\rho$ and $\rho v^i$). In the
relativistic case, however, the task to recover $(\rho, v^i, p)$ from
$(D,S^i,\tau)$ is much more difficult. Moreover, as state-of-the-art
SRHD codes are based on conservative schemes where the conserved
quantities are advanced in time, it is necessary to compute the
primitive variables from the conserved ones one (or even several)
times per numerical cell and time step making this procedure a crucial
ingredient of any algorithm.  (see Sect.~\ref{ss:lfq})

\subsection{SRHD as a hyperbolic system of conservation laws}
%           -----------------------------------------------
\label{ss:hscl}

    An important property of system (\ref{21}) is that it is
hyperbolic for causal EOS \cite{An89}. For hyperbolic systems of
conservation laws, the Jacobians $\partial {\bf F}^i({\bf u}) /
\partial {\bf u}$ have real eigenvalues and a complete set of
eigenvectors (see Sect.~\ref{ss:spectral}).  Information
about the solution propagates at finite velocities given by the
eigenvalues of the Jacobians. Hence, if the solution is known (in some
spatial domain) at some given time, this fact can be used to advance
the solution to some later time (initial value problem).  However, in
general, it is not possible to derive the exact solution for this
problem. Instead one has to rely on numerical methods which provide an
approximation to the solution. Moreover, these numerical methods must
be able to handle discontinuous solutions, which are inherent to
non--linear hyperbolic systems.

  The simplest initial value problem with discontinous data is called
a Riemann problem, where the one dimensional initial state consists of
two constant states separated by a discontinuity. The majority of
modern numerical methods, the so--called Godunov--type methods, are
based on exact or approximate solutions of Riemann problems.  Because
of its theoretical and numerical importance, we discuss the solution
of the special relativistic Riemann problem in the next subsection.

\subsection{Exact solution of the Riemann problem in SRHD}
%           -----------------------------------------------
\label{ss:esrp}

\begin{figure}
%\special{psfile=livrev_01_fig.ps hscale=75. vscale=75. hoffset=-25 voffset=-500
%         angle=0.}
%\vspace{16cm}
\centerline{
\epsfig{file=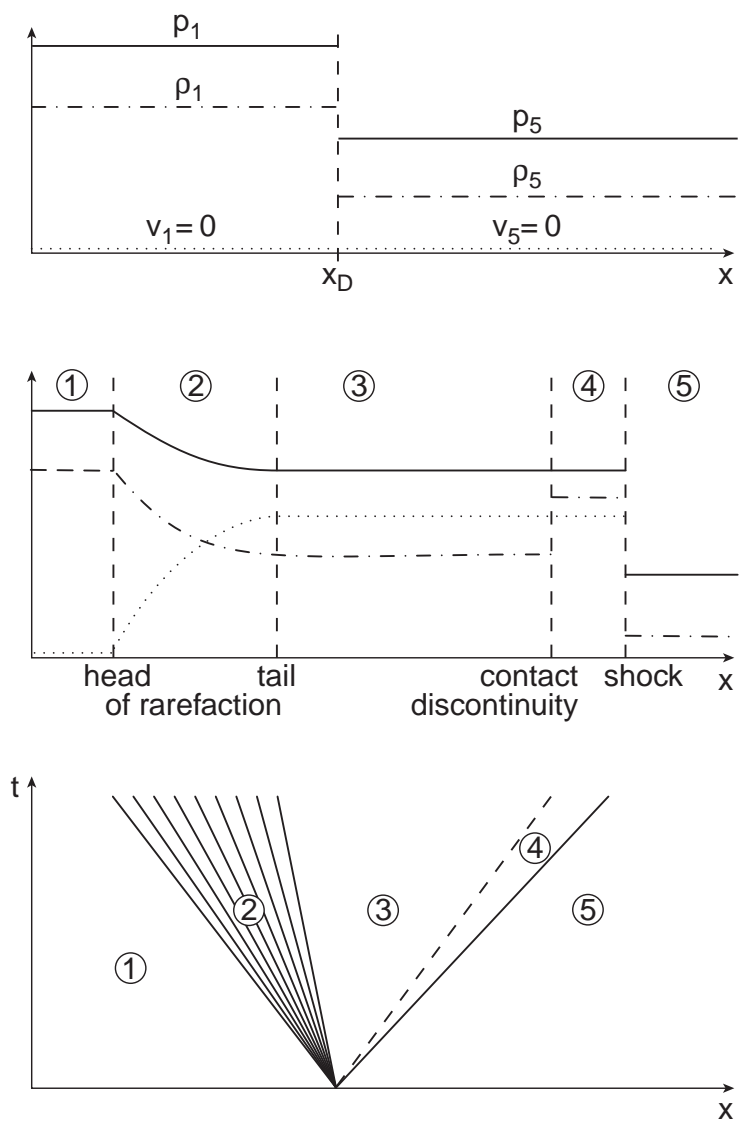, width=0.6\textwidth}}
\caption{\small Schematic solution of a Riemann problem in special
relativistic hydrodynamics.  The initial state at $t = 0$ (top figure)
consists of two constant states (1) and (5) with $p_1 > p_5$, $\rho_1
> \rho_5$ and $v_1 = v_2 = 0$ separated by a diaphragm at $x_{\rm
D}$. The evolution of the flow pattern once the diaphragm is removed
(middle figure) is illustrated in a space--time diagram (bottom
figure) with a shock wave (solid line) and a contact discontinuity
(dashed line) moving to the right. The bundle of solid lines
represents a rarefaction wave propagating to the left.
\label{f:rpbreakup}}
\end{figure}

  Let us first consider the one dimensional special relativistic flow
of an ideal gas with an adiabatic exponent $\gamma$ in the absence of
a gravitational field. The Riemann problem then consists of computing
the breakup of a discontinuity, which initially separates two
arbitrary constant states L (left) and R (right) in the gas (see
%% modified
Fig.~\ref{f:rpbreakup} with L $\equiv 1$ and R $\equiv 5$).  For
classical hydrodynamics the solution can be found, \eg in
\cite{CF76}. In the case of SRHD, the Riemann problem has been
considered by Mart\'{\i} \& M\"uller \cite{MM94}, who derived an exact
solution generalising previous results for particular initial data
\cite{To86}.

  The solution to this problem is self-similar, because it only
depends on the two constant states defining the discontinuity ${\bf
v}_{\rm L}$ and ${\bf v}_{\rm R}$, where ${\bf v} = (p,\rho,v)$, and
on the ratio $(x-x_0)/(t-t_0)$, where $x_0$ and $t_0$ are the initial
location of the discontinuity and the time of breakup, respectively.
Both in relativistic and classical hydrodynamics the discontinuity
decays into two elementary nonlinear waves (shocks or rarefactions)
which move in opposite directions towards the initial left and right
states. Between these waves two new constant states ${\bf v}_{\rm L*}$
and ${\bf v}_{\rm R*}$ 
%% modified
(note that ${\bf v}_{\rm L*} \equiv 3$ and ${\bf v}_{\rm R*} \equiv 4$
in Fig.~\ref{f:rpbreakup})
appear, which are separated from each other
through a contact discontinuity moving with the fluid. Across the
contact discontinuity the density exhibits a jump, whereas pressure
and velocity are continuous (see Figure~\ref{f:rpbreakup}).  As in the
classical case, the self-similar character of the flow through
rarefaction waves and the Rankine--Hugoniot conditions accross shocks
provide the relations to link the intermediate states ${\bf v}_{S*}$
($S=L,R$) with the corresponding initial states ${\bf v}_{S}$. They
also allow one to express the fluid flow velocity in the intermediate
states $v_{S*}$ as a function of the pressure $p_{S*}$ in these
states. Finally, the steadiness of pressure and velocity across the
contact discontinuity implies
\begin{equation}
  v_{\rm L*}(p_{*})=v_{\rm R*}(p_{*}) \, ,                                     
\label{39}                                                                     
\end{equation}   
where $p_*=p_{\rm L*}=p_{\rm R*}$, which closes the system.  The
functions $v_{S*}(p)$ are defined by
\begin{equation}                                                             
v_{S*}(p) = \left\{ \begin{array}{ll}                                          
                        {\cal R}^S(p) & \mbox{if $p \leq p_S$} \\              
                        {\cal S}^S(p) & \mbox{if $p > p_S$ \, ,}            
                        \end{array}                                            
                        \right.                                                
\end{equation}                                                         
where ${\cal R}^S(p)$ (${\cal S}^S(p)$) denotes the family of all
states which can be connected through a rarefaction (shock) with a
given state ${\bf v}_S$ ahead of the wave.

%%%%%%%%%%%%%%%%%%%%%%%%%%%%%%%%%%%%%%%%%%%%%%%%%%%%%%%%%%%%%%%%%%%%%%%%%%%%%%
%
%
%%%%%%%%%%%%%%%%%%%%%%%%%%%%%%%%%%%%%%%%%%%%%%%%%%%%%%%%%%%%%%%%%%%%%%%%%%%%%%

The fact that one Riemann invariant is constant through any
rarefaction wave provides the relation needed to derive the function
${\cal R}^S$
\begin{equation}                                                                           
{\cal R}^S(p)= \frac{(1+v_S)A_{\pm}(p)-(1-v_S)}                                 
                    {(1+v_S)A_{\pm}(p)+(1-v_S)}                                 
\label{41}                                                                      
\end{equation}                                                                            
with                                                                            
\begin{equation}                                                                            
A_{\pm}(p)=\left(\frac{\sqrt{\gamma-1}-c(p)}{\sqrt{\gamma-1}+c(p)}              
                 \frac{\sqrt{\gamma-1}+c_S}{\sqrt{\gamma-1}-c_S}                
           \right)^{\pm \frac{2}{\sqrt{\gamma -1}}}                             
\label{42}                                                                      
\end{equation}                                                                            
the $+$ ($-$) sign of $A_{\pm}$ corresponding to $S=L$ ($S=R$).  In
the above equation, $c_S$ is the sound speed of the state ${\bf v}_S$,
and $c(p)$ is given by
\begin{equation}                                                                            
c(p) = \left(\frac{\gamma (\gamma -1) p}{(\gamma - 1) \rho_S                    
(p/p_S)^{1/\gamma} + \gamma p}\right)^{1/2}  \, .                               
\end{equation}                                                                            
                                                                                
The family of all states ${\cal S}^S(p)$, which can be connected
through a shock with a given state ${\bf v}_S$ ahead of the wave, is
determined by the shock jump conditions.  One obtains
\begin{eqnarray}                                                                            
{\cal S}^S(p) & = &                                                                
\left(h_S W_S v_S \pm \frac{p-p_S}{j(p) \sqrt{1-V_{\pm}(p)^2}}\right) 
\nonumber \\          
 & & \left[h_S W_S + (p-p_S)                                                         
  \left(\frac{1}{\rho_S W_S} \pm \frac{v_S}{j(p) \sqrt{1-V_{\pm}(p)^2}}         
  \right)                                                                       
\right]^{-1}   \, ,                                                             
\label{43}                                                                      
\end{eqnarray}                                                                            
where the $+$ ($-$) sign corresponds to $S=R$ ($S=L$). $V_{\pm}(p)$
and $j(p)$ denote the shock velocity and the modulus of the mass flux
across the shock front, respectively.  They are given by
\begin{equation}                                                                            
V_{\pm}(p) = \frac{ \rho_S^2 W_S^2 v_S \pm j(p)^2                               
\sqrt{1+(\rho_S/j(p))^2}}{\rho_S^2 W_S^2 + j(p)^2}  \, ,                            
\label{44}                                                                      
\end{equation}                                                                            
and                                                                             
\begin{equation}                                                                            
j(p) = \sqrt{ \frac{p_S-p}{{\displaystyle \frac{h_S^2 - h(p)^2}{p_S -p}         
                                        - \frac{2h_S}{\rho_S}}} } \, ,              
\label{45}                                                                      
\end{equation}                                                                            
where the enthalpy $h(p)$ of the state behind the shock is the (unique)         
positive root of the quadratic equation                                         
\begin{equation}                                                                            
\left(1+\frac{(\gamma-1)(p_S-p)}{\gamma p} \right) h^2 -
\frac{(\gamma-1)(p_S-p)}{\gamma p} h +                                          
\frac{h_S (p_S-p)}{\rho_S} -h_S^2 = 0  \, ,                                     
\label{46}                                                                      
\end{equation}                                                                            
which is obtained from the Taub adiabat (the relativistic version of
the Hugoniot adiabat) for an ideal gas equation of state.

%%%%%%%%%%%%%%%%%%%%%%%%%%%%%%%%%%%%%%%%%%%%%%%%%%%%%%%%%%%%%%%%%%%%%%%%%%%%%%
%
\begin{figure}
%\special{psfile=livrev_02_fig.ps hscale=75. vscale=75. hoffset=-0
%         voffset=-260 angle=0.}
%\vspace{8cm}
\centerline{
\epsfig{file=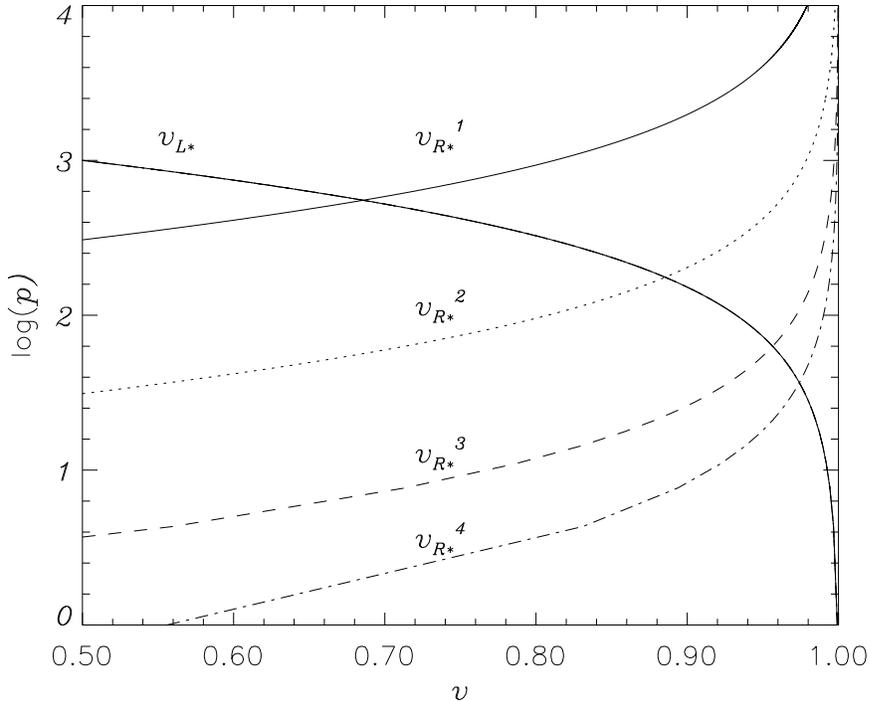, width=0.8\textwidth}}
\caption{\small Graphical solution in the $p$--$v$ plane of the
Riemann problems defined by the initial states $(p_{\rm L}=10^3,\,
\rho_{\rm L}=1,\, v_{\rm L}=0.5)$ and $(p_{\rm R}^i,\, \rho_{\rm
R}=1,\, v_{\rm R}=0)$ with $i = 1, 2, 3, 4$, where $p_{\rm R}^1=10^2$,
$p_{\rm R}^2=10$, $p_{\rm R}^3=1$, and $p_{\rm R}^4=10^{-1}$,
respectively.  The adiabatic index of the fluid is $5/3$ in all cases.
Note the asymptotic behavior of the functions as they approach $v=1$
(\ie the speed of light).
\label{f:pvdiagram}}
\end{figure}
%
%%%%%%%%%%%%%%%%%%%%%%%%%%%%%%%%%%%%%%%%%%%%%%%%%%%%%%%%%%%%%%%%%%%%%%%%%%%%%%
                                                                           
  The functions $v_{\rm L*}(p)$ and $v_{\rm R*}(p)$ are displayed in
Fig.~\ref{f:pvdiagram} in a $p$--$v$ diagram for a particular set of
Riemann problems. Once $p_*$ has been obtained, the remaining state
quantities and the complete Riemann solution, 
\begin{equation}
{\bf u} = {\bf u}(\frac{x-x_0}{t-t_0};{\bf u}_{\rm L},{\bf u}_{\rm R})),
\end{equation}
can easily be derived.

In Section\,\ref{ss:riemann} we provide a FORTRAN program called {\tt
RIEMANN}, which allows one to compute the exact solution of an
arbitrary special relativistic Riemann problem using the algorithm
just described.

The treatment of multidimensional special relativistic flows is
significantly more difficult than that of multidimensional Newtonian
flows. In SRHD all components (normal and tangential) of the flow
velocity are strongly coupled through the Lorentz factor, which
complicates the solution of the Riemann problem severely. For shock
waves, this coupling 'only' increases the number of algebraic jump
conditions, which must be solved. However, for rarefactions it implies
the solution of a system of ordinary differential equations
\cite{MM94}.

\section{HIGH-RESOLUTION SHOCK-CAPTURING \\ METHODS}
%        #######################################
\label{s:hrsc}

  The application of high-resolution shock-capturing (HRSC) methods
caused a revolution in numerical SRHD. These methods satisfy in a quite
natural way the basic properties required for any acceptable numerical
method: (i) high order of accuracy, (ii) stable and sharp description
of discontinuities, and (iii) convergence to the physically correct
solution.  Moreover, HRSC methods are conservative, and because of
their shock capturing property discontinuous solutions are treated
both consistently and automatically whenever and wherever they appear
in the flow.

  As HRSC methods are written in conservation form, the time evolution
of zone averaged state vectors is governed by some functions (the
numerical fluxes) evaluated at zone interfaces.  Numerical fluxes are
mostly obtained by means of an exact or approximate Riemann
solver. High resolution is usually achieved by using monotonic
polynomials in order to interpolate the approximate solutions within
numerical cells.

  Solving Riemann problems exactly involves time--consuming
computations, which are particularly costly in the case of
multidimensional SRHD due to the coupling of the equations through the
Lorentz factor (see Section\,\ref{ss:esrp}).  Therefore, as an
alternative, the usage of approximate Riemann solvers has been
proposed.

  In this Section we summarize the computation of the numerical fluxes
in a number of methods for numerical SRHD. Methods based on exact
Riemann solvers are discussed in Sections\,\ref{ss:rppm} and
\ref{ss:rglimm}, while those based on approximate solvers are
discussed in Sections\,(\ref{ss:twoshock} -- \ref{ss:marquina})
Readers not familiar with HRSC methods are referred to
Section\,\ref{ss:basicshrsc}, where the basic properties of these
methods as well as an outline of the recent developments are
described.

\subsection{Relativistic PPM}
%           ----------------
\label{ss:rppm}

  Mart\'{\i} \& M\"uller \cite{MM96} have used the procedure discussed
in Section\,\ref{ss:esrp} to construct an exact Riemann solver, which
they then incorporated in an extension of the PPM method \cite{CW84b}
for 1D SRHD. In their relativistic PPM method numerical fluxes are
calculated according to
\begin{equation}
{\widehat{\bf F}}^{\rm RPPM} = {\bf F}({\bf u}(0; {\bf u}_{\rm L}, 
                                                  {\bf u}_{\rm R})) \,,
\end{equation}
where ${\bf u}_L$ and ${\bf u}_R$ are approximations of the state
vector at the left and right side of a zone interface obtained by a
second--order accurate interpolation in space and time, and ${\bf
u}(0; {\bf u}_{\rm L}, {\bf u}_{\rm R})$ is the solution of the
Riemann problem defined by the two interpolated states at the position of
the initial discontinuity.

  The PPM interpolation algorithm described in \cite{CW84b} gives
monotonic conservative parabolic profiles of variables within a
numerical zone. In the relativistic version of PPM the original
interpolation algorithm is applied to zone averaged values of the
primitive variables ${\bf v} = (p, \rho, v)$, which are obtained from
zone averaged values of the conserved quantities ${\bf u}$.  For each
zone $j$, the quartic polynomial with zone--averaged values $a_{j-2}$,
$a_{j-1}$, $a_j$, $a_{j+1}$, and $a_{j+2}$ (where $a = \rho, p, v$) is
used to interpolate the structure inside the zone. In particular, the
values of $a$ at the left and right interface of the zone, $a_{L,j}$
and $a_{R,j}$, are obtained this way. These reconstructed values are
then modified such that the parabolic profile, which is uniquely
determined by $a_{L,j}$, $a_{R,j}$ and $a_j$, is monotonic inside the
zone.

  The time--averaged fluxes at an interface $j+1/2$ separating zones
$j$ and $j+1$ are computed from two spatially averaged states, ${\bf
v}_{j+\frac{1}{2}, L}$ and ${\bf v}_{j+\frac{1}{2}, R}$ at the left
and right side of the interface, respectively. These left and right
states are constructed taking into account the characteristic
information reaching the interface from both sides during the time
step. In the relativistic version of PPM the same procedure as in
\cite{CW84b} has been followed using the characteristic speeds and
Riemann invariants of the equations of relativistic hydrodynamics.

\subsection{Relativistic Glimm's method}
%           ---------------------------
\label{ss:rglimm}
                                                                  
  Wen \etal \cite{WP97} have extended Glimm's random choice method
\cite{Gl65} to 1D SRHD. They developed a first--order accurate
hydrodynamic code combining Glimm's method (using an exact Riemann
solver) with standard finite difference schemes.

  In the random choice method, given two adjacent states, ${\bf
u}_j^n$ and $ {\bf u}_{j+1}^n$, at time $t^n$, the value of the
numerical solution at time $t^{n+1/2}$ and position $x_{j+1/2}$ is given
by the exact solution ${\bf u}(x, t)$ of the Riemann problem evaluated
at a randomly chosen point inside zone $(j, j+1)$, \ie
\begin{equation}
  {\bf u}_{j+\frac{1}{2}}^{n+\frac{1}{2}} = 
  {\bf u} \left( \frac{(j+\xi_n) \Delta x}{(n+\frac{1}{2}) \Delta t};\,
                 {\bf u}_j^n ,\, {\bf u}_{j+1}^n \right) \, ,
\end{equation}
where $\xi_n$ is a random number in the interval $[0,1]$.

  Besides being conservative on average, the main advantages of
Glimm's method are that it produces both completely sharp shocks and
contact discontinuities, and that it is free of diffusion and
dispersion errors.

  Chorin \cite{Ch76} applied Glimm's method to the numerical solution
of homogeneous hyperbolic conservation laws. Colella \cite{Co82}
proposed an accurate procedure of randomly sampling the solution of
local Riemann problems and investigated the extension of Glimm's
method to two dimensions using operator splitting methods.

\subsection{Two-shock approximation for relativistic hydrodynamics}
%           ------------------------------------------------------
\label{ss:twoshock}

  This approximate Riemann solver is obtained from a relativistic
extension of Colella's method \cite{Co82} for classical fluid
dynamics, where it has been shown to handle shocks of arbitrary
strength \cite{Co82, WC84}. In order to construct Riemann solutions
in the two--shock approximation one analytically continues shock waves
towards the rarefaction side (if present) of the zone interface
instead of using an actual rarefaction wave solution. Thereby one gets
rid of the coupling of the normal and tangential components of the
flow velocity (see Section\,\ref{ss:esrp}), and the remaining minor
algebraic complications are the Rankine--Hugoniot conditions across
oblique shocks. Balsara \cite{Ba94} has developed an approximate
relativistic Riemann solver of this kind by solving the jump
conditions in the shocks' rest frames in the absence of transverse
velocities, after appropriate Lorentz transformations. Dai \& Woodward
\cite{DW97} have developed a similar Riemann solver based on the jump
conditions across oblique shocks making the solver more efficient.

  Table~\ref{t:twoshock} gives the converged solution for the
intermediate states obtained with both Balsara's and Dai's \&
Woodward's procedure for the case of the Riemann problems defined in
Section\,\ref{ss:blast} (involving strong rarefaction waves) together
with the exact solution. Despite the fact that both approximate
methods involve very different algebraic expressions, their results
differ by less than $2\%$.  However, the discrepancies are much larger
when compared with the exact solution (up to a $100\%$ error in the
density of the left intermediate state in Problem 2). The accuracy of
the two--shock approximation should be tested in the ultra-relativistic
limit, where the approximation can produce large errors in the Lorentz
factor (in the case of Riemann problems involving strong
rarefaction waves) with important implications for the fluid dynamics.
Finally, the suitability of the two--shock approximation for Riemann
problems involving transversal velocities still needs to be tested.

%%%%%%%%%%%%%%%%%%%%%%%%%%%%%%%%%%%%%%%%%%%%%%%%%%%%%%%%%%%%%%%%%%%%%%%%%%%%
%
\begin{table}[htb]
\begin{center}
\caption{\protect \small Pressure $p_*$, velocity $v_*$ and densities
$\rho_{\rm L*}$ (left), $\rho_{\rm R*}$ (right) for the intermediate
state obtained for the two--shock approximation of Balsara \cite{Ba94}
(B) and of Dai \& Woodward \cite{DW97} (DW) compared to the exact
solution (Exact) for the Riemann problems defined in
Sect.~\ref{ss:blast}.
\label{t:twoshock}}
\begin{tabular}{rcccc}
\\ \hline \\
Method     & $p_*$     & $v_*$    & $\rho_{\rm L*}$ & $\rho_{\rm R*}$ \\
\\ \hline \\
Problem\,1 &           &            &           &           \\
\\
B          & 1.440E+00 & 7.131E--01 & 2.990E+00 & 5.069E+00 \\
DW         & 1.440E+00 & 7.131E--01 & 2.990E+00 & 5.066E+00 \\
Exact      & 1.445E+00 & 7.137E--01 & 2.640E+00 & 5.062E+00 \\
\\ \hline \hline \\
Problem\,2 &           &            &            &           \\
\\
B          & 1.543E+01 & 9.600E--01 & 7.325E--02 & 1.709E+01 \\
DW         & 1.513E+01 & 9.608E--01 & 7.254E--02 & 1.742E+01 \\
Exact      & 1.293E+01 & 9.546E--01 & 3.835E--02 & 1.644E+01 \\
\\ \hline
\end{tabular}
\end{center}
\end{table}
%
%%%%%%%%%%%%%%%%%%%%%%%%%%%%%%%%%%%%%%%%%%%%%%%%%%%%%%%%%%%%%%%%%%%%%%%%%%%%%

\subsection{Roe-type relativistic solvers}
%           -----------------------------
\label{ss:roetype}

  Linearized Riemann solvers are based on the exact solution of
Riemann problems of a modified system of conservation equations
obtained by a suitable linearization of the original system. This idea
was put forward by Roe \cite{Ro81}, who developed a linearized Riemann
solver for the equations of ideal (classical) gas dynamics.  Eulderink
\etal \cite{Eu93, EM95} have extended Roe's Riemann solver to the
general relativistic system of equations in arbitrary spacetimes.
Eulderink uses a local linearization of the Jacobian matrices of the
system fulfilling the properties demanded by Roe in his original
paper.

Let ${\cal B} = \partial {\bf F} / \partial{\bf u}$ be the Jacobian
matrix associated with one of the fluxes ${\bf F}$ of the original
system, and ${\bf u}$ the vector of unknowns.  Then, the locally
constant matrix $\widetilde{\cal B}$, depending on ${\bf u}_{\rm L}$
and ${\bf u}_{\rm R}$ (the left and right state defining the local
Riemann problem) must have the following four properties:
\begin{itemize}
\item[] 1. It constitutes a linear mapping from the vector space ${\bf
u}$ to the vector space ${\bf F}$.
\item[] 2. As ${\bf u}_{\rm L} \rightarrow {\bf u}_{\rm R} \rightarrow
{\bf u},\; \widetilde{\cal B} ({\bf u}_{\rm L}, {\bf u}_{\rm R})
\rightarrow {\cal B}({\bf u})$.
\item[] 3. For any ${\bf u}_{\rm L}$, ${\bf u}_{\rm R},\; 
\widetilde{\cal B}({\bf u}_{\rm L},{\bf u}_{\rm R}) ({\bf u}_{\rm R}-
{\bf u}_{\rm L}) = {\bf F}({\bf u}_{\rm R}) - {\bf F}({\bf u}_{\rm L})$.
\item[] 4. The eigenvectors of $\widetilde{\cal B}$ are linearly independent.
\end{itemize}
Conditions\,1 and 2 are necessary if one is to recover smoothly the
linearized algorithm from the nonlinear version. Condition\,3
(supposing 4 is fulfilled) ensures that if a single discontinuity is
located at the interface, then the solution of the linearized problem
is the exact solution of the nonlinear Riemann problem.

  Once a matrix, $\widetilde{\cal B}$, satisfying Roe's conditions has
been obtained for every numerical interface, the numerical fluxes are
computed by solving the locally linear system. Roe's numerical flux is
then given by
\begin{equation}
 {\widehat{\bf F}}^{\rm ROE} = \frac{1}{2} 
   \left[{\bf F}({\bf u}_{\rm L}) + {\bf F}({\bf u}_{\rm R}) 
         - \sum_{p} |\widetilde{\lambda}^{(p)}| \widetilde{\alpha}^{(p)}
                     \widetilde{\bf r}^{(p)}\right] \, ,
\label{froe}
\end{equation}
with
\begin{equation}
  \widetilde{\alpha}^{(p)} = \widetilde{\bf l}^{(p)} \cdot 
                             ({\bf u}_{\rm R} - {\bf u}_{\rm L}) \, ,
\label{froe2}
\end{equation}
where $\widetilde{\lambda}^{(p)}$, $\widetilde{\bf r}^{(p)}$, and
$\widetilde{\bf l}^{(p)}$ are the eigenvalues and the right and left
eigenvectors of $\widetilde{\cal B}$, respectively ($p$ runs from 1 to
the number of equations of the system).

  Roe's linearization for the relativistic system of equations in a
general spacetime can be expressed in terms of the average state
\cite{Eu93, EM95}
\begin{equation}
  \widetilde{\bf w} = \frac{ {\bf w}_{\rm L} + {\bf w}_{\rm R}}
                             {k_{\rm L} + k_{\rm R}} 
\end{equation}
with
\begin{equation}
  {\bf w} = (ku^0, ku^1, ku^2, ku^3, k\frac{p}{\rho h}) 
\end{equation}
and
\begin{equation}
  k^2 = \sqrt{-g} \rho h \, ,
\end{equation}
where $g$ is the determinant of the metric tensor $g_{\mu \nu}$.  The
role played by the density ${\rho}$ in case of the Cartesian
non-relativistic Roe solver as a weight for averaging, is taken over in
the relativistic variant by $k$, which apart from geometrical factors
tends to $\rho$ in the non-relativistic limit. A Riemann solver for
special relativistic flows and the generalization of Roe's solver to
the Euler equations in arbitrary coordinate systems are easily deduced
from Eulderink's work. The results obtained in 1D test problems for
ultra-relativistic flows (up to Lorentz factors 625) in the presence of
strong discontinuities and large gravitational background fields
demonstrate the excellent performance of the Eulderink-Roe solver
\cite{EM95}.

  Relaxing condition\,3 above, Roe's solver is no longer exact for
shocks but still produces accurate solutions, and moreover, the
remaining conditions are fulfilled by a large number of averages.  The
1D general relativistic hydrodynamic code developed by Romero \etal
\cite{RI96} uses flux formula (\ref{froe}) with an arithmetic average
of the primitive variables at both sides of the interface. It has
successfully passed a long series of tests including the spherical
version of the relativistic shock reflection (see
Section\,\ref{ss:rsh}).

  Roe's original idea has been exploited in the so--called local
characteristic approach (see, \eg \cite{Ye89}). This approach relies
on a local linearization of the system of equations by defining at
each point a set of characteristic variables, which obey a system of
uncoupled scalar equations. This approach has proven to be very
successful, because it allows for the extension to systems of scalar
nonlinear methods. Based on the local characteristic approach are the
methods developed by Marquina \etal \cite{MM92} and Dolezal \& Wong
\cite{DW95}, which both use high--order reconstructions of the
numerical characteristic fluxes, namely PHM \cite{MM92} and ENO
\cite{DW95} (see Section~\ref{ss:basicshrsc}).

\subsection{Falle and Komissarov upwind scheme}
%           ----------------------------------
\label{ss:fksolver}

  Instead of starting from the conservative form of the hydrodynamic
equations, one can use a primitive--variable formulation in
quasi-linear form
\begin{equation}
  \frac{\partial {\bf v}}{\partial t} + 
  {\cal A} \frac{\partial {\bf v}}{\partial x} = 0 \, , 
\label{fkquasi}
\end{equation}
where ${\bf v}$ is any set of primitive variables. A local
linearization of the above system allows one to obtain the solution of
the Riemann problem, and from this the numerical fluxes needed to
advance a conserved version of the equations in time.

  Falle \& Komissarov \cite{FK96} have considered two different
algorithms to solve the local Riemann problems in SRHD by extending the
methods devised in \cite{Fa91}. In a first algorithm, the intermediate
states of the Riemann problem at both sides of the contact
discontinuity, ${\bf v}_{L*}$ and ${\bf v}_{R*}$, are obtained by
solving the system
\begin{equation}
  {\bf v}_{\rm L*} = {\bf v}_{\rm L} + b_{\rm L} {\bf r}^-_{\rm L}\,, \;\; 
  {\bf v}_{\rm R*} = {\bf v}_{\rm R} + b_{\rm R} {\bf r}^+_{\rm R}\,,
\end{equation}
where ${\bf r}^-_{\rm L}$ is the right eigenvector of ${\cal A}({\bf
v}_{\rm L})$ associated with sound waves moving upstream and ${\bf
r}^+_{\rm R}$ is the right eigenvector of ${\cal A}({\bf v}_{\rm R})$
of sound waves moving downstream. The continuity of pressure and of
the normal component of the velocity across the contact discontinuity
allows one to obtain the wave strengths $b_L$ and $b_R$ from the above
expressions, and hence the linear approximation to the intermediate
state ${\bf v}_*({\bf v}_{\rm L},{\bf v}_{\rm R})$.

  In the second algorithm proposed by Falle \& Komissarov \cite{FK96},
a linearization of system (\ref{fkquasi}) is obtained by constructing
a constant matrix $\widetilde{\cal A} ({\bf v}_{\rm L}, {\bf v}_{\rm R})
= {\cal A} (\frac{1}{2} ({\bf v}_{\rm L} + {\bf v}_{\rm R}))$. The
solution of the corresponding Riemann problem is that of a linear
system with matrix $\widetilde{\cal A}$, \ie
\begin{equation}
 {\bf v}_* = {\bf v}_{\rm L} + \sum_{\widetilde{\lambda}^{(p)}<0} 
             \widetilde{\alpha}^{(p)}\, \widetilde{\bf r}^{(p)} \, ,
\end{equation}
or, equivalently,
\begin{equation} 
  {\bf v}_* = {\bf v}_{\rm R} - \sum_{\widetilde{\lambda}^{(p)}>0}
              \widetilde{\alpha}^{(p)}\, \widetilde{\bf r}^{(p)})\, ,
\end{equation}
with 
\begin{equation}
  \widetilde{\alpha}^{(p)} = \widetilde{\bf l}^{(p)} \cdot 
                             ({\bf v}_{\rm R} - {\bf v}_{\rm L}) \, , 
\end{equation}
where $\widetilde{\lambda}^{(p)}$, $\widetilde{\bf r}^{(p)}$, and
$\widetilde{\bf l}^{(p)}$ are the eigenvalues and the right and left
eigenvectors of $\widetilde{\cal A}$, respectively ($p$ runs from 1 to
the number of equations of the system).
 
  In both algorithms, the final step involves the computation of the
numerical fluxes for the conservation equations
\begin{equation}
  {\widehat{\bf F}}^{\rm FK} =
     {\bf F}({\bf u}({\bf v}_*({\bf v}_{\rm L},{\bf v}_{\rm R})))\, .
\end{equation}

\subsection{Relativistic HLL Method} 
%           -----------------------
\label{ss:rhll}

Schneider \etal \cite{SK93} have proposed to use the method of Harten,
Lax \& van Leer (HLL hereafter \cite{HL83}) to integrate the equations
of SRHD.  This method avoids the explicit calculation of the
eigenvalues and eigenvectors of the Jacobian matrices and is based on
an approximate solution of the original Riemann problems with a single
intermediate state
\begin{equation}
{\bf u}^{\rm HLL}(x/t;{\bf u}_{\rm L},{\bf u}_{\rm R}) = 
                        \left\{ \begin{array}{ll}
                        {\bf u}_{\rm L} & \mbox{for $x<a_{\rm L}t$} \\
                        {\bf u}_* & \mbox{for $a_{\rm L}t \leq x \leq
                        a_{\rm R}t$} \\
                        {\bf u}_{\rm R} & \mbox{for $x>a_{\rm R}t$}
                        \end{array}                                            
                        \right.  \, ,                                          
\end{equation}
where $a_{\rm L}$ and $a_{\rm R}$ are lower and upper bounds for the
smallest and largest signal velocities, respectively. The intermediate
state ${\bf u}_*$ is determined by requiring consistency of the
approximate Riemann solution with the integral form of the
conservation laws in a grid zone.  The resulting integral average of
the Riemann solution between the slowest and fastest signals at some
time is given by
\begin{equation}
  {\bf u}_* = \frac{  a_{\rm R}{\bf u}_{\rm R}
                    - a_{\rm L}{\bf u}_{\rm L}
                    - {\bf F}({\bf u}_{\rm R})
                    + {\bf F}({\bf u}_{\rm L})
                   }{a_{\rm R}-a_{\rm L}} \, ,
\end{equation}
and the numerical flux by 
\begin{equation}
  {\widehat{\bf F}}^{\rm HLL} = 
     \frac{  a_{\rm R}^{+}{\bf F}({\bf u}_{\rm L}) 
           - a_{\rm L}^{-}{\bf F}({\bf u}_{\rm R}) 
           + a_{\rm R}^{+} a_{\rm L}^{-} ({\bf u}_{\rm R}-{\bf u}_{\rm L})
          }{a_{\rm R}^{+}-a_{\rm L}^{-}} \, ,
\end{equation}
where
\begin{equation}
  a_{\rm L}^{-} = {\rm min}\{0,a_{\rm L}\}\,, \;\;
  a_{\rm R}^{+} = {\rm max}\{0,a_{\rm R}\}\, .
\end{equation}

  An essential ingredient of the HLL scheme are good estimates for the
smallest and largest signal velocities. In the non-relativistic case,
Einfeldt \cite{Ei88} proposed calculating them based on the smallest
and largest eigenvalues of Roe's matrix. This HLL scheme with
Einfeldt's recipe is a very robust upwind scheme for the Euler
equations and possesses the property of being positively conservative.
The method is exact for single shocks, but it is very dissipative,
especially at contact discontinuities.

Schneider \etal \cite{SK93} have presented results in 1D
ultra-relativistic hydrodynamics using a version of the HLL method with
signal velocities given by
\begin{equation}
  a_{\rm R} = (\bar{v}+\bar{c}_s)/(1+\bar{v}\bar{c}_s) \, ,
\end{equation}
\begin{equation}
  a_{\rm L} = (\bar{v}-\bar{c}_s)/(1-\bar{v}\bar{c}_s) \, ,
\end{equation}
where $c_s$ is the relativistic sound speed, and where the bar denotes
the arithmetic mean between the initial left and right states. Duncan
\& Hughes \cite{DH94} have generalized this method to 2D SRHD and
applied it to the simulation of relativistic extragalactic jets.

\subsection{Marquina's flux formula}
%           -----------------------
\label{ss:marquina}

  Godunov--type schemes are indeed very robust in most situations
although they fail spectacularly on occasions. Reports on approximate
Riemann solver failures and their respective corrections (usually a
judicious addition of artificial dissipation) are abundant in the
literature \cite{Qu94}. Motivated by the search for a robust and
accurate approximate Riemann solver that avoids these common failures,
Donat \& Marquina \cite{DM96} have extended to systems a numerical
flux formula which was first proposed by Shu \& Osher \cite{SO89} for
scalar equations. In the scalar case and for characteristic wave speeds
which do not change sign at the given numerical interface, Marquina's
flux formula is identical to Roe's flux. Otherwise, the scheme
switches to the more viscous, entropy satisfying local Lax--Friedrichs
scheme \cite{SO89}. In the case of systems, the combination of Roe and
local--Lax--Friedrichs solvers is carried out in each characteristic
field after the local linearization and decoupling of the system of
equations \cite{DM96}. However, contrary to Roe's and other linearized
methods, the extension of Marquina's method to systems is not based on
any averaged intermediate state.

  Mart\'{\i} \etal have used this method in their simulations of
relativistic jets \cite{MM95, MM97}. The resulting numerical code has
been successfully used to describe ultra-relativistic flows in both one
and two spatial dimensions with great accuracy (a large set of test
calculations using Marquina's Riemann solver can be found in
Appendix~II of \cite{MM97}). Numerical experimentation in two
dimensions confirms that the dissipation of the scheme is sufficient
to eliminate the carbuncle phenomenon \cite{Qu94}, which appears in
high Mach number relativistic jet simulations when using other
standard solvers \cite{DF98}. Aloy \etal \cite{AI99a} have implemented
Marquina's flux formula in their three dimensional relativistic
hydrodynamic code GENESIS.  Font \etal \cite{FM99} have developed a 3D
general relativistic hydro code where the matter equations are
integrated in conservation form and fluxes are calculated with
Marquina's formula.

\subsection{Symmetric TVD schemes with nonlinear numerical dissipation}
%           ----------------------------------------------------------
\label{ss:syms}

  The methods discussed in the previous subsections are all based on
exact or approximate solutions of Riemann problems at cell interfaces
in order to stabilize the discretization scheme across strong shocks.
Another successful approach relies on the addition of nonlinear
dissipation terms to standard finite difference methods. The algorithm
of Davis \cite{Da84} is based on such an approach. It can be
interpreted as a Lax--Wendroff scheme with a conservative TVD
dissipation term. The numerical dissipation term is local, free of
problem dependent parameters and does not require any characteristic
information. This last fact makes the algorithm extremely simple when
applied to any hyperbolic system of conservation laws.

  A relativistic version of Davis' method has been used by Koide \etal
\cite{KN96, Ko97, NK98} in 2D and 3D simulations of relativistic
magneto--hydrodynamic jets with moderate Lorentz factors. Although the
results obtained are encouraging, the coarse grid zoning used in these
simulations and the relative smallness of the beam flow Lorentz factor
(4.56, beam speed $\approx 0.98c$) does not allow for a comparison
with Riemann--solver--based HRSC methods in the ultra-relativistic
limit.

\section{OTHER DEVELOPMENTS}
%        ##################
\label{s:other}

\subsection{Van Putten's approach}
%           ---------------------

  Relying on a formulation of Maxwell's equations as a hyperbolic
system in divergence form, van Putten \cite{vP91} has devised a
numerical method to solve the equations of relativistic ideal MHD in
flat spacetime \cite{vP93}. Here we only discuss the basic principles
of the method in one spatial dimension. In van Putten's approach, the
state vector ${\bf u}$ and the fluxes ${\bf F}$ of the conservation
laws are decomposed into a spatially constant mean (subscript 0) and a
spatially dependent variational (subscript 1) part
\begin{equation}
 {\bf u}(t,x) = {\bf u}_0(t) + {\bf u}_1(t,x)\, , \;\;
 {\bf F}(t,x) = {\bf F}_0(t) + {\bf F}_1(t,x)\, .
\end{equation}
The RMHD equations then become a system of evolution equations for the
integrated variational parts ${\bf u_1}^*$, which reads
\begin{equation}
 \frac{\partial {\bf u_1}^*}{\partial t} + {\bf F_1} = 0 \, ,
\label{vp}
\end{equation}
together with the conservation condition 
\begin{equation}
  \frac{d{\bf F}_0}{dt} = 0 \, .
\end{equation}
The quantities ${\bf u_1}^*$ are defined as
\begin{equation}
  {\bf u_1}^*(t,x) = \int^x {\bf u_1}(t,y) \,\, dy \, . 
\end{equation}  
They are continuous and standard methods can be used to integrate the
system (\ref{vp}). Van Putten uses a leapfrog method.

   The new state vector ${\bf u}(t,x)$ is then obtained from ${\bf
u_1}^*(t,x)$ by numerical differentiation.  This process can lead to
oscillations in the case of strong shocks and a smoothing algorithm
should be applied. Details of this smoothing algorithm and of the
numerical method in one and two spatial dimensions can be found in
\cite{vP92} together with results on a large variety of tests.

  Van Putten has applied his method to simulate relativistic
hydrodynamic and magneto hydrodynamic jets with moderately flow
Lorentz factors ($< 4.25$) \cite{vP93b, vP96}.

\subsection{Relativistic SPH}
%           ----------------
\label{ss:SPH}

Besides finite volume schemes, another completely different method is
widely used in astrophysics for integrating the hydrodynamic
equations. This method is Smoothed Particle Hydrodynamics, or SPH for
short (\cite{Lu77, GM77, Mo92}).  The fundamental idea of SPH is to
represent a fluid by a Monte Carlo sampling of its mass elements.  The
motion and thermodynamics of these mass elements is then followed as
they move under the influence of the hydrodynamics equations.  Because
of its Lagrangian nature there is no need within SPH for explicit
integration of the continuity equation, but in some implementations of
SPH this is done nevertheless for certain reasons. As both the
equation of motion of the fluid and the energy equation involve
continuous properties of the fluid and their derivatives, it is
necessary to estimate these quantities from the positions, velocities
and internal energies of the fluid elements, which can be thought of
as particles moving with the flow.  This is done by treating the
particle positions as a finite set of interpolating points where the
continuous fluid variables and their gradients are estimated by an
appropriately weighted average over neighbouring particles. Hence, SPH
is a free-Lagrange method, \ie spatial gradients are evaluated without
the use of a computational grid.

A comprehensive discussion of SPH can be found in the reviews of
Hernquist \& Katz \cite{HK89}, Benz \cite{Be90} and Monaghan
\cite{Mo85, Mo92}. The non-relativistic SPH equations are briefly
discussed in Section\,(\ref{ss:SPHeqs}).  The capabilities and limits
of SPH are explored, \eg in \cite{SM93, TT98}, and the stability of
the SPH algorithm is investigated in \cite{SH95}.

The SPH equations for special relativistic flows have been first
formulated by Monaghan \cite{Mo85}. For such flows the SPH equations
given in Section\,(\ref{ss:SPHeqs}) can be taken over except that each
SPH particle $a$ carries $\nu_a$ baryons instead of mass $m_a$
\cite{Mo85, CM97}. Hence, the rest mass of particle $a$ is given by
$m_a = m_0 \nu_a$, where $m_0$ is the baryon rest mass (if the fluid
is made of baryons).  Transforming the notation used in \cite{CM97} to
ours, the continuity equation, the momentum and the total energy
equations for particle $a$ are given by (unit of velocity is $c$)
\begin{equation}
 {d N_a \over d t} = -\sum_b  \nu_b (\vecva - \vecvb) 
                              \cdot \nabla_a W_{ab}  \; ,
\label{SPH-06}
\end{equation}
\begin{equation}
  {d \Shata \over d t}   =  - \sum_b \nu_b 
                              \rund{  {p_a \over N_a^2}
                                    + {p_b \over N_b^2}
                                    + \tpiab } \cdot \nabla_a W_{ab} \; ,
\label{SPH-07}
\end{equation}
and
\begin{equation}
  {d \thata \over d t}   =  - \sum_b \nu_b 
                              \rund{  {p_a \vecva \over N_a^2}
                                    + {p_b \vecvb \over N_b^2}
                                    + \tomab } \cdot \nabla_a W_{ab} \; ,
\label{SPH-08}
\end{equation}
respectively. Here, the summation is over all particles other than
particle $a$, and $d/dt$ denotes the Lagrangian time derivative. $N =
D / m_0$ is the baryon number density,
\begin{equation}
 \Shat \equiv \frac{\vecS}{N}  = m_0 h W \vecv  
\label{SPH-09}
\end{equation}
the momentum per particle, and 
\begin{equation}
 \that \equiv \frac{\tau}{N} + m_0 = m_0 h W - \frac{p}{N}
\label{SPH-10}
\end{equation}
the total energy per particle (all measured in the laboratory frame).
The momentum density $\vecS \equiv (S^1, S^2, S^3)^T$, the energy
density $\tau$ (measured in units of the rest mass energy density),
and the specific enthalpy $h$ are defined in
Section\,(\ref{ss:eqs}). $\tpiab$ and $\tomab$ are the SPH dissipation
terms, and $\nabla_a W_{ab}$ denotes the gradient of the kernel
$W_{ab}$ (see Section\,(\ref{ss:SPHeqs}) for more details).

Special relativistic flow problems have been simulated with SPH by
\cite{La89, KM90, Ma91, Ma93, CM97, SR99}. Extensions of SPH capable
of treating general relativistic flows have been considered by
\cite{KM90, LM93, SR99}. Concerning relativistic SPH codes the
artificial viscosity is the most critical issue. It is required to
handle shock waves properly, and ideally it should be predicted
by a relativistic kinetic theory for the fluid. However, unlike its
Newtonian analogue, the relativistic theory has not yet been developed
to the degree required to achieve this.  For Newtonian SPH Lattanzio
\etal \cite{LM86} have shown that a viscosity quadratic in the
velocity divergence is necessary in high Mach number flows. They
proposed a form such that the viscous pressure could be simply added
to the fluid pressure in the equation of motion and the energy
equation. Because this simple form of the artificial viscosity has
known limitations, they also proposed a more sophisticated form of the
artificial viscosity terms, which leads to a modified equation of
motion. This artificial viscosity works much better, but it cannot be
generalized to the relativistic case in a consistent way.  Utilizing
an equation for the specific internal energy both Mann \cite{Ma91} and
Laguna \etal \cite{LM93} use such an inconsistent formulation. Their
artificial visocity term is not included into the expression of the
specific relativistic enthalpy.  In a second approach, Mann \cite{Ma91} 
allows for a time--dependent smoothing length and
SPH particle mass, and further proposed a SPH variant based on the
total energy equation. Lahy \cite{La89} and Siegler \& Riffert
\cite{SR99} use a consistent artificial viscosity pressure added to
the fluid pressure. Siegler \& Riffert \cite{SR99} have also
formulated the hydrodynamic equations in conservation form.
  
Monaghan \cite{Mo97} incorporates concepts from Riemann solvers into
SPH. For this reason he also proposes to use a total energy equation
in SPH simulation instead of the commonly used internal energy
equation, which would involve time derivatives of the Lorentz factor
in the relativistic case.  Chow \& Monaghan \cite{CM97} have extended
this concept and have proposed an SPH algorithm, which gives good
results when simulating an ultra-relativistic gas.  In both cases the
intention was not to introduce Riemann solvers into the SPH algorithm,
but to use them as a guide to improve the artificial viscosity
required in SPH.

In Roe's Riemann solver \cite{Ro81}, as well as in its relativistic
variant proposed by Eulerdink \cite{Eu93, EM95} (see
Section\,\ref{ss:roetype}), the numerical flux is computed by solving a
locally linear system and depends on both the eigenvalues and (left
and right) eigenvectors of the Jacobian matrix associated to the
fluxes and on the jumps in the conserved physical variables (see
Eqs.\,\ref{froe} and \ref{froe2}).  Monaghan \cite{Mo97} realized that
an appropriate form of the dissipative terms $\tpiab$ and $\tomab$ for
the interaction between particles $a$ and $b$ can be obtained by
treating the particles as the equivalent of left and right states
taken with reference to the line joining the particles. The quantity
corresponding to the eigenvalues (wave propagation speeds) is an
appropriate signal velocity $v_{sig}$ (see below), and that equivalent
to the jump across characteristics is a jump in the relevant physical
variable.  For the artificial viscosity tensor, $\tpiab$, Monaghan
\cite{Mo97} assumes that the jump in velocity across characteristics
can be replaced by the velocity difference between $a$ and $b$ along
the line joining them.

With these considerations in mind Chow \& Monaghan \cite{CM97}
proposed for $\tpiab$ in the relativistic case the form
\begin{equation}
 \tpiab = -\frac{K v_{sig} (\Shatas - \Shatbs) \cdot \vecj}{
                 \overline{N}_{ab} } \; ,
\label{SPH-11}
\end{equation}
when particles $a$ and $b$ are approaching, and $\tpiab = 0$
otherwise.  Here $K = 0.5$ is a dimensionless parameter, which is
chosen to have the same value as in the non-relativistic case
\cite{Mo97}.  $\overline{N}_{ab} = (N_{a} + N_{b})/2$ is the average
baryon number density, which has to be present in (\ref{SPH-11}),
because the pressure terms in the summation of (\ref{SPH-01}) have an
extra density in the denominator arising from the SPH interpolation.
Furthermore,
\begin{equation}
 \vecj = \frac{\vecrab}{|\vecrab|}
\label{SPH-12}
\end{equation}
is the unit vector from $b$ to $a$, and 
\begin{equation}
  \Shats  =  m_0 h \Ws \vecv  \; ,
\label{SPH-13}
\end{equation}
where
\begin{equation}
  \Ws  = \frac{1}{\sqrt{1 - (\vecv \cdot \vecj)^2} } \; .
\label{SPH-14}
\end{equation}
Using instead of $\Shat$ (see Eq.~\ref{SPH-09}) the modified momentum
$\Shats$, which involves the line of sight velocity $\vecv \cdot
\vecj$, guarantees that the viscous dissipation is positive definite
\cite{CM97}. 

The dissipation term in the energy equation is derived in a similar
way and is given by \cite{CM97}
\begin{equation}
 \tomab = -\frac{K v_{sig} (\thatas - \thatbs)\, \vecj}{\overline{N}_{ab}}\; ,
\label{SPH-15}
\end{equation}
if $a$ and $b$ are approaching, and $\tomab = 0$ otherwise.  $\tomab$
involves the energy $\thats$, which is identical to $\that$ (see
Eq.~\ref{SPH-10}) except that $W$ is replaced by $\Ws$.

To determine the signal velocity Chow \& Monaghan \cite {CM97} (and
Monaghan \cite{Mo97} in the non-relativistic case) start from the
(local) eigenvalues, and hence the wave velocities $(v \pm c_s)/(1 \pm
vc_s)$ and $v$ of one--dimensional relativistic hydrodynamic
flows. Again considering particles $a$ and $b$ as the left and right
states of a Riemann problem with respect to motions along the line
joining the particles, the appropriate signal velocity is the speed of
approach (as seen in the computing frame) of the signal sent from $a$
towards $b$ and that from $b$ to $a$. This is the natural speed for
the sharing of physical quantities, because when information about the
two states meets it is time to construct a new state. This speed of
approach should be used when determining the size of the time step by
the Courant condition (for further details see \cite{CM97}).

Chow \& Monaghan \cite{CM97} have demonstrated the performance of
their Riemann problem guided relativistic SPH algorithm by calculating
several shock tube problems involving ultra-relativistic speeds up to
$v = 0.9999$. The algorithm gives good results, but finite volume
schemes based on Riemann solvers give more accurate results and can
handle even larger speeds (see Section\,\ref{s:tests}).

\subsection{Relativistic beam scheme}
%           -------------------------
\label{ss:beam}

  Sanders \& Prendergast \cite{SP74} proposed an explicit scheme to
solve the equilibrium limit of the non-relativistic Boltzmann equation,
\ie the Euler equations of Newtonian fluid dynamics. In their
so--called beam scheme the Maxwellian velocity distribution function
is approximated by several Dirac delta functions or discrete beams of
particles in each computational cell, which reproduce the appropriate
moments of the distribution function. The beams transport mass,
momentum and energy into adjacent cells, and their motion is followed
to first--order accuracy. The new (\ie time advanced) macroscopic
moments of the distribution function are used to determine the new
local non-relativistic Maxwell distribution in each cell. The entire
process is then repeated for the next time step.  The CFL stability
condition requires that no beam of gas travels farther than one cell
in one time step.  This beam scheme, although being a particle method
derived from a microscopic kinetic description, has all the desirable
properties of modern characteristic--based wave propagating methods
based on a macroscopic continuum description.

The non--relativistic scheme of Sanders \& Prendergast \cite{SP74} has
been extended to relativistic flows by Yang \etal \cite{YC97}. They
replaced the Maxwellian distribution function by its relativistic
analogue, \ie by the more complex J\"uttner distribution function,
which involves modified Bessel functions. For three--dimensional
flows the J\"uttner distribution function is approximated by seven
delta functions or discrete beams of particles, which can viewed as
dividing the particles in each cell into seven distinct groups.  In
the local rest frame of the cell these seven groups represent
particles at rest and particles moving in $\pm x, \pm y$ and $\pm z$
directions, respectively. 

Yang \etal \cite{YC97} show that the integration scheme for the beams
can be cast in the form of an upwind conservation scheme in terms of
numerical fluxes. They further show that the beam scheme not only
splits the state vector but also the flux vectors, and has some
entropy--satisfying mechanism embedded as compared with approximate
relativistic Riemann solver \cite{DW95, SK93} based on Roe's method
\cite{Ro81}. The simplest relativistic beam scheme is only
first--order accurate in space, but can be extended to higher--order
accuracy in a straightforward manner.  Yang \etal consider three
high--order accurate variants (TVD2, ENO2, ENO3) generalizing their
approach developed in \cite{YH92, YH95} for Newtonian gas dynamics,
which is based on the essentially non-oscillatory (ENO) piecewise
polynomial reconstruction scheme of Harten \etal \cite{HE87}.

Yang \etal \cite{YC97} present several numerical experiments including
relativistic one--dimen\-sional shock tube flows and the simulation of
relativistic two--dimensional Kelvin--Helmholtz instabilities.  The
shock tube experiments consist of a mildly relativistic shock tube,
relativistic shock heating of a cold flow, the relativistic blast wave
interaction of Woodward \& Colella \cite {WC84} (see
Section\,\ref{sss:RBWI}), and the perturbed relativistic shock tube
flow of Shu \& Osher \cite{SO89}.

\section{SUMMARY OF METHODS}
%        ##################
\label{s:sum_meth}

  This Section contains a summary of all the methods reviewed in the two 
preceding sections as well as several FCT and artificial viscosity codes.
The main characteristic of the codes (dissipation algorithm, spatial and 
temporal orders of accuracy, reconstruction techniques) are listed in two 
tables (Table~\ref{t:methods} for HRSC codes; Table~\ref{t:methods_other} for 
other approaches).

%%%%%%%%%%%%%%%%%%%%%%%%%%%%%%%%%%%%%%%%%%%%%%%%%%%%%%%%%%%%%%%%%%%%%%%
%
{\small

\begin{table}[]
\begin{center}
\caption{High-resolution shock-capturing methods. All the codes rely on a 
conservation form of the RHD equations with the exception of ref.~\cite{WP97}}
\bigskip
\label{t:methods}
{\footnotesize

\begin{tabular}{ll}
\hline\\[-0.8ex]
Code                                 & Basic characteristics \\[0.6ex]
\hline\\[-0.4ex]

Roe type-l \cite{MI91,RI96,FM99}     & Riemann solver of Roe type with arithmetic averaging; monotonicity \\
                                     & preserving, linear reconstruction of primitive variables; 2nd order time \\
                                     & stepping (\cite{MI91,RI96}: predictor-corrector;
                                       \cite{FM99}: standard scheme) \\
[0.6ex]
Roe-Eulderink \cite{Eu93}            & Linearized Riemann solver based on Roe averaging; 2nd order \\
                                     & accuracy in space and timere \\
[0.6ex]
HLL-l \cite{SK93}                    & Harten-Lax-van Leer approximate Riemann solver; monotonic linear \\
                                     & reconstruction of conserved/primitive variables; 2nd order accuracy \\
                                     & in space and time \\
[0.6ex]
LCA-phm  \cite{MM92}                 & Local linearization and decoupling of the system; PHM reconstruction \\
                                     & of characteristic fluxes; 3rd
                                     order TVD preserving RK method for \\
                                     & time stepping \\
[0.6ex]
LCA-eno \cite{DW95}                  & Local linearization and decoupling of the system; high order ENO \\
                                     & reconstruction of
                                     characteristic split fluxes; high
                                     order TVD \\
                                     & preserving RK methods for time stepping \\
[0.6ex]
rPPM  \cite{MM96}                    & Exact (ideal gas) Riemann solver; PPM reconstruction of primitive \\
                                     & variables; 2nd order accuracy
                                     in time by averaging states in
                                     the \\
                                     & domain of dependence of zone interfaces \\
[0.6ex]
Falle-Komissarov \cite{FK96}         & Approximate Riemann solver based on local linearizations of the \\
                                     & RHD equations in primitive
                                     form; monotonic linear
                                     reconstruction \\
                                     & of $p$, $\rho$  and $u^i$; 2nd order predictor-corrector time stepping \\
[0.6ex]
MFF-ppm  \cite{MM97,AI99a}           & Marquina flux formula for numerical flux computation; PPM \\
                                     & reconstruction of primitive variables; 2nd and 3rd order TVD \\
                                     & preserving RK methods for time stepping \\
[0.6ex]
MFF-eno/phm \cite{DF98}              & Marquina flux formula for numerical flux computation; upwind biased \\
                                     & ENO/PHM reconstruction of characteristic fluxes; 2nd and 3rd order \\
                                     & TVD preserving RK methods for time stepping \\
[0.6ex]
MFF-l \cite{FM99}                    & Marquina flux formula for numerical flux computation; monotonic \\
                                     & linear reconstruction of
                                     primitive variables; standard 2nd
                                     order \\
                                     & finite difference algorithms for time stepping \\
[0.6ex]
Flux split \cite{FM99}               & TVD flux-split second order method \\
[0.5ex]
sTVD \cite{KN96}                     & Davis (1984) symmetric TVD scheme with nonlinear numerical \\
                                     & dissipation; 2nd order accuracy in space and time \\
[0.6ex]
rGlimm \cite{WP97}                   & Glimm's method applied to RHD equations in primitive form; 1st \\ 
                                     & order accuracy in space and time \\
[0.6ex]
rBS \cite{YC97}                      & Relativistic beam scheme solving equilibrium limit of relativistic \\
                                     & Boltzmann equation; distribution function approximated by discrete \\
                                     & beams of particles reproducing appropriate moments; 1st and 2nd \\
                                     & order TVD, 2nd and 3rd order ENO schemes \\
[0.6ex]\hline \\
\end{tabular}
}
\end{center}
\end{table}
}
%
%%%%%%%%%%%%%%%%%%%%%%%%%%%%%%%%%%%%%%%%%%%%%%%%%%%%%%%%%%%%%%%%%%%%%%%%%%%%%

%%%%%%%%%%%%%%%%%%%%%%%%%%%%%%%%%%%%%%%%%%%%%%%%%%%%%%%%%%%%%%%%%%%%%%%%%%%%%%
%
{\small

\begin{table}[]
\begin{center}
\caption{Code characteristics}
\bigskip
\label{t:methods_other}
{\footnotesize
\begin{tabular}{ll}
\hline\\[-0.8ex]
Code                                         & Basic characteristics \\[0.6ex]
\hline\\[-0.2ex]

\multicolumn{2}{c}{Artificial viscosity}\\[0.6ex]

\hline\\[-0.4ex]

AV-mono \cite{CW84,HS84,MW89}       & Non-conservative formulation of the RHD equations (transport\\
                                    & differencing, internal energy equation); artificial viscosity \\
                                    & extra term in the momentum flux; monotonic 2nd order \\
                                    & transport differencing; explicit time stepping \\
[0.6ex]
cAV-implicit \cite{NW86}            & Non-conservative formulation of the RHD equations; internal \\
                                    & energy equation; consistent formulation of artificial viscosity; \\
                                    & adaptive mesh and implicit time stepping \\
[0.6ex]
\hline\\[-0.2ex]

\multicolumn{2}{c}{Flux corrected transport}\\[0.6ex]

\hline\\[-0.4ex]

FCT-lw \cite{Du91}                  & Non-conservative formulation of the RHD equations (transport\\
                                    & differencing, equation for $\rho h W$); explicit 2nd order \\
                                    & Lax-Wendroff scheme with FCT algorithm \\
[0.6ex]
SHASTA-c \cite{SK93,DB93,DB94}      & FCT algorithm based on SHASTA \cite{BB73}; advection of \\
                                    & conserved variables \\
[0.6ex]
\hline\\[0.2ex]

\multicolumn{2}{c}{van Putten's approach}\\[0.6ex]

\hline\\[-0.4ex]

van Putten \cite{vP93}              & Ideal RMHD equations in constraint-free, divergence form; \\
                                    & evolution of integrated variational parts of conserved quantities; \\
                                    & smoothing algorithm in numerical differentiation step; \\
                                    & leap-frog method for time stepping \\
[0.6ex]
\hline\\[-0.2ex]

\multicolumn{2}{c}{Smooth particle hydrodynamics}\\[0.6ex]

\hline\\[-0.4ex]

SPH-AV-0 \cite{Ma91}(SPH0), \cite{LM93} & Specific internal energy equation; artificial viscosity extra terms\\ 
                                    & in momentum and energy equations; 2nd order time stepping \\
                                    & (\cite{Ma91}: predictor-corrector; \cite{LM93}: RK method) \\
[0.6ex]
SPH-AV-1 \cite{Ma91}(SPH1)          & Time derivatives in SPH equations include variations in smoothing \\
                                    & length and mass per particle; Lorentz factor terms treated more \\
                                    & consistently; otherwise same as SPH-AV-0 \\
[0.6ex]
SPH-AV-c \cite{Ma91}(SPH2)          & Total energy equation; otherwise same as SPH-AV-1 \\
[0.6ex]
SPH-cAV-c \cite{SR99}               & RHD equations in conservation form; consistent formulation of \\
                                    & artificial viscosity \\
[0.6ex]
SPH-RS-c \cite{CM97}                & RHD equations in conservation form; dissipation terms constructed \\
                                    & in analogy to terms in Riemann solver based methods \\
[0.6ex]
\hline
\end{tabular}
}
\end{center}
\end{table}
}
%
%%%%%%%%%%%%%%%%%%%%%%%%%%%%%%%%%%%%%%%%%%%%%%%%%%%%%%%%%%%%%%%%%%%%%%%%%%%%%

\section{TEST BENCH}
%        ##########
\label{s:tests}

\subsection{Relativistic shock heating in planar, cylindrical 
%           -------------------------------------------------
            and spherical geometry}
%           ----------------------
\label{ss:rsh}

  Shock heating of a cold fluid in planar, cylindrical or spherical
geometry has been used since the early developments of numerical
relativistic hydrodynamics as a test case for hydrodynamic codes,
because it has an analytical solution (\cite{BM76} in planar symmetry;
\cite{MM97} in cylindrical and spherical symmetry), and because it
involves the propagation of a strong relativistic shock wave.

\begin{figure}
%\special{psfile=livrev_03_fig.ps hscale=100. vscale=100. hoffset=20
%          voffset=-300 angle=0.}  
%\vspace{10cm} 
\centerline{
\epsfig{file=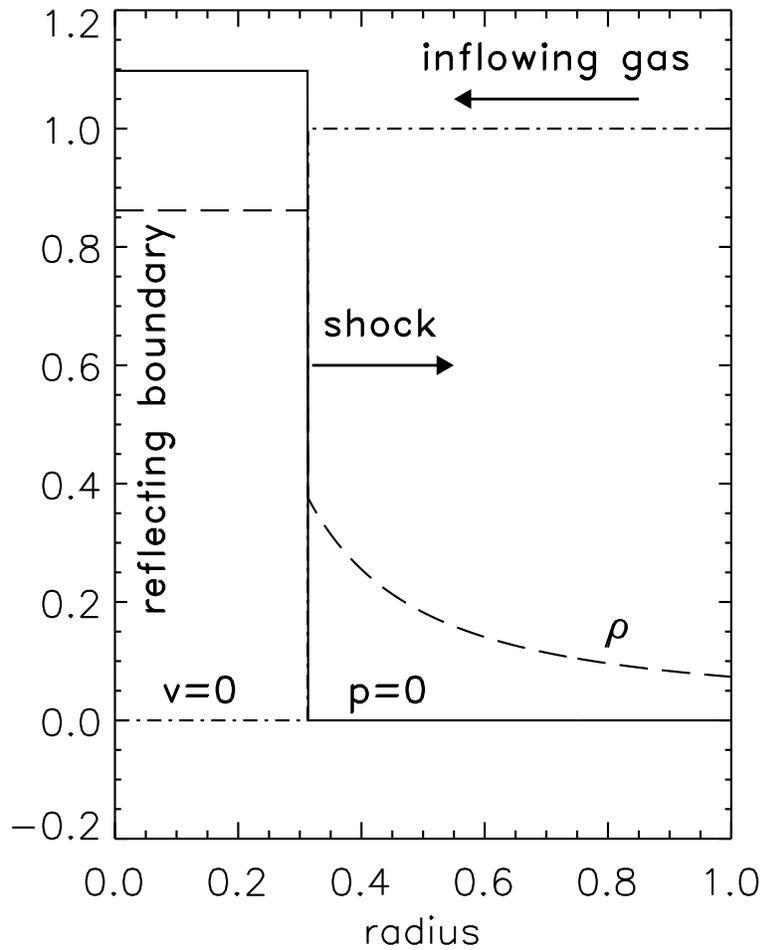, width=0.7\textwidth}}
\caption{Schematic solution of the shock heating problem in spherical
geometry. The initial state consists of a spherically symmetric flow
of cold ($p=0$) gas of unit rest mass density having a coordinate
velocity $v_1=-1$ everywhere.  A shock is generated at the center of the
sphere, which propagates upstream with constant speed. The post--shock
state is constant and at rest.  The pre--shock state, where the flow
is self--similar, has a density which varies as $\rho = (1 + t/r)^2$
with time $t$ and radius $r$.
\label{f:shockheating_an}}
\end{figure}

  In planar geometry, an initially homogeneous, cold (\ie $\varepsilon
\approx 0$) gas with coordinate velocity $v_1$ and Lorentz factor
$W_1$ is supposed to hit a wall, while in the case of cylindrical and
spherical geometry the gas flow converges towards the axis or the
center of symmetry. In all three cases the reflection causes
compression and heating of the gas as kinetic energy is converted into
internal energy.  This occurs in a shock wave, which propagates
upstream. Behind the shock the gas is at rest ($v_2 = 0$).  Due to
conservation of energy across the shock the gas has a specific
internal energy given by
\begin{equation}
  \varepsilon_2 = W_1 - 1 \, .
\end{equation}
The compression ratio of shocked and unshocked gas, $\sigma$, follows from
\begin{equation}
 \sigma = \frac{\gamma + 1}{\gamma - 1} + 
          \frac{\gamma}{\gamma - 1}\, \varepsilon_2 \, ,
\end{equation}
where $\gamma$ is the adiabatic index of the equation of state.  The
shock velocity is given by
\begin{equation}
  V_s = \frac{(\gamma-1) W_1 |v_1|}{W_1 + 1} \, .
\end{equation}
In the unshocked region ($r \in [V_st, \infty [$\,) the pressureless
gas flow is self-similar and has a density distribution given by
\begin{equation}
  \rho (t,r) = \left(1+\frac{|v_1|t}{r}\right)^{\alpha} \rho_0 \, ,
\end{equation}
where $\alpha = 0, 1, 2$ for planar, cylindrical or spherical
geometry, and where $\rho_0$ is the density of the inflowing gas at
infinity (see Fig.\,\ref{f:shockheating_an}).

%%%%%%%%%%%%%%%%%%%%%%%%%%%%%%%%%%%%%%%%%%%%%%%%%%%%%%%%%%%%%%%%%%%%%%%%%%%%%%%%
%
%
%%%%%%%%%%%%%%%%%%%%%%%%%%%%%%%%%%%%%%%%%%%%%%%%%%%%%%%%%%%%%%%%%%%%%%%%%%%%%%%%

  In the Newtoninan case the compression ratio $\sigma$ of shocked and
unshocked gas cannot exceed a value of $\sigma_{\rm max} = (\gamma +
1)/(\gamma - 1)$ independently of the inflow velocity. This is
different for relativistic flows, where $\sigma$ grows linearly with
the flow Lorentz factor and becomes infinite as the inflowing gas
velocity approaches to speed of light.

  The maximum flow Lorentz factor achievable for a hydrodynamic code
with acceptable errors in the compression ratio $\sigma$ is a measure
of the code's quality.  Table\,\ref{t:shockheat} contains a summary of
the results obtained for the shock heating test by various authors.

  Explicit finite--difference techniques based on a non--conservative 
formulation of the hydrodynamic equations and on non-consistent artificial 
viscosity \cite{CW84,HS84} are able to handle flow Lorentz factors up to 
$\approx 10$ with moderately large errors ($\sigma_{\rm error} \approx 1 - 3\%$) 
at best \cite{WM88, MW89}. Norman \& Winkler \cite{NW86} got very good results 
($\sigma_{\rm error} \approx 0.01$\% for a flow Lorentz factor of 10 using 
consistent artificial viscosity terms and an implicit adaptive--mesh method.

  The performance of explicit codes improved significantly when
numerical methods based on Riemann solvers were introduced \cite{MI91,
MM92, Eu93, SK93, EM95, MM96, FK96}. For some of these codes the
maximum flow Lorentz factors is only limited by the precision by
which numbers are represented on the computer used for the simulation
\cite{DW95, WP97, AI99a}. 

  Schneider \etal \cite{SK93} have compared the accuracy of a code based on the 
relativistic HLL Riemann solver with different versions of relativistic FCT 
codes for inflow Lorentz factors in the range 1.6 to 50. They found that the 
error in $\sigma$ was reduced by a factor of two when using HLL.  

  Within SPH methods, Chow \& Monaghan \cite{CM97} have obtained 
results comparable to those of HRSC methods ($\sigma_{\rm error} <2\,10^{-3}$) 
for flow Lorentz factors up to 70, using a relativistic SPH code with 
Riemann solver guided dissipation. Sieglert \& Riffert \cite{SR99}
have succeded in reproducing the post-shock state accurately for inflow Lorentz
factors of 1000 with a code based on a consistent formulation of artificial
viscosity. However, the disspation introduced by SPH methods at the shock 
transition is very large (10-12 particles in the code of ref.~\cite{SR99}; 
20-24 in the code of ref.~\cite{CM97}) compared with the typical dissipation
of HRSC methods (see below).

%%%%%%%%%%%%%%%%%%%%%%%%%%%%%%%%%%%%%%%%%%%%%%%%%%%%%%%%%%%%%%%%%%%%%%%%%%%%%%
%
{\small

\begin{table}[]
\begin{center}
\caption{\protect \small Summary of relativistic shock heating test
calculations by various authors in planar ($\alpha = 0$), cylindrical
($\alpha = 1$), and spherical ($\alpha = 2$) geometry.  $W_{\rm max}$
and $\sigma_{\rm error}$ are the maximum inflow Lorentz factor and
compression ratio error extracted from tables and figures of the
corresponding reference.  $W_{\rm max}$ should only be considered as
indicative of the maximum Lorentz factor achievable by every
method. Methods are described in Sects.~\ref{s:hrsc} and \ref{s:other} and their
basic properties summarized in Sect.~\ref{s:sum_meth} 
(Tables~\ref{t:methods},\ref{t:methods_other}).
\label{t:shockheat}}
\begin{tabular}{lclcc}
\\ \hline \\
References & $\alpha$ & Method & $W_{\rm max}$ & $\sigma_{\rm error}$ [\%] \\
\\ \hline \\
Centrella \& Wilson (1984) \cite{CW84}    & 0 & AV-mono          & 2.29          & $\approx 10$ \\  
Hawley \etal (1984)        \cite{HS84}    & 0 & AV-mono          & 4.12          & $\approx 10$ \\
Norman \& Winkler (1986)   \cite{NW86}    & 0 & cAV-implicit     & 10.0          & 0.01         \\
McAbee \etal (1989)        \cite{MW89}    & 0 & AV-mono          & 10.0          & 2.6          \\
Mart\'{\i} \etal (1991)    \cite{MI91}    & 0 & Roe type-l       & 23            & 0.2          \\
Marquina \etal (1992)      \cite{MM92}    & 0 & LCA-phm          & 70            & 0.1          \\
Eulderink (1993)           \cite{Eu93}    & 0 & Roe-Eulderink    & 625           & $\leq 0.1^{a}$\\
Schneider \etal (1993)     \cite{SK93}    & 0 & HLL-l            & $10^6$        & 0.2$^{b}$    \\
                                          & 0 & SHASTA-c         & $10^6$        & 0.5$^{b}$    \\
Dolezal \& Wong (1995)     \cite{DW95}    & 0 & LCA-eno          & 7.0 $\, 10^5$ & $\leq 0.1^{a}$\\
Mart\'{\i} \& M\"uller (1996) \cite{MM96} & 0 & rPPM             & 224           & 0.03         \\
Falle \& Komissarov (1996) \cite{FK96}    & 0 & Falle-Komissarov & 224           & $\leq 0.1^{a}$\\
Romero \etal (1996)        \cite{RI96}    & 2 & Roe type-l       & 2236          & 2.2          \\
Mart\'{\i} \etal (1997)    \cite{MM97}    & 1 & MFF-ppm          & 70            & 1.0          \\
Chow \& Monaghan (1997)    \cite{CM97}    & 0 & SPH-RS-c         & 70            & 0.2          \\
Wen \etal (1997)           \cite{WP97}    & 2 & rGlimm           & 224           & $10^{-9}$    \\
Donat \etal (1998)         \cite{DF98}    & 0 & MFF-eno          & 224           & $\leq 0.1^{a}$\\
Aloy \etal (1999)          \cite{AI99a}   & 0 & MFF-ppm          & 2.4 $\, 10^5$ & 3.5$^{c}$    \\
Sieglert \& Riffert (1999) \cite{SR99}    & 0 & SPH-cAV-c        & 1000          & $\leq 0.1^{a}$\\
\\ \hline 
\end{tabular}
\end{center}
$^{a}$ Estimated from figures. \\
$^{b}$ For $W_{\rm max}$ = 50. \\
$^{c}$ Including points at shock transition.\\
\end{table}
}
%
%%%%%%%%%%%%%%%%%%%%%%%%%%%%%%%%%%%%%%%%%%%%%%%%%%%%%%%%%%%%%%%%%%%%%%%%%%%%%%

%%%%%%%%%%%%%%%%%%%%%%%%%%%%%%%%%%%%%%%%%%%%%%%%%%%%%%%%%%%%%%%%%%%%%%%%%%%%%%
%
\begin{figure}
%
%%%---MOVIE---  livrev_mov1.mpg
%
%\vspace{3cm}
\centerline{
\epsfig{file=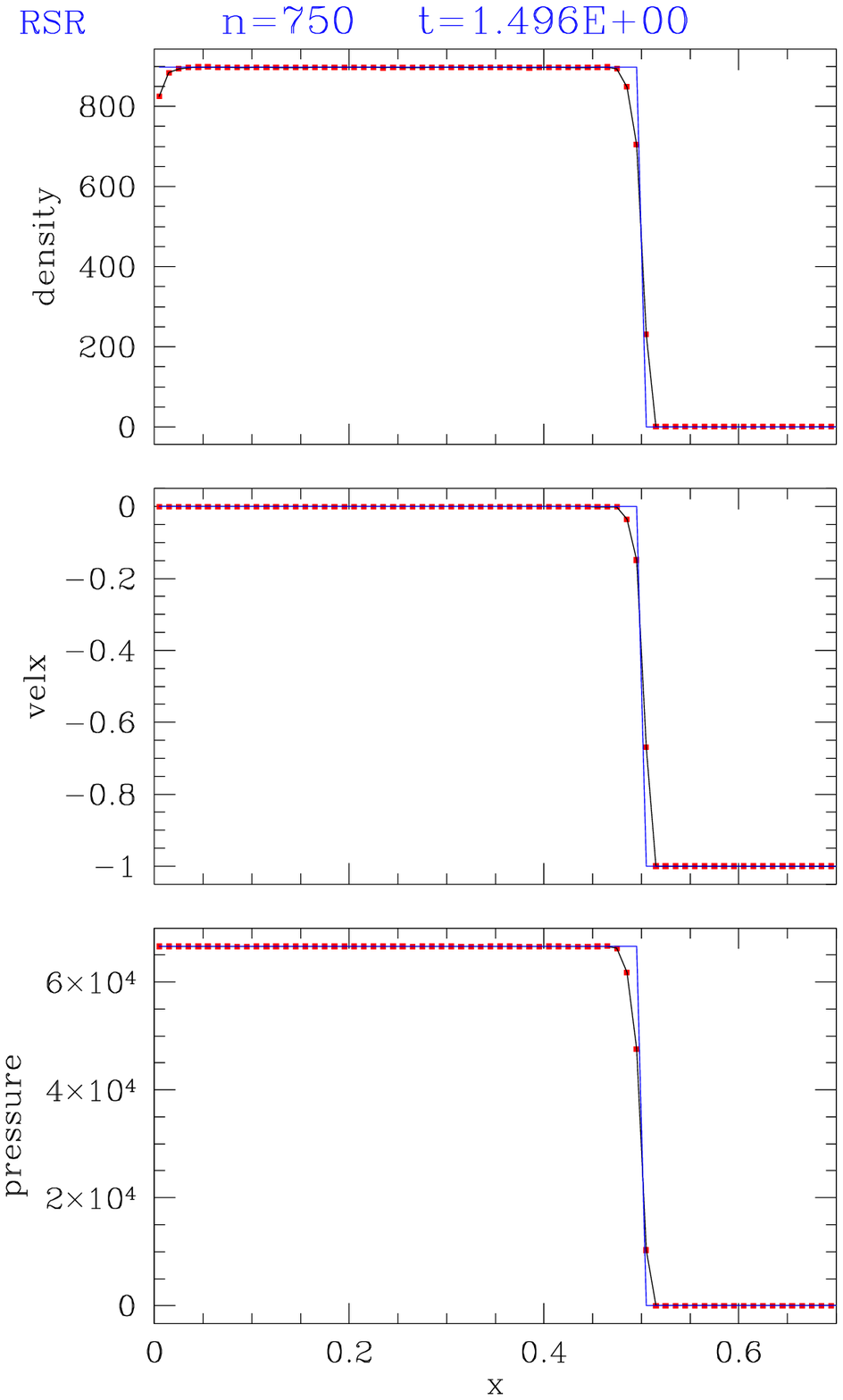, width=0.6\textwidth}}
\caption{\small MPEG movie 
({\it only final frame of movie is displayed!})
showing the evolution of the density distribution for the shock
heating problem with an inflow velocity $v_1 = -0.99999c$ in Cartesian
coordinates.  The reflecting wall is located at $x=0$.  The adiabatic
index of the gas is 4/3. For numerical reasons, the specific internal
energy of the inflowing cold gas is set to a small finite value
($\varepsilon_1 = 10^{-7}\,W_1$).  The final frame of the movie also
shows the analytical solution (blue lines).  The simulation has been
performed on an equidistant grid of 100 zones.}
\label{shock_heating_p}
\end{figure}
%
%%%%%%%%%%%%%%%%%%%%%%%%%%%%%%%%%%%%%%%%%%%%%%%%%%%%%%%%%%%%%%%%%%%%%%%%%%%%%%

  The performance of a HRSC method based on a relativistic Riemann
solver is illustrated by means of a MPEG movie 
({\it final frame of movie is displayed in Fig.\,\ref{shock_heating_p}})
for the planar shock heating problem for an inflow velocity $v_1 =
-0.99999c$ ($W_1 \approx 223$). These results are obtained with the
relativistic PPM code of \cite{MM96}, which uses an exact Riemann
solver based on the procedure described in Section~\ref{ss:esrp}.

  The shock wave is resolved by three zones and there are no
post--shock numerical oscillations. The density increases by a factor
$\approx 900$ across the shock.  Near $x=0$ the density distribution
slightly undershoots the analytical solution (by $\approx 8\%$) due to
the numerical effect of wall heating. The profiles obtained for other
inflow velocities are qualitatively similar. The mean relative error
of the compression ratio $\sigma_{\rm error} < 10^{-3}$, and, in
agreement with other codes based on a Riemann solver, the accuracy of
the results does not exhibit any significant dependence on the Lorentz
factor of the inflowing gas.

  Some authors have considered the problem of shock heating in
cylindrical or spherical geometry using adapted coordinates to test
the numerical treatment of geometrical factors \cite{RI96, MM97,
WP97}.  Aloy \etal \cite{AI99a} have considered the spherically
symmetric shock heating problem in 3D Cartesian coordinates as a test
case for both the directional splitting and the symmetry properties of
their code GENESIS. The code is able to handle this test up to inflow Lorentz 
factors of the order of 700.

  In the shock reflection test conventional schemes often give
numerical approximations which exhibit a consistent $O(1)$ error for
the density and internal energy in a few cells near the reflecting
wall. This 'overheating', as it is known in classical hydrodynamics
\cite{No87}, is a numerical artifact which is considerably reduced
when Marquina's scheme is used \cite{DF98}.  

\subsection{Propagation of relativistic blast waves}
%           ---------------------------------------
\label{ss:blast}

\begin{figure}
%\special{psfile=livrev_05_fig.ps hscale=100. vscale=100. hoffset=-30
%         voffset=-300 angle=0.}
%\vspace{10cm}
\centerline{
\epsfig{file=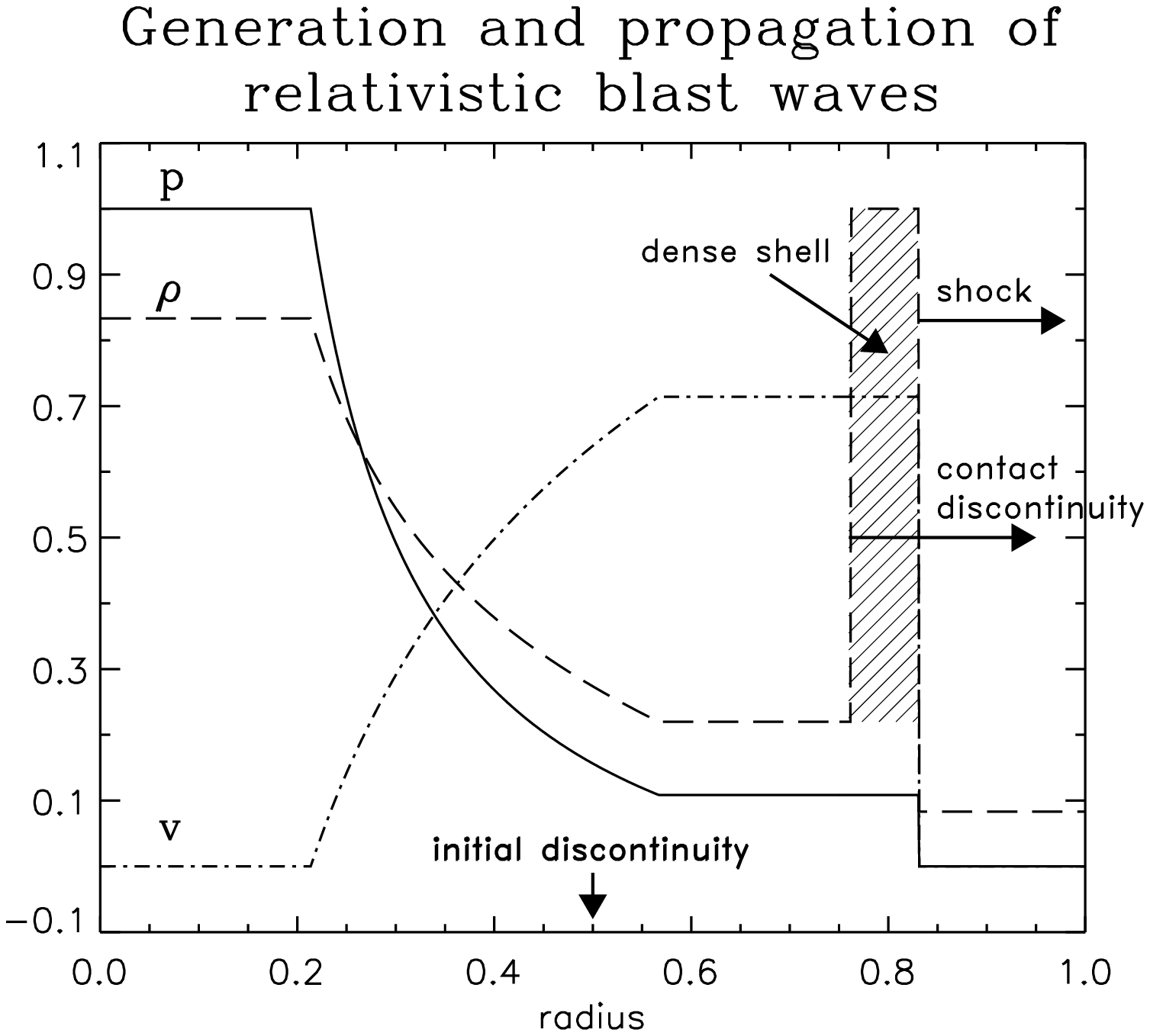, width=0.9\textwidth}}
\caption{\small Generation and propagation of a relativistic blast
wave (schematic). The large pressure jump at a discontinuity initially
located at $r=05$ gives rise to a blast wave and a dense shell of
material propagating at relativistic speeds. For appropriate initial
conditions both the speed of the leading shock front and the velocity
of the shell approach the speed of light producing very narrow
structures.
\label{f:blast_an}}
\end{figure}

  Riemann problems with large initial pressure jumps produce blast
waves with dense shells of material propagating at relativistic speeds
(see Fig.\,\ref{f:blast_an}). For appropriate initial conditions, both
the speed of the leading shock front and the velocity of the shell
material approach the speed of light producing very narrow
structures. The accurate description of these thin, relativistic
shells involving large density contrasts is a challenge for any
numerical code.  Some particular blast wave problems have become
standard numerical tests. Here we consider the two most common of
these tests. The initial conditions are given in
Table\,\ref{t:blasts}.

%%%%%%%%%%%%%%%%%%%%%%%%%%%%%%%%%%%%%%%%%%%%%%%%%%%%%%%%%%%%%%%%%%%%%%%%%%%%%%
%
%
%%%%%%%%%%%%%%%%%%%%%%%%%%%%%%%%%%%%%%%%%%%%%%%%%%%%%%%%%%%%%%%%%%%%%%%%%%%%%%

%%%%%%%%%%%%%%%%%%%%%%%%%%%%%%%%%%%%%%%%%%%%%%%%%%%%%%%%%%%%%%%%%%%%%%%%%%%%%%
%
{\small

\begin{table}[htb]
\begin{center}
\caption{\small Initial data (pressure $p$, density $\rho$, velocity
$v$) for two common relativistic blast wave test problems. The decay
of the initial discontinuity leads to a shock wave (velocity $v_{\rm
shock}$, compression ratio $\sigma_{\rm shock}$) and the formation of
a dense shell (velocity $v_{\rm shell}$, time--dependent width $w_{\rm
shell}$) both propagating to the right. The gas is assumed to be ideal
with an adiabatic index $\gamma = 5/3$.
\label{t:blasts}}
\begin{tabular}{crrrr}
\\ \hline \\
             & \multicolumn{2}{c}{{\bf Problem\,1}} & \multicolumn{2}{c}
             {{\bf Problem\,2}} \\
\\
             & Left  & Right & Left & Right \\
\\ \hline \\
$p$          & 13.33 & 0.00 & 1000.00 & 0.01 \\
$\rho$       & 10.00 & 1.00 &    1.00 & 1.00 \\
$v$          &  0.00 & 0.00 &    0.00 & 0.00 \\
\\ \hline \\
$v_{\rm shell}$     & \multicolumn{2}{c}{0.72} & \multicolumn{2}{c}{0.960} \\
$w_{\rm shell}$     & \multicolumn{2}{c}{0.11\,$t$} & 
                      \multicolumn{2}{c}{0.026\,$t$} \\
$v_{\rm shock}$     & \multicolumn{2}{c}{0.83} & \multicolumn{2}{c}{0.986} \\
$\sigma_{\rm shock}$& \multicolumn{2}{c}{5.07} & \multicolumn{2}{c}{10.75} \\
\\ \hline
\end{tabular}
\end{center}
\end{table}
}
%
%%%%%%%%%%%%%%%%%%%%%%%%%%%%%%%%%%%%%%%%%%%%%%%%%%%%%%%%%%%%%%%%%%%%%%%%%%%%%%

  Problem\,1 was a demanding problem for relativistic hydrodynamic
codes in the mid eighties \cite{CW84, HS84}, while Problem\,2 is a
challenge even for today's state-of-the-art codes. The analytical
solution of both problems can be obtained with program {\tt RIEMANN}
(see Sect.\,\ref{ss:riemann}).

\subsubsection{Problem\,1}
%              .........

  In Problem\,1, the decay of the initial discontinuity gives rise to
a dense shell of matter with velocity $v_{\rm shell} = 0.72$ ($W_{\rm
shell} = 1.38$) propagating to the right. The shell trailing a shock
wave of speed $v_{\rm shock} = 0.83$ increases its width, $w_{\rm
shell}$, according to $w_{\rm shell} = 0.11 t$, \ie at time $t = 0.4$
the shell covers about 4\% of the grid ($0 \le x \le 1$).
Tables\,\ref{t:problem1_other},\ref{t:problem1_hrsc} give a summary of 
the references where this test was considered for non-HRSC and HRSC methods,
respectively.

%%%%%%%%%%%%%%%%%%%%%%%%%%%%%%%%%%%%%%%%%%%%%%%%%%%%%%%%%%%%%%%%%%%%%%%%%%%%%%
%
{\small

\begin{table}[]
\begin{center}
\caption{\protect \small Summary of references where the blast wave
Problem\,1 (defined in Table\,\ref{t:blasts}) has been considered in
1D, 2D and 3D, respectively. Methods are described in Sects.~\ref{s:hrsc} and 
\ref{s:other} and their basic properties summarized in Sect.~\ref{s:sum_meth}
(Tables~\ref{t:methods},\ref{t:methods_other}).}
\label{t:problem1_other}
\begin{tabular}{lcll}
\\ \hline \\
References                                & Dim.   & Method        & Comments  \\
\\ \hline \\
Centrella \& Wilson (1984) \cite{CW84}    & 1D     & AV-mono       & Stable profiles without oscillations \\
                                          &        &               & Velocity overestimated by $7\%$ \\
Hawley \etal (1984)        \cite{HS84}    & 1D     & AV-mono       & Stable profiles without oscillations \\
                                          &        &               & $\rho_{\rm shell}$ overestimated by $16\%$ \\
Dubal (1991)$^{a}$         \cite{Du91}    & 1D     & FCT-lw        & 10-12 zones at the CD \\
                                          &        &               & Velocity overestimated by 4.5\% \\
Mann (1991)                \cite{Ma91}    & 1D     & SPH-AV-0,1,2  & Sistematic errors in the rarefaction \\
                                          &        &               & wave and the constant states \\
                                          &        &               & Large amplitud spikes at the CD \\ 
                                          &        &               & Excessive smearing at the shell \\
Laguna \etal (1993)        \cite{LM93}    & 1D     & SPH-AV-0      & Large amplitud spikes at the CD \\
                                          &        &               & $\rho_{\rm shell}$ overestimated by $5\%$ \\
van Putten (1993)$^{b}$    \cite{vP93}    & 1D     & van Putten    & Stable profiles \\
                                          &        &               & Excessive smearing, specially at the \\
                                          &        &               & CD ($\approx 50$ zones) \\
Schneider \etal (1993)     \cite{SK93}    & 1D     & SHASTA-c      & Non monotonic intermediate states \\
                                          &        &               & $\rho_{\rm shell}$ underestimated by $10\% $ with \\
                                          &        &               & 200 zones \\
Chow \& Monaghan (1997)    \cite{CM97}    & 1D     & SPH-RS-c      & Stable profiles without spikes \\
                                          &        &               & Excessive smearing at the CD and \\
                                          &        &               & at the shock \\
Siegler \& Riffert (1999)  \cite{SR99}    & 1D     & SPH-cAV-c     & Correct constant states \\ 
                                          &        &               & Large amplitud spikes at the CD \\
                                          &        &               & Excessive smearing at the shock \\
                                          &        &               & transition ($\approx 20$ zones) \\
\\ \hline
\end{tabular}
\end{center}
$^{a}$ For a Riemann problem with slightly different initial conditions. \\
$^{b}$ For a Riemann problem with slightly different initial conditions
       including a nonzero transverse magnetic field. \\
\end{table}
}
%
%%%%%%%%%%%%%%%%%%%%%%%%%%%%%%%%%%%%%%%%%%%%%%%%%%%%%%%%%%%%%%%%%%%%%%%%%%%%%%

%%%%%%%%%%%%%%%%%%%%%%%%%%%%%%%%%%%%%%%%%%%%%%%%%%%%%%%%%%%%%%%%%%%%%%%%%%%%%%
%
{\small

\begin{table}[]
\begin{center}
\caption{\protect \small Summary of references where the blast wave
Problem\,1 (defined in Table\,\ref{t:blasts}) has been considered in
1D, 2D and 3D, respectively. Methods are described in Sects.~\ref{s:hrsc} and
\ref{s:other} and their basic properties summarized in Sect.~\ref{s:sum_meth}
(Tables~\ref{t:methods},\ref{t:methods_other}).}
\label{t:problem1_hrsc}
\begin{tabular}{lcll}
\\ \hline \\
References                                & Dim.   & Method        & Comments$^{a}$ \\
\\ \hline \\
Eulderink (1993)           \cite{Eu93}    & 1D     & Roe-Eulderink & Correct $\rho_{\rm shell}$ with 500 zones \\
                                          &        &               & 4 zones at the CD\\
Schneider \etal (1993)     \cite{SK93}    & 1D     & HLL-l         & $\rho_{\rm shell}$ underestimated by $10\%$\\
                                          &        &               & with 200 zones \\
Mart\'{\i} \& M\"uller (1996) \cite{MM96} & 1D     & rPPM          & Correct $\rho_{\rm shell}$ with 400 zones \\
                                          &        &               & 6 zones at the CD \\
Mart\'{\i} \etal (1997)    \cite{MM97}    & 1D, 2D & MFF-ppm       & Correct $\rho_{\rm shell}$ with 400 zones \\
                                          &        &               & 6 zones at the CD \\
Wen \etal (1997)           \cite{WP97}    & 1D     & rGlimm        & No difussion at discontinuities \\
Yang \etal (1997)          \cite{YC97}    & 1D     & rBS           & Stable profiles \\

Donat \etal (1998)         \cite{DF98}    & 1D     & MFF-eno       & Correct $\rho_{\rm shell}$ with 400 zones \\
                                          &        &               & 8 zones at the CD \\
Aloy \etal (1999)          \cite{AI99a}   & 3D     & MFF-ppm       & \\
Font \etal (1999)          \cite{FM99}    & 1D, 3D & MFF-l         & Correct $\rho_{\rm shell}$ with 400 zones \\
                                          &        &               & 12-14 zones at the CD \\
                                          & 1D, 3D & Roe type-l    & Correct $\rho_{\rm shell}$ with 400 zones \\
                                          &        &               & 12-14 zones at the CD \\
                                          & 1D, 3D & Flux split    & $\rho_{\rm shell}$ overestimated by $5\%$ \\
                                          &        &               & 8 zones at the CD \\
\\ \hline
\end{tabular}
\end{center}
$^{a}$ All the methods produce stable profiles without numerical oscillations.
Comments correspond to 1D cases. \\
\end{table}
}
%
%%%%%%%%%%%%%%%%%%%%%%%%%%%%%%%%%%%%%%%%%%%%%%%%%%%%%%%%%%%%%%%%%%%%%%%%%%%%%%

  Using artificial viscosity techniques Centrella \& Wilson
\cite{CW84} were able to reproduce the analytical solution with a 7\%
overshoot in $v_{\rm shell}$, whereas Hawley \etal \cite{HS84} got a
16\% error in the shell density.

  The results obtained with early relativistic SPH codes \cite{Ma91} were
affected by systematic errors in the rarefaction wave and the constant states,
large amplitud spikes at the contact discontinuity and large smearing.
Smaller systematic errors and spikes are obtained
with Laguna \etal 's (1993) code \cite{LM93}. This code also leads to a large overshoot
in the shell's density. Much cleaner states are obtained with the methods of Chow \&
Monaghan (1997) \cite{CM97} and Siegler \& Riffert (1999) \cite{SR99}, both based on 
conservative formulations of the SPH equations. In the case of Chow \& Monaghan's (1997) 
method \cite{CM97}, the spikes at the contact discontinuity dissapear but at the cost
of an excessive smearing. Shock profiles with relativistic SPH codes are more smeared out 
than with HRSC methods covering typically more than 10 zones.

  Van Putten has considered a similar initial value problem with
somewhat more extreme conditions ($v_{\rm shell} \approx 0.82c$,
$\sigma_{\rm shock} \approx 5.1$) and with a transversal magnetic
field. For suitable choices of the smoothing parameters his results
are accurate and stable, although discontinuities appear to be more
smeared than with typical HRSC methods (6--7 zones for the strong
shock wave; $\approx 50$ zones for the contact discontinuity).

\begin{figure}
%
%%%---MOVIE---  livrev_mov2.mpg
%
%\vspace{3cm}
\centerline{
\epsfig{file=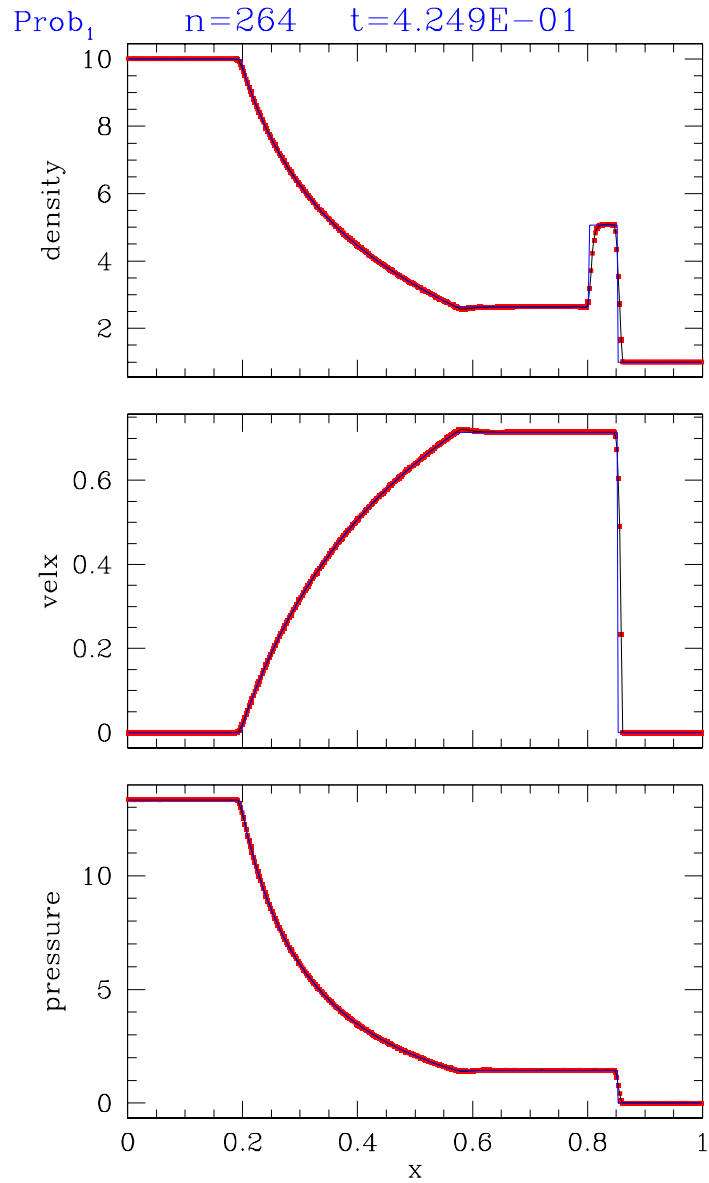, width=0.6\textwidth}}
\caption{\protect \small MPEG movie 
({\it only final frame of movie is displayed!})
showing the evolution of the density distribution for the relativistic
blast wave Problem\,1 defined in Table\,\ref{t:blasts}.  The final
frame of the movie also shows the analytical solution (blue lines).
The simulation has been performed with relativistic PPM on an
equidistant grid of 400 zones.}
\label{problem_1}
\end{figure}

  MPEG movie\,2
({\it final frame of movie is displayed in Fig.\,\ref{problem_1}})
shows the Problem\,1 blast wave evolution obtained with a modern HRSC
method (the relativistic PPM method introduced in
Section\,\ref{ss:rppm}). The grid has 400 equidistant zones, and the
relativistic shell is resolved by 16 zones. Because of both the high
order accuracy of the method in smooth regions and its small numerical
diffusion (the shock is resolved with 4--5 zones only) the density of
the shell is accurately computed (errors less than 0.1\%). Other codes
based on relativistic Riemann solvers \cite{EM95} give similar results
(see Table~\ref{t:problem1_hrsc}).  The relativistic HLL method
\cite{SK93} underestimates the density in the shell by about 10\% in a
200 zone calculation.

%%%%%%%%%%%%%%%%%%%%%%%%%%%%%%%%%%%%%%%%%%%%%%%%%%%%%%%%%%%%%%%%%%%%%%%%%%%%%
%
%
%%%%%%%%%%%%%%%%%%%%%%%%%%%%%%%%%%%%%%%%%%%%%%%%%%%%%%%%%%%%%%%%%%%%%%%%%%%%%

\subsubsection{Problem\,2}
%              .........

  Problem\,2 was first considered by Norman \& Winkler\cite{NW86}. The
flow pattern is similar to that of Problem\,1, but more extreme.
Relativistic effects reduce the post--shock state to a thin dense
shell with a width of only about 1\% of the grid length at $t =
0.4$. The fluid in the shell moves with $v_{\rm shell} = 0.960$ (\ie
$W_{\rm shell} = 3.6$), while the leading shock front propagates with
a velocity $v_{\rm shock} = 0.986$ (\ie $W_{\rm shock}=6.0$). The jump in
density in the shell reaches a value of 10.6. Norman \& Winkler
\cite{NW86} obtained very good results with an adaptive grid of 400
zones using an implicit hydro--code with artificial viscosity. Their
adaptive grid algorithm placed 140 zones of the available 400 zones
within the blast wave thereby accurately capturing all features of the
solution.

  Several HRSC methods based on relativistic Riemann solvers have used
Problem\,2 as a standard test \cite{MI91,MM92,MM96,FK96,WP97,DF98}
%.
(see Table\,\ref{t:problem2}).
% gives a summary of the references where this test was considered.

%%%%%%%%%%%%%%%%%%%%%%%%%%%%%%%%%%%%%%%%%%%%%%%%%%%%%%%%%%%%%%%%%%%%%%%%%%%%%%
%
{\small

\begin{table}[]
\begin{center}
\caption{\protect \small Summary of references where the blast wave
Problem\,2 (defined in Table\,\ref{t:blasts}) has been considered.
Methods are described in Sects.~\ref{s:hrsc} and \ref{s:other} and their basic 
properties summarized in Sect.~\ref{s:sum_meth} 
(Tables~\ref{t:methods},\ref{t:methods_other}).}
\label{t:problem2}
\begin{tabular}{lll}
\\ \hline \\
References                                   & Method           & $\sigma/\sigma_{\rm exact}$\\
\\ \hline \\
Norman \& Winkler (1986)   \cite{NW86}       & cAV-implicit     & 1.00       \\
Dubal (1991)               \cite{Du91}$^{a}$ & FCT-lw           & 0.80       \\
Mart\'{\i} et al. (1991)   \cite{MI91}       & Roe type-l       & 0.53       \\
Marquina et al. (1992)     \cite{MM92}       & LCA-phm          & 0.64       \\
Mart\'{\i} \& M\"uller (1996) \cite{MM96}    & rPPM             & 0.68       \\
Falle \& Komissarov (1996) \cite{FK96}       & Falle-Komissarov & 0.47       \\
Wen et al. (1997)          \cite{WP97}       & rGlimm           & 1.00       \\
Chow \& Monaghan (1997)    \cite{CM97}       & SPH-RS-c         & 1.16$^{b}$ \\
Donat et al. (1998)        \cite{DF98}       & MFF-phm          & 0.60       \\
\\ \hline
\end{tabular}
\end{center}
$^{a}$ For a Riemann problem with slightly different initial conditions. \\
$^{b}$ At $t=0.15$. \\
\end{table}
}
%
%%%%%%%%%%%%%%%%%%%%%%%%%%%%%%%%%%%%%%%%%%%%%%%%%%%%%%%%%%%%%%%%%%%%%%%%%%%%%%

\begin{figure}
%
%%%---MOVIE---  livrev_mov3.mpg
%
%\vspace{3cm}
\centerline{
\epsfig{file=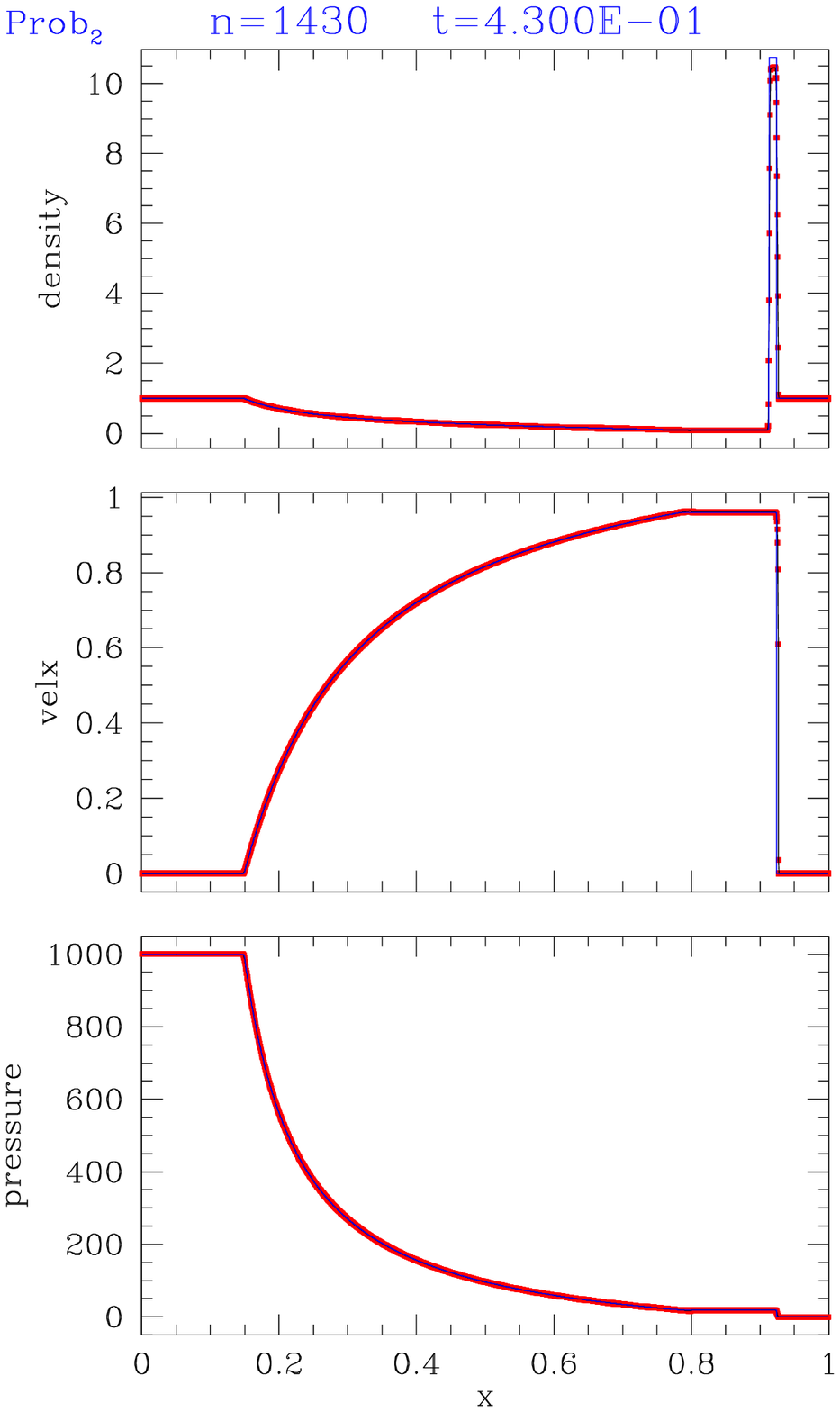, width=0.6\textwidth}}
\caption{\protect \small MPEG movie 
({\it only final frame of movie is displayed!})
showing the evolution of the density distribution for the relativistic
blast wave Problem\,2 defined in Table\,\ref{t:blasts}.  The final
frame of the movie also shows the analytical solution (blue lines).
The simulation has been performed with relativistic PPM on an
equidistant grid of 2000 zones.}
\label{problem_2}
\end{figure}

\begin{figure}
%\vspace{16cm}
%\special{psfile=livrev_08_fig.ps hoffset=-100 voffset=-240}
\centerline{
\epsfig{file=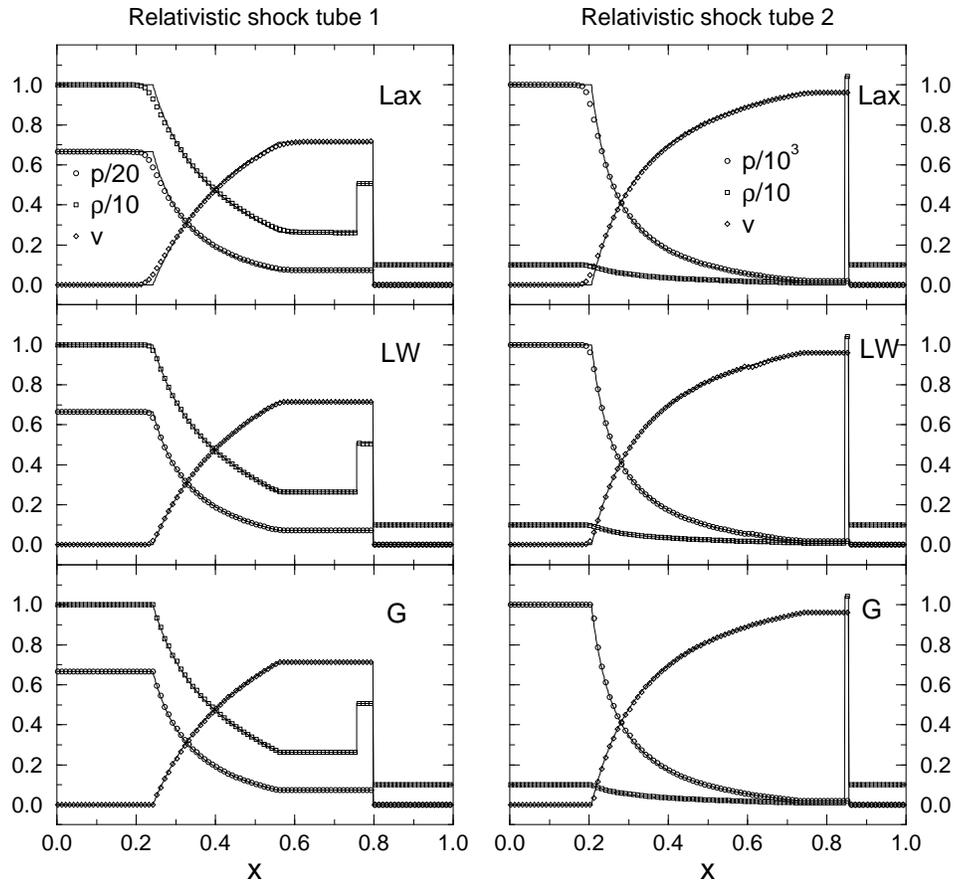, width=0.7\textwidth} \qquad}
\bigskip
\caption{\protect \small Results from \cite{WP97} for the relativistic
blast wave Problems\,1 (left column) and 2 (right column),
respectively. Relativistic Glimm's method is only used in regions with steep 
gradients. Standard finite difference schemes are applied in the smooth 
remaining part of the computational domain. In the above plots, Lax and LW
stand respectively for Lax and Lax-Wendroff methods; G refers to pure Glimm's
method.}
\label{wpl1}
\end{figure}

  MPEG movie\,3
({\it final frame of movie is displayed in Fig.\,\ref{problem_2}})
shows the Problem\,2 blast wave evolution obtained with the
relativistic PPM method introduced in Section\,\ref{ss:rppm}) on a
grid of 2000 equidistant zones.  At this resolution the relativistic
PPM code obtains a converged solution. The method of Falle \&
Komissarov \cite{FK96} requires a seven--level adaptive grid
calculation to achieve the same the finest grid spacing corresponding
to a grid of 3200 zones. As their code is free of numerical diffusion
and dispersion, Wen \etal \cite{WP97} are able to handle this problem
with high accuracy (see Fig\,\ref{wpl1}).  At lower resolution (400
zones) the relativistic PPM method only reaches 69\% of the
theoretical shock compression value (54\% in case of the second--order
accurate upwind method of Falle \& Komissarov \cite{FK96}; 60\% with
the code of Donat \etal \cite{DF98}).
 
%%%%%%%%%%%%%%%%%%%%%%%%%%%%%%%%%%%%%%%%%%%%%%%%%%%%%%%%%%%%%%%%%%%%%%%%%%%%%%
%
%
%%%%%%%%%%%%%%%%%%%%%%%%%%%%%%%%%%%%%%%%%%%%%%%%%%%%%%%%%%%%%%%%%%%%%%%%%%%%%%

%%%%%%%%%%%%%%%%%%%%%%%%%%%%%%%%%%%%%%%%%%%%%%%%%%%%%%%%%%%%%%%%%%%%%%%%%%%%%%
%
%
%%%%%%%%%%%%%%%%%%%%%%%%%%%%%%%%%%%%%%%%%%%%%%%%%%%%%%%%%%%%%%%%%%%%%%%%%%%%%%

  Chow \& Monaghan \cite{CM97} have considered Problem\,2 to test
their relativistic SPH code. Besides a 15\% overshoot in the shell's
density, the code produces a non--causal blast wave propagation speed
(\ie $v_{\rm shock} > 1$).

%NEW VERSION (CH)

\subsubsection{Collision of two relativistic blast waves}
%              .........................................
\label{sss:RBWI}

  The collision of two strong blast waves was used by Woodward \& Colella 
\cite{WC84} to compare the performance of several numerical methods in classical 
hydrodynamics. In the relativistic case, Yang \etal \cite{YC97} considered this 
problem to test the high-order extensions of the relativistic beam scheme, 
whereas Mart\'{\i} \& M\"uller \cite{MM96} used it to evaluate the performance 
of their relativistic PPM code. In this last case, the original boundary 
conditions were changed (from reflecting to outflow) to avoid the reflection 
and subsequent interaction of rarefaction waves allowing for a comparison with 
an analytical solution. In the following we summarize the results on this test
obtained by Mart\'{\i} \& M\"uller in \cite{MM96}.

\begin{figure}
%\vspace{16cm}
%\special{psfile=livrev_09_fig.ps hscale=85. vscale=85. hoffset=0 
%         voffset=-20}
\centerline{
\epsfig{file=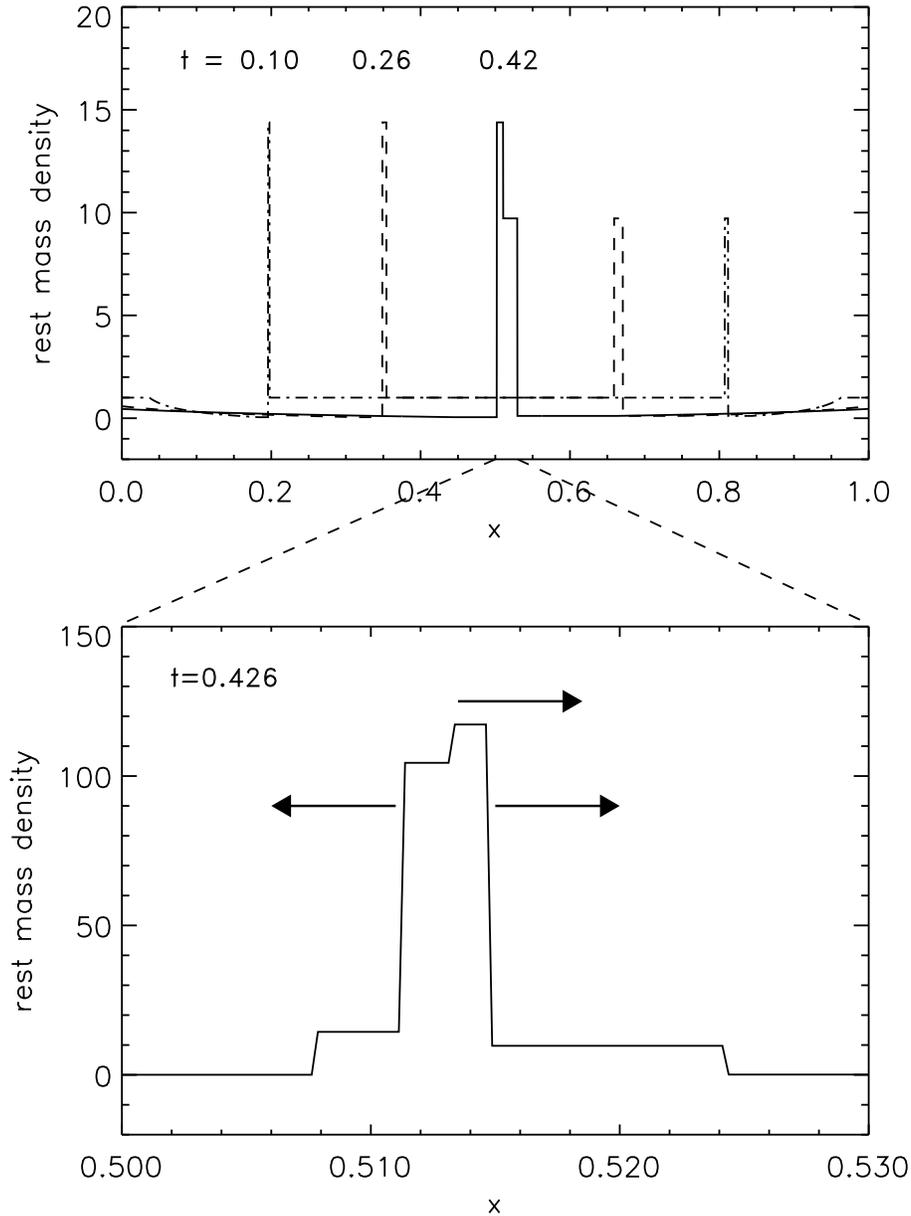, width=0.8\textwidth}}
\caption{\small The top panel shows a sequence of snapshots of the
density profile for the colliding relativistic blast wave problem up
to the moment when the waves begin to interact. The density profile of
the new states produced by the interaction of the two waves is shown
in the bottom panel (note the change in scale on both axes with
respect to the top panel).}
\label{collbw-f}
\end{figure}

  The initial data corresponding to this test, consisting in three constant 
states with large pressure jumps at the discontinuities separating the states 
(at $x = 0.1$ and $x = 0.9$), as well as the properties of the blast waves 
created by the decay of the initial discontinuities, are listed in 
Table~\ref{t:blastcol}. The propagation velocity of the two blast waves is 
slower than in the Newtonian case, but very close to the speed of light (0.9776 
and $-0.9274$ for the shock wave propagating to the right and left, 
respectively). Hence, the shock interaction occurs later (at $t = 0.420$) than 
in the Newtonian problem (at $t = 0.028$). Top panel in Fig.~\ref{collbw-f} 
shows four snapshots of the density distribution including the moment of the 
collision of the blast waves at $t = 0.420$ and $x = 0.5106$. At the time of 
collision the two shells have a width of $\Delta x = 0.008$ (left shell) and 
$\Delta x = 0.019$ (right shell), respectively, \ie the whole interaction 
takes place in a very thin region (about 10 times smaller than in the Newtonian 
case where $\Delta x \approx 0.2$).

%%%%%%%%%%%%%%%%%%%%%%%%%%%%%%%%%%%%%%%%%%%%%%%%%%%%%%%%%%%%%%%%%%%%%%%%%%%%%%
%
{\small

\begin{table}[htb]
\begin{center}
\caption{\small Initial data (pressure $p$, density $\rho$, velocity
$v$) for the two relativistic blast wave collision test problem. The decay
of the initial discontinuities (at $x = 0.1$ and $x = 0.9$) leads to the 
formation of two shock waves (velocities $v_{\rm shock}$, compression ratios 
$\sigma_{\rm shock}$) leading two dense shells (velocities $v_{\rm shell}$, 
time--dependent widths $w_{\rm shell}$) moving in opposite directions. The gas 
is assumed to be ideal with an adiabatic index $\gamma = 1.4$.
\label{t:blastcol}}
\begin{tabular}{crrrr}
\\ \hline \\
             & Left    & \multicolumn{2}{c}{Middle} & Right\\
\\ \hline \\
$p$          & 1000.00 & \multicolumn{2}{r}{0.01} & 100.00 \\
$\rho$       & 1.00    & \multicolumn{2}{r}{1.00}   & 1.00 \\
$v$          & 0.00    & \multicolumn{2}{r}{0.00}   & 0.00 \\
\\ \hline \\
$v_{\rm shell}$     & \multicolumn{2}{c}{\,\,\,\,\,\,\,\,\,\,0.957}            & \multicolumn{2}{c}{$-$0.882} \\
$w_{\rm shell}$     & \multicolumn{2}{c}{\,\,\,\,\,\,\,\,\,\,\,\,\,0.021\,$t$} & \multicolumn{2}{c}{\,\,\,\,\,\,\,\,0.045\,$t$} \\
$v_{\rm shock}$     & \multicolumn{2}{c}{\,\,\,\,\,\,\,\,\,\,0.978}            & \multicolumn{2}{c}{$-$0.927} \\
$\sigma_{\rm shock}$& \multicolumn{2}{c}{\,\,\,\,14.39}                        & \multicolumn{2}{c}{\,\,9.72} \\
\\ \hline
\end{tabular}
\end{center}
\end{table}
}
%
%%%%%%%%%%%%%%%%%%%%%%%%%%%%%%%%%%%%%%%%%%%%%%%%%%%%%%%%%%%%%%%%%%%%%%%%%%%%%%

%%%%%%%%%%%%%%%%%%%%%%%%%%%%%%%%%%%%%%%%%%%%%%%%%%%%%%%%%%%%%%%%%%%%%%%%%%%%%%
%
%
%%%%%%%%%%%%%%%%%%%%%%%%%%%%%%%%%%%%%%%%%%%%%%%%%%%%%%%%%%%%%%%%%%%%%%%%%%%%%%

  The collision gives rise to a narrow region of very high density (see lower 
panel of Fig.~\ref{collbw-f}) bounded by two shocks moving at speeds 0.088 
(shock at the left) and 0.703 (shock at the right) and large compression ratios 
(7.26 and 12.06, respectively) well above the classical limit for strong
shocks (6.0 for $\gamma = 1.4$). The solution just described applies until 
$t = 0.430$ when the next interaction takes place.

  The complete analytical solution before and after the collision up 
to time $t = 0.430$ can be obtained following Appendix\,II in \cite{MM96}.

%%%%%%%%%%%%%%%%%%%%%%%%%%%%%%%%%%%%%%%%%%%%%%%%%%%%%%%%%%%%%%%%%%%%%%%%%%%%%%
%
\begin{figure}
%
%%%---MOVIE---  livrev_mov4.mpg
%
%\vspace{6cm}
\centerline{
\epsfig{file=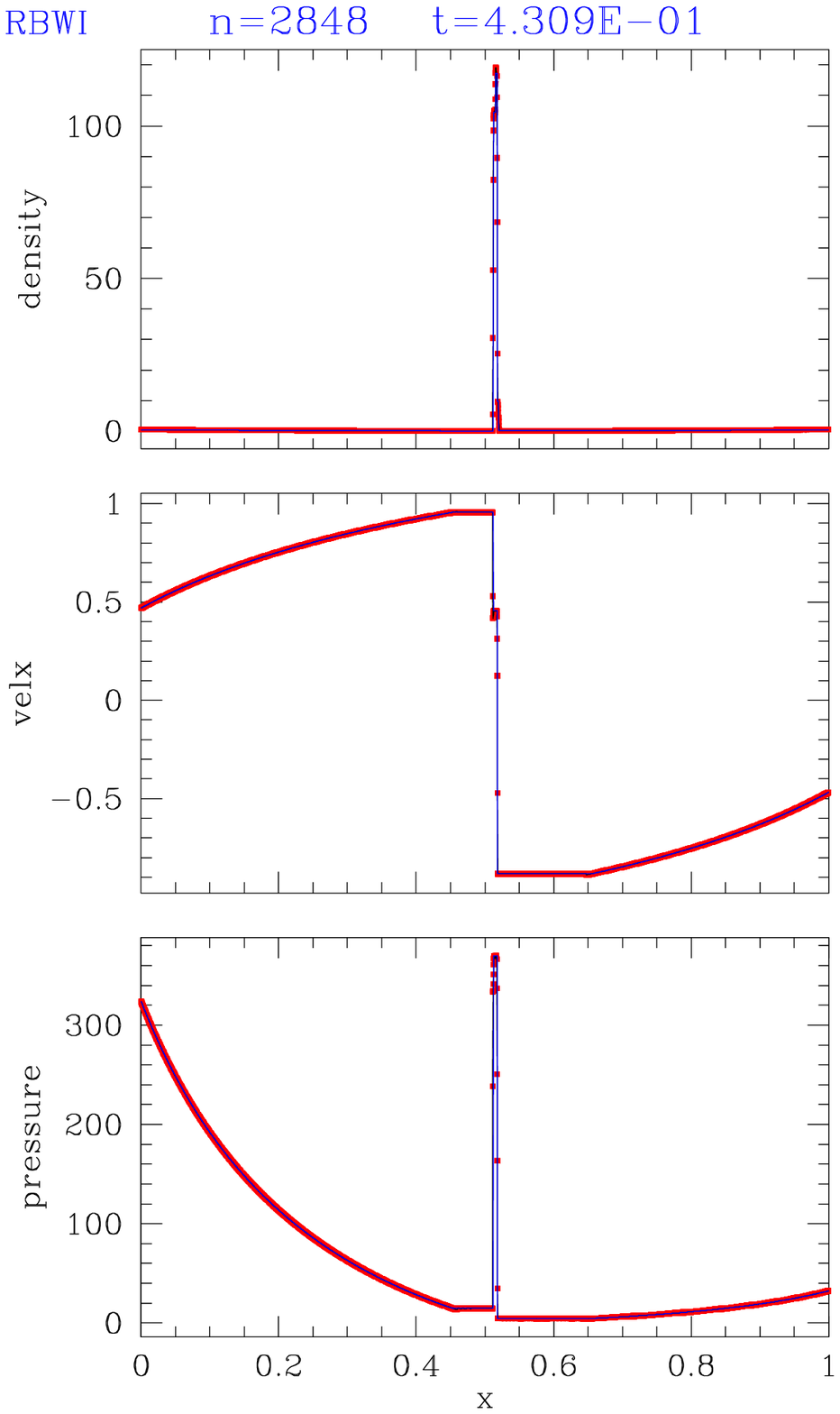, width=0.6\textwidth}}
\caption{\small MPEG movie 
({\it only final frame of movie is displayed!})
showing the evolution of the density distribution for the colliding
relativistic blast wave problem up to the interaction of the waves.
The final frame of the movie also shows the analytical solution (blue
lines).  The computation has been performed with relativistic PPM on
an equidistant grid of 4000 zones.}
\label{collbw-m1}
\end{figure}
%
%%%%%%%%%%%%%%%%%%%%%%%%%%%%%%%%%%%%%%%%%%%%%%%%%%%%%%%%%%%%%%%%%%%%%%%%%%%%%%

%%%%%%%%%%%%%%%%%%%%%%%%%%%%%%%%%%%%%%%%%%%%%%%%%%%%%%%%%%%%%%%%%%%%%%%%%%%%%%
%
\begin{figure}
%
%%%---MOVIE---  livrev_mov5.mpg
%
%\vspace{6cm}
\centerline{
\epsfig{file=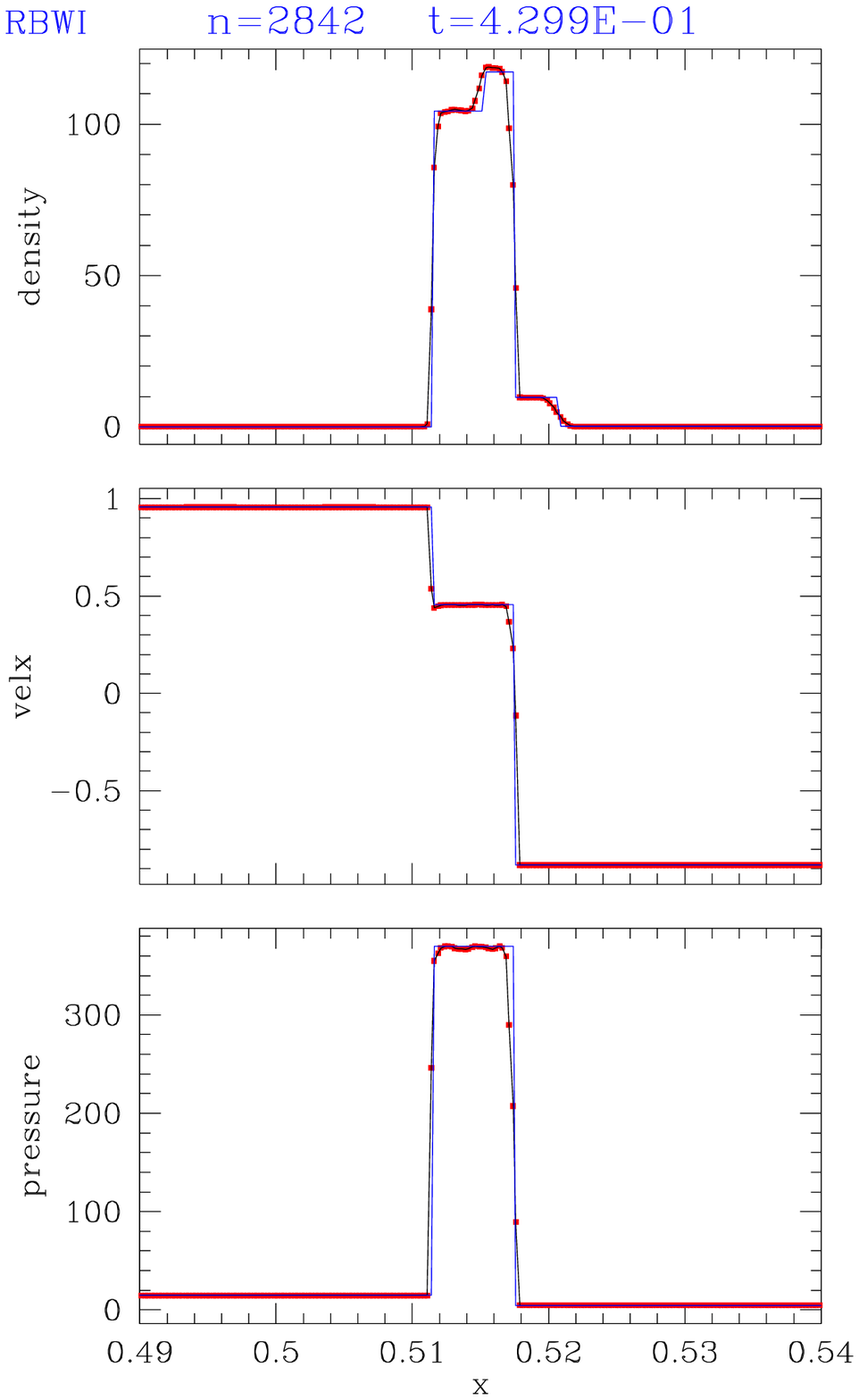, width=0.6\textwidth}}
\caption{\small MPEG movie 
({\it only final frame of movie is displayed!})
showing the evolution of the density distribution for the colliding
relativistic blast wave problem around the time of interaction of the
waves at an enlarged spatial scale.  The final frame of the movie also
shows the analytical solution (blue lines).  The computation has been
performed with relativistic PPM on an equidistant grid of 4000 zones.}
\label{collbw-m2}
\end{figure}
%
%%%%%%%%%%%%%%%%%%%%%%%%%%%%%%%%%%%%%%%%%%%%%%%%%%%%%%%%%%%%%%%%%%%%%%%%%%%%%%

  MPEG movie\,4
({\it final frame of movie is displayed in Fig.\,\ref{collbw-m1}}) 
shows the evolution of the density up to
the time of shock collision at $t = 0.4200$. The movie was obtained
with the relativistic PPM code of Mart\'{\i} \& M\"uller
\cite{MM96}. The presence of very narrow structures involving large
density jumps requires very fine zoning to resolve the states
properly. For the movie a grid of 4000 equidistant zones was
used. The relative error in the density of the left (right) shell is
always less than $2.0\%$ ($0.6\%$), and is about $1.0\%$ ($0.5\%$) at
the moment of shock collision. Profiles obtained with the relativistic
Godunov method (first--order accurate, not shown) show relative errors in the
density of the left (right) shell of about $50\%$ ($16\%$) at $t =
0.20$. The errors drop only slightly to about $40\%$ ($5\%$) at the
time of collision ($t = 0.420$).

  MPEG movie\,5
({\it final frame of movie is displayed in Fig.\,\ref{collbw-m2}}) 
shows the numerical solution after the interaction has
occurred. Compared to MPEG movie\,4
({\it final frame of movie is displayed in Fig.\,\,\ref{collbw-m1}}) 
a very different scaling for the x-axis had to be used to display the
narrow dense new states produced by the interaction. Obviously, the
relativistic PPM code resolves the structure of the collision region
satisfactorily well, the maximum relative error in the density
distribution being less than $2.0\%$.  When using the first--order
accurate Godunov method instead, the new states are strongly smeared
out and the positions of the leading shocks are wrong.

\section{APPLICATIONS}
%        ############
\label{s:appl}

\subsection{Astrophysical jets}
%           ------------------

  The most compelling case for a special relativistic phenomenon are the 
ubiquitous jets in extragalactic radio sources associated with active galactic 
nuclei. In the commonly accepted standard model \cite{BB84}, flow velocities as 
large as 99\% (in some cases even beyond) of the speed of light are required to 
explain the apparent superluminal motion observed in many of these sources. 
Models which have been proposed to explain the formation of relativistic jets, 
involve accretion onto a compact central object, such as a neutron star or 
stellar mass black hole in the galactic micro--quasars GRS\,1915+105 \cite{MR94} 
and GRO\,J1655-40 \cite{TJ95}, or a rotating super massive black hole in an 
active galactic nucleus, which is fed by interstellar gas and gas from tidally 
disrupted stars.  

  Inferred jet velocities close to the speed of light suggest that jets are 
formed within a few gravitational radii of the event horizon of the black hole. 
Moreover, very--long--baseline interferometric (VLBI) radio observations reveal
that jets are already collimated at subparsec scales.  Current theoretical 
models assume that accretion disks are the source of the bipolar outflows which 
are further collimated and accelerated via MHD processes (see, \eg \cite{Bl90}). 
There is a large number of parameters which are potentially important for jet 
powering: the black hole mass and spin, the accretion rate and the type of 
accretion disk, the properties of the magnetic field and of the environment.

  At parsec scales the jets, observed via their synchrotron and inverse Compton 
emission at radio frequencies with VLBI imaging, appear to be highly collimated 
with a bright spot (the core) at one end of the jet and a series of components 
which separate from the core, sometimes at superluminal speeds. In the standard 
model \cite{BK79}, these speeds are interpreted as a consequence of relativistic 
bulk motions in jets propagating at small angles to the line of sight with 
Lorentz factors up to 20 or more. Moving components in these jets, usually 
preceded by outbursts in emission at radio wavelengths, are interpreted in terms 
of traveling shock waves.

  Finally, the morphology and dynamics of jets at kiloparsec scales are 
dominated by the interaction of the jet with the surrounding extragalactic 
medium, the jet power being responsible for dichotomic morphologies (the so 
called Fanaroff--Riley I and II classes \cite{FR74}, FR\,I and FR\,II, 
respectively). Whereas current models \cite{Bi96, La96} interpret FR\,I 
morphologies as the result of a smooth deceleration from relativistic to 
non--relativistic, transonic speeds on kpc scales due to a slower shear layer, 
flux asymmetries between jets and counter--jets in the most powerful radio 
galaxies (FRII) and quasars indicate that relativistic motion extends up to kpc 
scales in these sources, although with smaller values of the overall bulk speeds
\cite{Br94}.

  Although MHD and general relativistic effects seem to be crucial for
a successful launch of the jet (for a review see, \eg \cite{Ca98}),
purely hydrodynamic, special relativistic simulations are adequate to
study the morphology and dynamics of relativistic jets at distances
sufficiently far from the central compact object (\ie at parsec scales
and beyond). The development of relativistic hydrodynamic codes based
on HRSC techniques (see Sections\,\ref{s:hrsc} and \ref{s:other}) has
triggered the numerical simulation of relativistic jets at parsec and
kiloparsec scales.

%%%%%%%%%%%%%%%%%%%%%%%%%%%%%%%%%%%%%%%%%%%%%%%%%%%%%%%%%%%%%%%%%%%%%%%%%%%%%%
%
\begin{figure}
%\special{psfile=livrev_12_fig.ps hscale=80. vscale=80. hoffset=-40 
%         voffset=-600 angle=0.}
%\vspace{14cm}
\centerline{
\epsfig{file=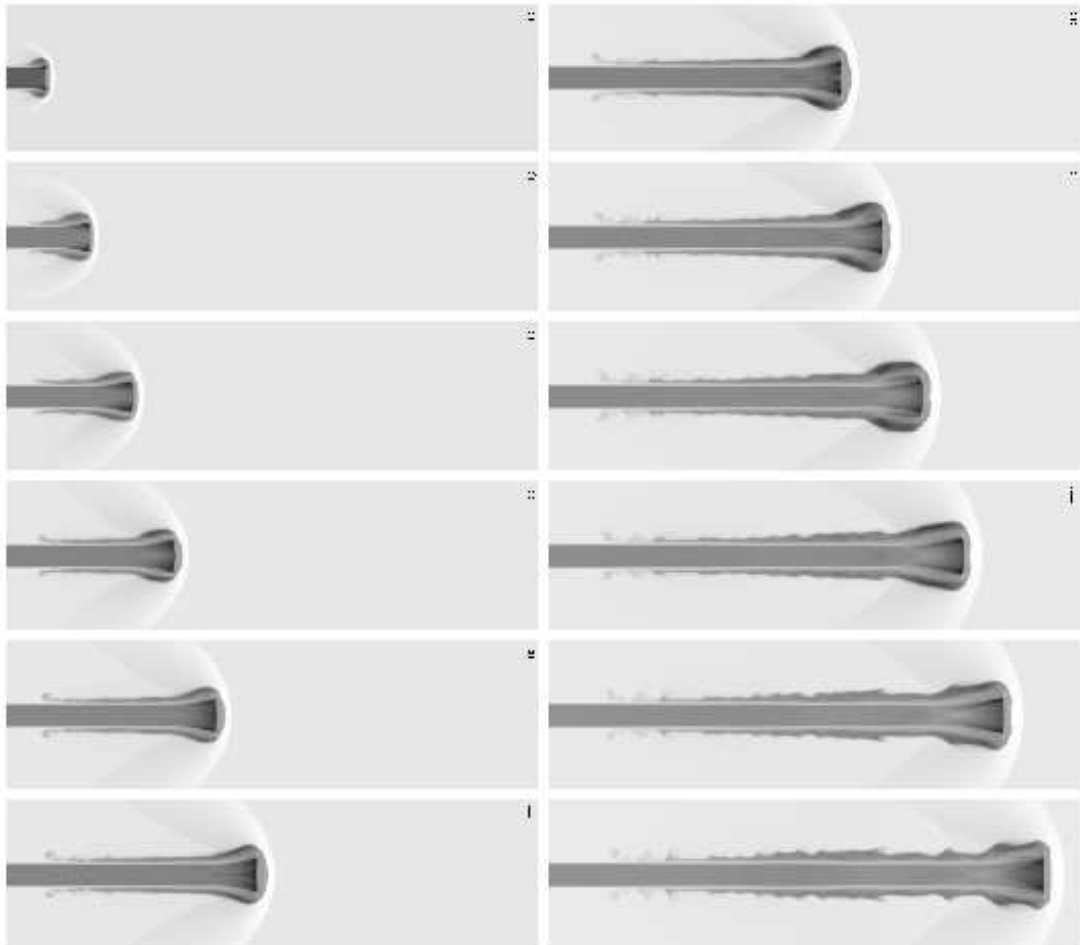, width=0.9\textwidth}}
\caption{\protect \small Time evolution of a light, relativistic (beam
flow velocity equal to 0.99) jet with large internal energy. The
logarithm of the proper rest--mass density is plotted in grey scale
the maximum value corresponding to white and the minimum to black.}
\label{hotjet}
\end{figure}
%
%%%%%%%%%%%%%%%%%%%%%%%%%%%%%%%%%%%%%%%%%%%%%%%%%%%%%%%%%%%%%%%%%%%%%%%%%%%%%%

%%%%%%%%%%%%%%%%%%%%%%%%%%%%%%%%%%%%%%%%%%%%%%%%%%%%%%%%%%%%%%%%%%%%%%%%%%%%%%
%
\begin{figure}
%\special{psfile=livrev_13_fig.ps hscale=100. vscale=100. hoffset=-100 
%         voffset=-600 angle=0.}
%\vspace{14cm}
\centerline{
\epsfig{file=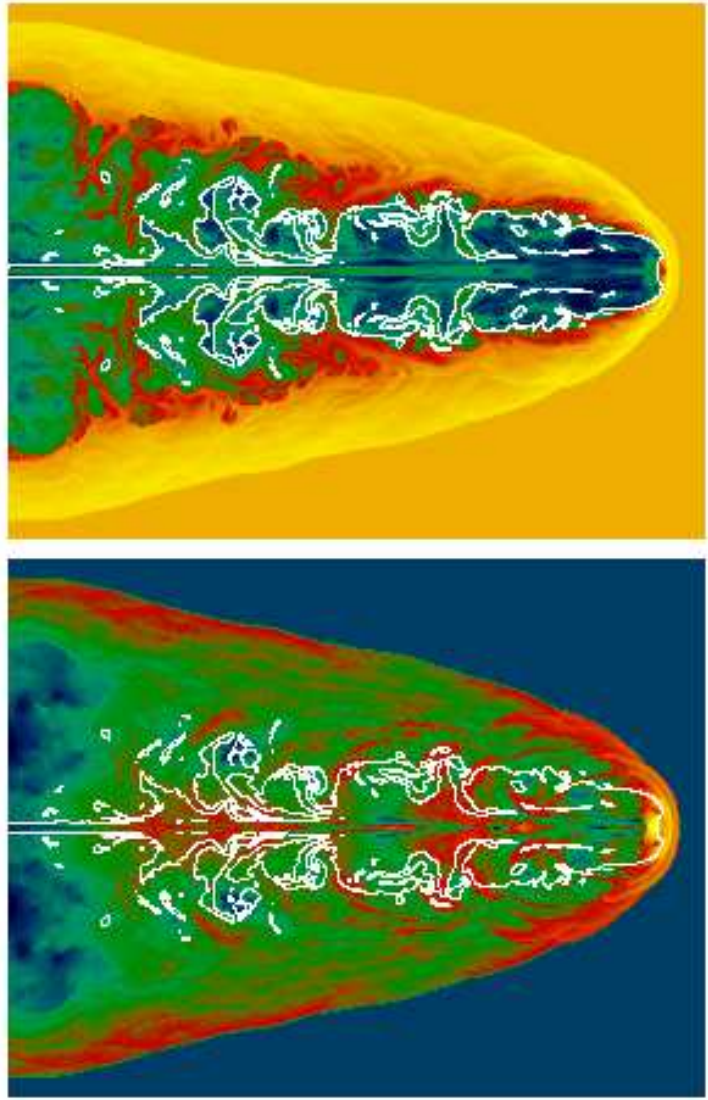, width=0.6\textwidth}}
\caption{\protect \small Logarithm of the proper rest--mass density
and energy density (from top to bottom) of an evolved, powerful jet
propagating through the intergalactic medium.  The white contour
encompasses the jet material responsible for the synchrotron
emission.}
\label{jetlong}
\end{figure}
%
%%%%%%%%%%%%%%%%%%%%%%%%%%%%%%%%%%%%%%%%%%%%%%%%%%%%%%%%%%%%%%%%%%%%%%%%%%%%%%

%%%%%%%%%%%%%%%%%%%%%%%%%%%%%%%%%%%%%%%%%%%%%%%%%%%%%%%%%%%%%%%%%%%%%%%%%%%%%%
%
\begin{figure}
%\special{psfile=livrev_14_fig.ps hscale=100. vscale=100. hoffset=-40 
%         voffset=-200 angle=0.}
%\vspace{8cm}
\centerline{
\epsfig{file=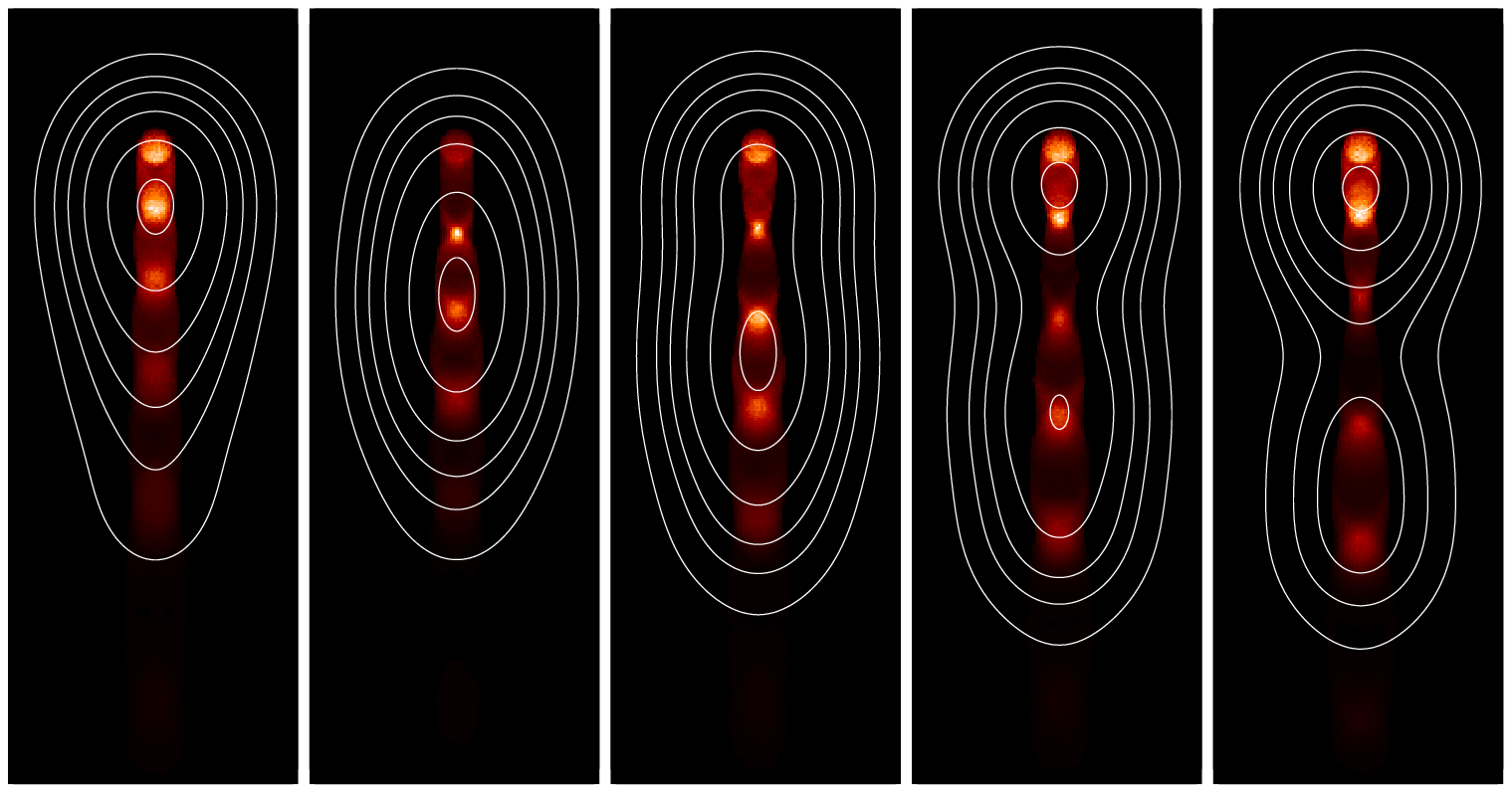, width=0.9\textwidth}}
\caption{\protect \small Computed radio maps of a compact relativistic
jet showing the evolution of a superluminal component (from left to
right). Two resolutions are shown: present VLBI resolution (white
contours) and resolution provided by the simulation (black/white
images).}
\label{super}
\end{figure}
%
%%%%%%%%%%%%%%%%%%%%%%%%%%%%%%%%%%%%%%%%%%%%%%%%%%%%%%%%%%%%%%%%%%%%%%%%%%%%%%

  At kiloparsec scales, the implications of relativistic flow speeds
and/or relativistic internal energies for the morphology and dynamics
of jets have been the subject of a number of papers in recent years
\cite{MM94a, DH94, MM95, MM97, KoF98}. Beams
with large internal energies show little internal structure and
relatively smooth cocoons allowing the terminal shock (the hot spot in
the radio maps) to remain well defined during the evolution. Their
morphologies resemble those observed in naked quasar jets like 3C273
\cite{DM85}. Fig.\,\ref{hotjet} shows several snapshots of the time
evolution of a light, relativistic jet with large internal energy. The
dependence of the beam's internal structure on the flow speed suggests
that relativistic effects may be relevant for the understanding of the
difference between slower, knotty BL Lac jets and faster, smoother
quasar jets \cite{GM94}.

  Highly supersonic models, in which kinematic relativistic effects
due to high beam Lorentz factors dominate, have extended overpressured
cocoons. These overpressured cocoons can help to confine the jets
during the early stages of their evolution \cite{MM95} and even cause
their deflection when propagating through non--homogeneous
environments \cite{PM97}. The cocoon overpressure causes the formation
of a series of oblique shocks within the beam in which the synchrotron
emission is enhanced. In long term simulations (see
Fig.\,\ref{jetlong}), the evolution is dominated by a strong
deceleration phase during which large lobes of jet material (like the
ones observed in many FR\,IIs, \eg Cyg~A \cite{CP96}) start to inflate
around the jet's head. These simulations reproduce some properties
observed in powerful extragalactic radio jets (lobe inflation, hot
spot advance speeds and pressures, deceleration of the beam flow along
the jet) and can help to constrain the values of basic parameters
(such as the particle density and the flow speed) in the jets of real
sources.

  The development of multidimensional relativistic hydrodynamic codes
has allowed, for the first time, the simulation of parsec scale jets
and superluminal radio components \cite{GM97, KF97, MH97}. The
presence of emitting flows at almost the speed of light enhances the
importance of relativistic effects in the appearance of these sources
(relativistic Doppler boosting, light aberration, time delays). Hence,
one should use models which combine hydrodynamics and synchrotron
radiation transfer when comparing with observations.  In these models,
moving radio components are obtained from perturbations in steady
relativistic jets. Where pressure mismatches exist between the jet and
the surrounding atmosphere reconfinement shocks are produced. The
energy density enhancement produced downstream from these shocks can
give rise to stationary radio knots as observed in many VLBI
sources. Superluminal components are produced by triggering small
perturbations in these steady jets which propagate at almost the jet
flow speed. One example of this is shown in Fig.\,\ref{super} (see
also \cite{GM97}), where a superluminal component (apparent speed
$\approx 7$ times the speed of light) is produced from a small
variation of the beam flow Lorentz factor at the jet inlet. The
dynamic interaction between the induced traveling shocks and the
underlying steady jet can account for the complex behavior observed in
many sources \cite{GM98}.

  The first magnetohydrodynamic simulations of relativistic jets have
been already undertaken in 2D \cite{KN96,Ko97} and 3D \cite{NK97,NK98} to
study the implications of ambient magnetic fields in the morphology and
bending properties of relativistic jets. However, despite the impact of these
results in specific problems like ,\eg, the understanding of the misalignement
of jets between pc and kpc scales, these 3D simulations have not addressed
the effects on the jet structure and dynamics of the third spatial degree of
freedom. This has been the aim of the work undertaken by
Aloy \etal \cite{AI99b}. 

  Finally, Koide \etal \cite{KS98} have developed a general relativistic
MHD code and applied it to the problem of jet formation from black hole
accretion disks. Jets are formed with a two-layered shell structure consisting 
of a fast gas pressure driven jet (Lorentz factor $\approx 2$) in the inner part 
and a slow magnetically driven outflow in the outer part both of which being 
collimated by the global poloidal magnetic field penetrating the disk. 

\subsection{Gamma-Ray Bursts (GRBs)}
%           ----------------------
A second phenomenon which involves flows with velocities very close to
the speed of light are gamma-ray bursts (GRBs). Although known
observationally since over 30 years, until recently their distance
(``local'' or ``cosmological'') has been, and their nature still is a
matter of controversial debate \cite{FM95, Me95, Pi97, Pi99}.  GRBs do
not repeat except for a few soft gamma--ray repeaters. They are
detected with a rate of about one event per day, and their duration
varies from milliseconds to minutes. The duration of the shorter
bursts and the temporal substructure of the longer bursts implies a
geometrically small source (less than $\sim c \cdot 1 {\rm msec} \sim
100\,$km), which in turn points towards compact objects, like neutron
stars or black holes. The emitted gamma--rays have energies in the
range 30\,keV to 2\,MeV.

Concerning the distance of GRB sources major progress has occured
through the observations by the BATSE detector on board the Compton
Gamma--Ray Observatory (GRO), which have proven that GRBs are
distributed isotropically over the sky \cite{MF92}. Even more
important the detection and the rapid availability of accurate
coordinates ($\sim$ arc minutes) of the fading X--ray counterparts of
GRBs by the BeppoSAX spacecraft beginning in 1997 \cite{CX97, PX98}
has allowed for subsequent successful ground based observations of
faint GRB afterglows at optical and radio wavelength. In case of
GRB\,990123 the optical, X--ray and gamma--ray emission was detected
for the first time almost simultaneously (optical observations began
22 seconds after the onset of the GRB) \cite{BB99, AB99}.  From
optical spectra thus obtained redshifts of several gamma--ray bursts
have been determined, \eg GRB\,970508 ($z = 0.835$ \cite{Me97, PB98}),
GRB\,971214 ($z = 3.42$ \cite{KD98}), GRB\,980703 ($z = 0.966$
\cite{DK98}) and GRB\,990123 ($1.60 \leq z < 2.05$ \cite{An99}), which
confirm that (at least some) GRBs occur at cosmological
distances. Assuming isotropic emission the inferred total energy of
cosmological GRBs emitted in form of gamma--rays ranges from several
$10^{51}\,$erg to $3\,10^{53}\,$erg (for GRB\,971214) \cite{CT99}, and
exceeds $10^{54}\,$erg for GRB\,990123 \cite{An99, BB99}.  Updated
information on GRBs localized with BeppoSAX, BATSE/RXTE(PCA) or
BATSE/RXTE(ASM) can be obtained from a web site maintained by Greiner
\cite{Gr99}.

The compact nature of the GRB source, the observed flux and the
cosmological distance taken together imply a large photon density.
Such a source has a large optical depth for pair production. This is,
however, inconsistent with the optically thin source indicated by the
non--thermal gamma--ray spectrum, which extends well beyond the pair
production threshold at 500\,keV. This problem can be resolved by
assuming an ultra-relativistic expansion of the emitting region, which
eliminates the compactness constraint. The bulk Lorentz factor
required are then $W > 100$ (see, \eg \cite{Pi99})

In April 1998 the pure cosmological origin of GRBs was challenged by
the detection of the Type Ib/c supernova SN\,1998bw \cite{GV98a,
GV98b} within the 8\,arc minute error box of GRB\,980425 \cite{SF98,
PA99}. Its explosion time is consistent with that of the GRB, and
relativistic expansion velocities are derived from radio observations
of SN\,1998bw \cite{KF98}. BeppoSAX detected two fading X--ray sources
within the error box, one being positionally consistent with the
supernova and a fainter one not consistent with the position of
SN\,1998bw \cite{PA99}. Taken together these facts suggest a
relationship between GRBs and SNe\,Ib/c, \ie core collapse supernovae
of massive stellar progenitors which have lost their hydrogen and
helium envelopes \cite{GV98b, Iw98, WE98}. As the host galaxy
ESO\,184-82 of SN\,1998bw is only at a redshift of $z = 0.0085$
\cite{TS98} and as GRB\,980425 was not extraordinarily bright,
GRB--supernovae are more than four orders of magnitude fainter ($E
{\rm tot}_{\gamma} = 7\,10^{47}\,$erg for GRB\,980425 \cite{CT99})
than a typical cosmological GRB. However, the observation of the
second fading X--ray source within the error box of GRB\,980425 and
unrelated with SN\,1998bw still causes some doubts on the GRB
supernova connection, although the probability of chance coincidence
of GRB\,980425 and SN\,1998bw is extremely low \cite{PA99}.

In order to explain the energies released in a GRB various
catastrophic collapse events have been proposed including
neutron--star/neutron--star mergers \cite{Pa86, Go86, Ei89},
neutron--star/black--hole mergers \cite{MH93}, collapsars \cite{Wo93,
MF99} and hypernovae \cite{Pa98}. These models all rely on a common
engine, namely a stellar mass black hole which accretes several solar
masses of matter from a disk (formed during a merger or by a
non--spherical collapse) at a rate of $\sim 1\,\ms\,$s$^{-1}$
\cite{PW98}. A fraction of the gravitational binding energy released
by accretion is converted into neutrino and anti--neutrino pairs,
which in turn annihilate into electron--positron pairs. This creates a
pair fireball, which will also include baryons present in the
environment surrounding the black hole. Provided the baryon load of
the fireball is not too large, the baryons are accelerated together
with the e$^+\,$e$^-$ pairs to ultra-relativistic speeds with Lorentz
factors $> 10^2$ \cite{CR78, PS93, Pi99}. The existence of such
relativistic flows is supported by radio observations of GRB\,980425
\cite{KF98}. It has been further argued that the rapid temporal decay
of several GRB afterglows is inconsistent with spherical (isotropic)
blast wave models, and instead is more consistent with the evolution
of a relativistic jet after it slows down and spreads laterally
\cite{SP99}. Independent of the flow pattern the bulk kinetic energy of
the fireball then is thought to be converted into gamma-rays via
cyclotron radiation and/or inverse Compton processes (see, \eg
\cite{Me95, Pi99}).

One--dimensional numerical simulations of spherically symmetric
relativistic fireballs have been performed by several authors to model
GRB sources \cite{PS93, PM98, PM99}.  Multi-dimensional modelling of
ultra--relativistic jets in the context of GRBs has for the first time
been attempted by Aloy \etal \cite{AM99}. Using a collapsar progenitor
model of MacFadyen \& Woosley \cite{MF99} they have simulated the
propagation of an axisymmetric jet through the mantle and envelope of
a collapsing massive star ($10\,\ms$) using the GENESIS special
relativistic hydrodynamic code \cite{AI99a}. The jet forms as a
consequence of an assumed energy deposition of $10^{51}\,$erg/sec
within a 30 degree cone around the rotation axis. At break-out, \ie
when the jets reaches the surface of the stellar progenitor, the
maximum Lorentz factor of the jet flow is about 20.

\section{CONCLUSION}
%        ##########
\label{s:concl}

\subsection{Evaluation of the methods}
%           -------------------------

  An assessment of the quality of the numerical methods should
consider, at least, the following aspects: (i) accuracy and robustness
in describing high Lorentz factor flows with strong shocks; (ii)
effort required to extend to multi dimensions; (iii) effort required
to extend to RMHD and GRHD. 
%% modified
In Table\,(\ref{t:perform-methods}) we have summarized these aspects
of numerical methods for SRHD.

  Since their introduction in numerical RHD at the beginning of
nineties, HRSC methods have demonstrated their ability to describe
accurately (stable and without excessive smearing) relativistic flows
of arbitrarily large Lorentz factors and strong discontinuities
reaching the same quality as in classical hydrodynamics. In addition
(as it is the case for classical flows, too), HRSC methods show the
best performance compared to any other method (\eg artificial
viscosity, FCT or SPH).

  Despite of the latter fact, a lot of effort has been put into
improving these non HRSC methods.  Using a consistent formulation of
artificial viscosity has significantly enhanced the capability of
finite difference schemes \cite{NW86} as well as of relativistic SPH
\cite{SR99} to handle strong shocks without spurious post--shock
oscillations.  However, this comes at the price of a large numerical
dissipation at shocks. Concerning relativistic SPH recent
investigations using a conservative formulation of the hydrodynamic
equations \cite{CM97,SR99} have reached an unprecedented accuracy with
respect to previous simulations, although some issues still
remain. Besides the strong smearing of shocks, the description of
contact discontinuities and of thin structures moving at
ultrarelativistic speeds needs to be improved (see
Sect.\,\ref{ss:blast}).

  Concerning FCT techniques, those codes based on a conservative
formulation of the RHD equations have been able to handle relativistic
flows with discontinuities at all flow speeds, although the quality of
the results is below that of HRSC methods in all cases \cite{SK93}.

  The extension to multi dimensions is simple for most relativistic
codes. Finite difference techniques are easily extended using
directional splitting. Note, however, that HRSC methods based on exact
solutions of the Riemann problem \cite{MM96,WP97} first require the
development of a multidimensional version of the relativistic Riemann
solver. The adapting--grid, artificial viscosity, implicit code of
Norman \& Winkler \cite{NW86} and the relativistic Glimm method of Wen
\etal \cite{WP97} are restricted to one dimensional flows. Note that
Glimm's method produces the best results in all the tests analyzed in
Sect.\,\ref{s:tests}.

  The symmetric TVD scheme proposed by Davis \cite{Da84} and extended
to GRMHD (see below) by Koide \etal \cite{KN96} combines several 
characteristics
making it very attractive. It is written in conservation form and is
TVD, \ie it is converging to the physical solution. In addition, it is
independent of spectral decompositions, which allows for a simple
extension to RMHD.  Quite similar statements can be made about the
approach proposed by van Putten \cite{vP93}. In contrast to FCT
schemes (which are also easily extended to general systems of
equations), both Koide \etal's and van Putten's method are very stable
when simulating mildly relativistic flows (maximum Lorentz factors
$\approx 4$) with discontinuities. Their only drawback is an excessive
smearing of the latter. A comparison of Davis' method with Riemann
solver based methods would be desirable.

%%%%%%%%%%%%%%%%%%%%%%%%%%%%%%%%%%%%%%%%%%%%%%%%%%%%%%%%%%%%%%%%%%%%%%%%%%%%%%
%
{\small

\begin{table}[]
\begin{center}
\caption{\protect \small Evaluation of numerical methods for
SRHD. Methods have been categorized for clarity}
\label{t:perform-methods}
\begin{tabular}{lccccc}

\hline

\multicolumn{1}{c}{Method} & Ultrarelativistic & Handling of            & Extension to several      & \multicolumn{2}{c}{Extension to} \\
                           & regime            & discontinuities $^{a}$ & spatial dimensions $^{b}$ & GRHD & RMHD \\

\hline

AV-mono                    & $\times$           & O, SE           & $\surd$                  & $\surd$ & $\times$ \\
cAV-implicit               & $\surd$            & $\surd$         & $\times$                        & $\times$ & $\times$    \\
HRSC $^{c}$                & $\surd$            & $\surd$         & $\surd$ $^{d}$            & $\surd$ $^{e}$ & $\times$ $^{f}$ \\
rGlimm                     & $\surd$            & $\surd$         & $\times$                        & $\times$ & $\times$    \\
sTVD                       & $\surd$ $^{g}$      & D               & $\surd$                  & $\surd$ & $\surd$ \\
van Putten                 & $\surd$ $^{g}$      & D               & $\surd$                  & $\times$ & $\surd$          \\
FCT                        & $\surd$            & O               & $\surd$                  & $\times$ & $\times$            \\
SPH                        & $\surd$            & D, O            & $\surd$                  & $\surd$ $^{h}$ & $\times$ $^{i}$  \\

\hline
\end{tabular}
\end{center}

$^{a}$ D: excessive dissipation; O: oscillations; SE: systematic errors. \\
$^{b}$ all finite difference methods are extended by directional splitting.\\
$^{c}$ contains all the methods listed in Table\,\ref{t:methods} with
       exception of rGlimm \cite{WP97} and sTVD \cite{KN96}. \\
$^{d}$ rPPM \cite{MM96} requires an exact relativistic Riemann solver
       with non-zero transverse speeds. \\
$^{e}$ there exist GRHD extensions of several HRSC methods based on
       linearized Riemann solvers.  The procedure developed by Pons
       \etal \cite{PF98} allows any SRHD Riemann solver to be applied
       to GRHD flows. \\
$^{f}$ except HLL which requires spectral decomposition of RMHD
       equations or solution of RMHD Riemann problem.  Van Putten
       \cite{vP98} has studied the characteristic structure of the
       RMHD equations in (constraint free) divergence form as a first
       step to extend modern HRSC methods to RMHD. Komissarov
       \cite{Ko99} has developed a multidimensional RMHD code based on
       a linearized Riemann solver. \\
$^{g}$ needs confirmation. \\
$^{h}$ codes of refs.\,\cite{LM93,SR99}. \\
$^{i}$ There is one code which considered such an extension \cite{Ma91},
       but the results are not completely satisfactory. \\

\end{table}

}
%
%%%%%%%%%%%%%%%%%%%%%%%%%%%%%%%%%%%%%%%%%%%%%%%%%%%%%%%%%%%%%%%%%%%%%%%%%%%%%

\subsection{Further developments}
%           --------------------
\label{ss:fdevelop}

The directions of future developments in this field of research are
quite obvious. They can be divided into four main categories:

\subsubsection{Incorporation of realistic microphysics} 
Up to now most astrophysical SRHD simulations have assumed matter
whose thermodynamic properties can be described by an inviscid ideal
equation of state with a constant adiabatic index.  This
simplification may have been appropriate in the first generation of
SRHD simulations, but it clearly must be given up when aiming at a
more realistic modelling of astrophysical jets, gamma--ray burst
sources or accretion flows onto compact objects. For these phenomena a
realistic equation of state should include contributions from
radiation ($\gamma=4/3$ ``fluid''), allow for the formation of
electron--positron pairs at high temperatures, allow the ideal gas
contributions to be arbitrarily degenerate and/or relativistic.
Effects due to cooling, heat conduction, nuclear transmutations and
viscosity may have to be considered, too.  When simulating
relativistic heavy ion collisions the use of a realistic equation of
state is essential for an adequate description of the
phenomenon. However, as these simulations have been performed with FCT
based difference schemes (see, \eg \cite{SH97}), this poses no
specific numerical problem. The simulation of flows obeying an
elaborated microphysics with HRSC methods needs in some cases the
extension of the present relativistic Riemann solvers to handle
general equations of state. This is the case of the Roe-Eulderink
method (extendable by the procedure developed in the classical case by
Glaister \cite{Gl88}), and rPPM and rGlimm both relying on a exact
solution of the Riemann problem for ideal gases with constant
adiabatic exponent (which can also be extended following the procedure
of Colella \& Glaz \cite{CG85} for classical hydrodynamics). We expect
the second generation of SRHD codes being capable of treating general
equations of state and various source/sink terms routinely.

\subsubsection{Coupling of SRHD schemes with AMR} 
Modelling astrophysical phenomena often involves an enormous range of
length scales and time scales to be covered in the simulations. In two
and definitely in three spatial dimensions many such simulations
cannot be performed with sufficient spatial resolution on a static
equidistant or non--equidistant computational grid, but they will
require dynamic, adaptive grids. In our opinion the most promising
approach in this direction will be the coupling of SRHD solvers with
the adaptive mesh refinement (AMR) techniques \cite{BC89}. AMR
automatically increases the grid resolution near flow discontinuities
or in regions of large gradients (of the flow variables) until a
prescribed accuracy of the difference approximation is achieved. A
SRHD simulation of a relativistic jet based on a combined HLL--AMR
scheme was performed by Duncan \& Hughes \cite{DH94}. Plewa \etal
\cite{PM97} have modelled the deflection of highly supersonic jets
propagating through non--homogeneous environments using the HRSC
scheme of Mart\'{\i} \etal \cite{MM97} combined with the AMR
implementation AMRA of Plewa \cite{Plew}.  Komissarov \& Falle
\cite{KF97} have combined their numerical scheme with the adaptive
grid code, Cobra, which has been developed by Mantis Numerics Ltd. for
industrial applications \cite{FG93}, and which uses a hierarchy of
grids with a constant refinement factor of two between subsequent grid
levels.

\subsubsection{General relativistic hydrodynamics (GRHD)}
Up to now only very few attempts have been made to extend HRSC methods
to GRHD and all of these have used linearized Riemann solvers
\cite{MI91, EM95, RI96, BF97, FM99}. In the most recent of these
approaches Font \etal \cite{FM99} have developed a 3D general
relativistic HRSC hydrodynamic code where the matter equations are
integrated in conservation form and fluxes are calculated with
Marquina's formula.  A very interesting and powerful procedure has
been proposed by Pons \etal \cite{PF98}, which allows one to exploit
all the developments in the field of special relativistic Riemann
solvers in general relativistic hydrodynamics.  The procedure relies
on a local change of coordinates at each zone interface such that the
spacetime metric is locally flat. In that locally flat spacetime any
special relativistic Riemann solver can be used to calculate the
numerical fluxes, which are then transformed back. Finer grids and
improved time advancing methods will be required in regions where
large gradients or temporal variations of the gravitational field are
encountered. The numerical implementation is simple and
computationally inexpensive.

The characteristic formulations of the Einstein field equations are
able to handle the long term numerical description of single black
hole space times in vacuum \cite{BG97}. In order to include matter in
such an scenario, Papadopoulos \& Font \cite{PF99} have generalized
HRSC methods to cope with the hydrodynamic equations in this null
foliation.

Other developments in GRHD in the past included finite element methods
for simulating spherically symmetric collapse in general relativity
\cite{Ma3a}, general relativistic pseudo--spectral codes based on the
(3+1) ADM formalism \cite{AD62} for computing radial perturbations
\cite{Go92} and 3D gravitational collapse of neutron stars
\cite{BF96}, and general relativistic SPH \cite{Ma91}. The potential
of these methods for the future is unclear, as none of them is
specifically appropriate for ultrarelativistic speeds and strong shock
waves which are characteristic of most astrophysical applications.

Peitz \& Appl \cite{PA98} have addressed the difficult issue of
non--ideal GRHD, which is of particular importance, \eg for the
simulation of accretion discs around compact objects, rotating
relativistic fluid configurations, and the evolution of density
fluctuations in the early universe. They have accounted for
dissipative effects by applying the theory of extended causal
thermodynamics, which eliminates the causality violating infinite
signal speeds arising from the conventional Navier--Stokes equation.
Peitz \& Appl have not implemented their model numerically yet.

\subsubsection{Relativistic magneto--hydrodynamics (RMHD)}
The inclusion of magnetic effects is of great importance in many
astrophysical flows. The formation and collimation process of
(relativistic) jets most likely involves dynamically important
magnetic fields and occurs in strong gravitational fields. The same is
likely to be true for accretion discs around black holes.
Magneto--relativistic effects even play a non negligible role in the
formation of proto--stellar jets in regions close to the light
cylinder \cite{Ca98}. Thus, relativistic MHD codes are a very
desirable tool in astrophysics. The non-trivial task of developing
such a kind of code is considerably simplified by the fact that
because of the high conductivity of astrophysical plasmas one must
only consider ideal RMHD in most applications.

Special relativistic 2D MHD test problems with Lorentz factors up to
$\sim3$ have been investigated by Dubal \cite{Du91} with a code based on
FCT tecniques (see Sect.~\ref{s:other}). In a series
of papers Koide and coworkers \cite{KN96, Ko97, NK97, NK98, KS98} have
investigated relativistic magnetized jets using a symmetric TVD scheme
(see Sect.~\ref{s:hrsc}).  Koide, Nishikawa \& Mutel \cite{KN96} simulated a 
2D RMHD slab jet, whereas Koide \cite{Ko97} investigated the effect of an
oblique magnetic field on the propagation of a relativistic slab jet.
Nishikawa \etal \cite{NK97, NK98} extended these simulations to 3D and
considered the propagation of a relativistic jet with a Lorentz
factor $W = 4.56$ along an aligned and an oblique external magnetic field. 
The 2D and 3D simulations published up to now only cover the very early
propagation of the jet (up to 20 jet radii) and are performed with
moderate spatial resolution on an equidistant Cartesian grid (up to
101 zones per dimension, \ie 5 zones per beam radius).  

Van Putten \cite{vP92,vP93} has proposed a method for accurate and stable
numerical simulations of RMHD in the presence of dynamically
significant magnetic fields in two dimensions and up to moderate
Lorentz factors. The method is based on MHD in divergence form using a
2D shock--capturing method in terms of a pseudo--spectral smoothing
operator (see Sect.~\ref{s:other}. He applied this method to 2D blast waves
\cite{vP94} and astrophysical jets \cite{vP93b,vP96}.

Steps towards the extension of linearized Riemann solvers to ideal RMHD have 
already been taken. Romero \cite{RI97} has derived an analytical 
expression for the spectral decomposition of the Jacobian in the case of a
planar relativistic flow field permeated by a transversal magnetic field 
(nonzero field component only orthogonal to flow direction). Van Putten
\cite{vP98} has studied the characteristic structure of the RMHD equations in 
(constraint free) divergence form. Finally, Komissarov \cite{Ko99} has presented 
a robust Godunov--type scheme for RMHD, which is based on a linear Riemann
solver, has second--order accuracy in smooth regions, enforces magnetic flux 
conservation, and which can cope with ultrarelativistic flows. 

We end with the simulations performed by Koide, Shibata \& Kudoh 
\cite{KS98} on magnetically driven axisymmetric jets from black hole accretion 
disks. Their GRMHD code is an extension of the special relativistic MHD code 
developed by Koide \etal \cite{KN96, Ko97, NK97}. The necessary modifications 
of the code were quite simple, because in the (nonrotating) black
hole's Schwarzschild spacetime the GRMHD equations are identical to
the SRMHD equations in general coordinates, except for the gravitational force 
terms and the geometric factors of the lapse function. With the pioneering work 
of Koide, Shibata \& Kudoh the epoch of exciting GRMHD simulations has just 
begun.

\section{ADDITIONAL INFORMATION}
%        ######################
\label{s:additional}

\subsection{Algorithms to recover primitive quantities}
%           ------------------------------------------    
\label{ss:lfq}

The expressions relating the primitive variables $(\rho, v^i, p)$ to
the conserved quantities $(D, S^i, \tau)$ depend explicitly on the
equation of state $p(\rho, \varepsilon)$ and simple expressions are
only obtained for simple equations of state (\ie ideal gas).

A function of pressure, whose zero represents the pressure in the
physical state, can easily be obtained from Eqs.\,(\ref{4}--\ref{6}),
(\ref{lorentz}) and (\ref{12}):
\begin{equation}
  f(\bar{p}) = p \left( \rho_\ast (\bar{p}),\varepsilon_\ast (\bar{p})
                 \right)  -  \bar{p}
\label{16}
\end{equation}
with $\rho_\ast (\bar{p})$ and $\varepsilon_\ast (\bar{p})$ given by
\begin{equation}
  \rho_\ast (\bar{p}) = \frac{D}{W_\ast (\bar{p})} \, ,
\label{17}
\end{equation}
and
\begin{equation}
  \varepsilon_\ast (\bar{p}) = \frac{\tau + D\,[1 - W_\ast (\bar{p})] +
                                      \bar{p}\,[1 - W_\ast (\bar{p})^2])
                                    }{D\,W_\ast (\bar{p})} \, ,
\label{18}
\end{equation}
where
\begin{equation}
  W_\ast (\bar{p}) = \frac{1}{{\displaystyle \sqrt{1 - v_\ast^i(\bar{p})
                               v_{\ast \, i}(\bar{p})}}}  \,
\label{19}
\end{equation}
and
\begin{equation}
  v_\ast^i(\bar{p}) = \frac{S^i}{\tau + D + \bar{p}} \, .
\end{equation}
The root of (\ref{16}) can be obtained by means of a nonlinear
root--finder (\eg a one--dimensional Newton--Raphson iteration). For
an ideal gases with a constant adiabatic exponent such a procedure has
proven to be very successful in a large number of tests and
applications \cite{MI91, MM96, MM97}. The derivative of $f$ with
respect to $\bar{p}$, $f^\prime$, can be approximated by \cite{AI99a}
\begin{equation}
f^\prime = v_\ast^i(\bar{p}) v_{\ast \, i}(\bar{p}) c_{s \,
           \ast}(\bar{p})^2 - 1 \, , 
\label{fapprox} 
\end{equation}
where $c_{s\ast}$ is the sound speed which can efficiently be computed
for any EOS. Moreover, approximation (\ref{fapprox}) tends to the
exact derivative as the solution is approached.

Eulderink \cite{Eu93, EM95} has also developed several procedures to
calculate the primitive variables for an ideal EOS with a constant
adiabatic index.  One procedure is based on finding the physically
admissible root of a fourth--order polynomial of a function of the
specific enthalpy. This quartic equation can be solved analytically by
the exact algebraic quartic root formula although this computation is
rather expensive. The root of the quartic can be found much more
efficiently using a one--dimensional Newton--Raphson iteration.
Another procedure is based on the use of a six--dimensional
Newton--Kantorovich method to solve the whole nonlinear set of
equations.

Also for ideal gases with constant gamma, Schneider \etal \cite{SK93}
transform system (\ref{4}--\ref{6}), (\ref{lorentz}) and (\ref{12})
algebraically into a fourth--order polynomial in the modulus of the
flow speed, which can be solved analytically or by means of iterative
procedures.

For a general EOS, Dean \etal \cite{DB94} and Dolezal \& Wong
\cite{DW95} proposed the use of iterative algorithms for $v^2$ and
$\rho$, respectively.

\subsection{Spectral decomposition of the 3D SRHD equations}
%           -----------------------------------------------
\label{ss:spectral}

The full spectral decomposition (right and left eigenvectors) of the Jacobian 
matrices associated to the SRHD system in 3D has been first derived by Donat 
\etal \cite{DF98}. Previously, Mart\'{\i} \etal. \cite{MI91} obtained the 
spectral decomposition in 1D SRHD and Eulderink \cite{Eu93} and Font \etal 
\cite{FI94}, the (eigenvalues and) right eigenvectors in 3D. The Jacobians are 
given by 
\begin{equation}
  {\cal B}^{i} = \frac{\partial{\bf F}^{i}({\bf u})}
                      {\partial \bf u \rm} \, ,
\label{B}
\end{equation}
where the state vector ${\bf u}$ and the flux vector ${\bf F}^{i}$ are
defined in (\ref{22}) and (\ref{23}), respectively. In the following
we explicitly give both the eigenvalues and the right and left
eigenvectors of the Jacobi matrix ${\cal B}^{x}$ only (the cases $i =
y, z$ are easily obtained by symmetry considerations).

\noindent
The eigenvalues of matrix ${\cal B}^x({\bf u})$ are

\begin{eqnarray}
\lambda_{\pm} = \frac{1}{1- v^2 c_s^2}
                \left\{v^{x}(1-c_{s}^{2})  {\pm} c_{s}
                       \sqrt{(1-{\rm v}^{2}) 
                             [1-v^x v^x-(v^2-v^x v^x)c_{s}^{2}]}
                \right\}  \, ,
\label{lambdapm}
\end{eqnarray}
and
\begin{equation}
  \lambda_0 = v^x \mbox{\,\,\,\,(triple)} \, .
\label{lambda0}
\end{equation}
A complete set of right-eigenvectors is 
\begin{eqnarray}
  {\bf r}_{0,1} =  \left( {\displaystyle{\frac{{\cal K}}{h W}}},
  v^x, v^y, v^z, 1 - {\displaystyle{\frac{{\cal K}}{h W}}} \right)
\end{eqnarray}
\begin{eqnarray}
  {\bf r}_{0,2} =  \left( W v^y , 2 h W^2 v^x v^y,
  h(1+2 W^2 v^y v^y), 2 h W^2 v^y v^z, 2 h W^2 v^y - W v^y\right)
\end{eqnarray}
\begin{eqnarray}
  {\bf r}_{0,3} = \left( W v^z , 2 h W^2 v^x v^z,
  2 h W^2 v^y v^z, h(1+2 W^2 v^z v^z),  2 h W^2 v^z - W v^z\right)
\end{eqnarray}
\begin{eqnarray}
  {\bf r}_{\pm} = (1, h W {\cal A}_{\pm} {\lambda}_{\pm}, h W v^y,
  h W v^z, h W {\cal A}_{\pm} - 1)
\end{eqnarray}
where
\begin{eqnarray}
 {\cal K} \equiv {\displaystyle{\frac{\tilde{\kappa}}
                                     {\tilde{\kappa} - c_s^2}} \, , \quad 
 {\tilde \kappa}= \frac{\kappa}{\rho}} \, , \quad
 {\cal A}_{\pm} \equiv {\displaystyle{\frac{1 - v^x v^x}
                                     {1 - v^x {\lambda}_{\pm}}}} \, .
\end{eqnarray}
The corresponding complete set of left-eigenvectors is
\begin{eqnarray*}
  {\bf l}_{0,1} = {\displaystyle{\frac{W}{{\cal K} - 1}} }
  (h - W, W v^x, W v^y, W v^z, -W)
\end{eqnarray*}
\begin{eqnarray*}
  {\bf l}_{0,2} = {\displaystyle{\frac{1}{h (1 - v^x v^x)}} }
  (- v^y, v^x v^y, 1 - v^x v^x, 0, -v^y)
\end{eqnarray*}
\begin{eqnarray*}
  {\bf l}_{0,3} = {\displaystyle{\frac{1}{h (1 - v^x v^x)}} }
  (- v^z, v^x v^z, 0, 1 - v^x v^x, -v^z)
\end{eqnarray*}
\[
  {\bf l}_{\mp} = ({\pm} 1){\displaystyle{\frac{h^2}{\Delta}}}
  \left[\begin{array}{c}
        h W {\cal A}_{\pm} (v^x - {\lambda}_{\pm}) -
        v^x - W^2 (v^2 - v^x v^x) (2 {\cal K} - 1)
        (v^x - {\cal A}_{\pm} {\lambda}_{\pm}) +
        {\cal K} {\cal A}_{\pm} {\lambda}_{\pm}
\\ \\
        1 + W^2 (v^2 - v^x v^x) (2 {\cal K} - 1) (1 - {\cal A}_{\pm}) -
        {\cal K} {\cal A}_{\pm}  
\\ \\
        W^2 v^y (2 {\cal K} -1) {\cal A}_{\pm} (v^x -{\lambda}_{\pm}) 
\\ \\
        W^2 v^z (2 {\cal K} -1) {\cal A}_{\pm} (v^x -{\lambda}_{\pm}) 
\\ \\
        -v^x - W^2 (v^2 - v^x v^x) (2 {\cal K} - 1)
        (v^x - {\cal A}_{\pm} {\lambda}_{\pm}) +
        {\cal K} {\cal A}_{\pm} {\lambda}_{\pm}
  \end{array} \right]
\]
where $\Delta$ is the determinant of the matrix of right-eigenvectors,
\ie
\begin{eqnarray}
  \Delta = h^3 W ({\cal K} - 1)\, (1 - v^x v^x) \,
           ({\cal A}_{+} {\lambda}_{+} - {\cal A}_{-} {\lambda}_{-}) \, .
\end{eqnarray}
For an ideal gas equation of state ${\cal K} = h$, \ie ${\cal K} > 1$,
and hence $\Delta \neq 0$ for $|v^x| < 1$.

\subsection{Program {\tt RIEMANN}}
%           --------------------
\label{ss:riemann}

{\tt
      PROGRAM RIEMANN \\
C
C     This program computes the solution of a 1D  \\
c     relativistic Riemann problem with \\
C     initial data UL if X<0.5 and UR if X>0.5 \\
C     in the whole spatial domain [0, 1] \\
      ... \\
      END
}

\subsection{Basics of HRSC methods and recent developments}
%           ----------------------------------------------
\label{ss:basicshrsc}
  
In this section we introduce the basic notation of finite
differencing and summarize recent advances in the development of HRSC
methods for hyperbolic systems of conservation laws. The content of
this section is not specific to SRHD, but applies to hydrodynamics in
general.

In order to simplify the notation and taking into account that most
powerful results have been derived for scalar conservation laws in one
spatial dimension, we will restrict ourselves to the initial value
problem given by the equation
\begin{equation}
  \frac{\partial u}{\partial t} + \frac{\partial f(u)}{\partial x} = 0 
\label{scalarcl}
\end{equation}  
with the initial condition $u(x,t=0) = u_0(x)$.

In hydrodynamic codes based on finite difference or finite volume
techniques, equation (\ref{scalarcl}) is solved on a discrete
numerical grid $(x_j,t^n)$ with
\begin{equation}
x_j = (j - 1/2) \Delta x, \,\,\,\, j=1,2,\ldots \, ,
\end{equation}
and
\begin{equation} 
t^n = n \Delta t, \,\,\,\, n=0,1,2,\ldots \, ,
\end{equation}
where $\Delta t$ and $\Delta x$ are the time step and the zone size,
respectively. A difference scheme is a time--marching procedure
allowing one to obtain approximations to the solution at the new time,
$u_j^{n+1}$, from the approximations in previous time steps.  Quantity
$u_j^n$ is an approximation to $u(x_j,t^n)$ but, in the case of a
conservation law, it is often preferable to view it as an
approximation to the average of $u(x, t)$ within a zone $[x_{j-1/2},
x_{j+1/2}]$ (\ie as a zone average), where $x_{j \pm 1/2} = (x_j +
x_{j \pm 1})/2$. Hence
\begin{equation}
  \bar{u}_j^n = \frac{1}{\Delta x} \int_{x_{j-1/2}}^{x_{j+1/2}} 
                                         u(x,t^n) dx \, ,
\end{equation}
which is consistent with the integral form of the conservation law.

Convergence under grid refinement implies that the global error
$||E_{\Delta x}||$, defined as
\begin{equation}
  ||E_{\Delta x}|| = \Delta x \sum_j |\bar{u}_j^n-u_j^n|,
\end{equation}
tends to zero as $\Delta x \rightarrow 0$. For hyperbolic systems of
conservation laws methods in conservation form are preferred as they
guarantee that if the numerical solution converges, it converges to a
weak solution of the original system of equations (Lax--Wendroff
theorem \cite{LW60}).  Conservation form means that the algorithm can
be written as
\begin{equation}
  u_j^{n+1} = u_j^n - \frac{\Delta t}{\Delta x} 
              \left(\hat{f}(u_{j-r}^n,u_{j-r+1}^n,\ldots,u_{j+q}^n) -
                    \hat{f}(u_{j-r-1}^n,u_{j-r}^n,\ldots,u_{j+q-1}^n) 
              \right) \,
\label{un+1}
\end{equation}
where $q$ and $r$ are positive integers, and $\hat{f}$ is a consistent
(\ie $\hat{f} (u, u, \ldots, u) = f(u)$) numerical flux function.

  The Lax--Wendroff theorem cited above does not establish whether the
method converges. To guarantee convergence, some form of stability is
required, as for linear problems (Lax equivalence theorem
\cite{RM67}).  In this context the notion of total--variation
stability has proven to be very successful, although powerful results
have only been obtained for scalar conservation laws. The total
variation of a solution at $t = t^n$, TV($u^n$), is defined as
\begin{equation}
  {\rm TV}(u^n) = \sum_{j=0}^{+\infty} |u_{j+1}^n - u_j^n| \, .
\end{equation}
A numerical scheme is said to be TV--stable, if TV($u^n$) is bounded
for all $\Delta t$ at any time for each initial data. One can then
prove the following convergence theorem for non--linear, scalar
conservation laws \cite{Le92}: For numerical schemes in conservation
form with consistent numerical flux functions, TV--stability is a
sufficient condition for convergence.

  Modern research has focussed on the development of high--order,
accurate methods in conservation form, which satisfy the condition of
TV--stability. The conservation form is ensured by starting with the
integral version of the partial differential equations in conservation
form (finite volume methods). Integrating the PDE over a finite
space--time domain $[x_{j-1/2}, x_{j+1/2}] \times [t^n,t^{n+1}]$ and
comparing with (\ref{un+1}), one recognizes that the numerical flux
function $\hat{f}_{j+1/2}$ is an approximation to the time--averaged
flux across the interface, \ie
\begin{equation}
  \hat{f}_{j+1/2} \approx \frac{1}{\Delta t} \int_{t^n}^{t^{n+1}}
                                             f(u(x_{j+1/2},t)) dt \, .
\end{equation}
Note that the flux integral depends on the solution at the zone
interface, $u(x_{j+1/2}, t)$, during the time step.  Hence, a possible
procedure is to calculate $u(x_{j+1/2}, t)$ by solving Riemann
problems at every zone interface to obtain
\begin{equation}
  u(x_{j+1/2},t) = u(0;u_j^n,u_{j+1}^n) \, .
\end{equation}
This is the approach followed by an important subset of
shock--capturing methods, called Godunov--type methods \cite{HL83,
Ei88} after the seminal work of Godunov \cite{Go59}, who first used an
exact Riemann solver in a numerical code. These methods are written in
conservation form and use different procedures (Riemann solvers) to
compute approximations to $u(0; u_j^n, u_{j+1}^n)$. The book of Toro
\cite{To97} gives a comprehensive overview of numerical methods based
on Riemann solvers.  The numerical dissipation required to stabilize
an algorithm across discontinuities can also be provided by adding
local conservative dissipation terms to standard finite--difference
methods. This is the approach followed in the symmetric TVD schemes
developed in \cite{Da84, Ro84, Ye87}.

  High--order of accuracy is usually achieved by using conservative
monotonic polynomial functions to interpolate the approximate
solution within zones.  The idea is to produce more accurate left and
right states for the Riemann problem by substituting the mean values
$u_j^n$ (that give only first--order accuracy) by better
representations of the true flow near the interfaces, let say
$u_{j+1/2}^{\rm L}$, $u_{j+1/2}^{\rm R}$.  The FCT algorithm
\cite{BB73} constitutes an alternative procedure where higher accuracy
is obtained by adding an anti--diffusive flux term to the first--order
numerical flux. The interpolation algorithms have to preserve the
TV--stability of the scheme. This is usually achieved by using
monotonic functions which lead to the decrease of the total variation
(total--variation--diminishing schemes, TVD \cite{Ha84}). High--order
TVD schemes were first constructed by van Leer \cite{vL79}, who
obtained second--order accuracy by using monotonic piecewise linear
slopes for cell reconstruction. The piecewise parabolic method (PPM)
\cite{CW84b} provides even higher accuracy. The TVD property implies
TV--stability, but can be too restrictive. In fact, TVD methods
degenerate to first--order accuracy at extreme points
\cite{OC84}. Hence, other reconstruction alternatives have been
developed where some growth of the total variation is allowed. This is
the case for the total--variation--bounded (TVB) schemes \cite{Sh87},
the essentially non-oscillatory (ENO) schemes \cite{HE87} and the
piecewise--hyperbolic method (PHM) \cite{Ma94}.

\subsection{Newtonian SPH equations}
%           -----------------------
\label{ss:SPHeqs}

Following Monaghan \cite{Mo97} the SPH equation of motion for a
particle $a$ with mass $m$ and velocity $\vecv$ is given by
\begin{equation}
  {d \vecv_a \over d t}   =  - \sum_b m_b \rund{  {p_a \over \rho_a^2}
                                               + {p_b \over \rho_b^2}
                                               + \tpiab }
                                       \nabla_a W_{ab} \; ,
\label{SPH-01}
\end{equation}
where the summation is over all particles other than particle $a$, $p$
is the pressure, $\rho$ is the density, and $d / dt$ denotes the
Lagrangian time derivative.  $\tpiab$ is the artificial viscosity
tensor, which is required in SPH to handle shock waves. It poses a
major obstacle in extending SPH to relativistic flows (see, \eg
\cite{Is87, CM97}). $W_{ab}$ is the interpolating kernel, and
$\nabla_a W_{ab}$ denotes the gradient of the kernel taken with
respect to the coordinates of particle $a$.  

The kernel is a function of $|\vecr_a-\vecr_b|$ (and of the SPH
smoothing length $h_{\rm SPH}$), \ie its gradient is given by
\begin{equation}
  \nabla_a W_{ab} =  \vecrab F_{ab} \; ,
\label{SPH-02}
\end{equation}
where $F_{ab}$ is a scalar function which is symmetric in $a$ and $b$,
and $\vecrab$ is a shorthand for $(\vecra - \vecrb)$.  Hence, the
forces between particles are along the line of centers.  

Various types of spherically symmetric kernels have been suggested
over the years \cite{Mo85, Be90}. Among those the spline kernel of
Monaghan \& Lattanzio \cite{ML85}, mostly used in current SPH--codes,
yields the best results. It reproduces constant densities exactly in
1D, if the particles are placed on a regular grid of spacing $h_{\rm
SPH}$, and has compact support.

In the Newtonian case $\tpiab$ is given by \cite{Mo97}
\begin{equation}
  \tpiab =  -\alpha \, \frac{h_{\rm SPH} \, \vecvab \cdot
                              \vecrab}{\overline{\rho}_{ab}\, |\vecrab|^2}
                  \rund{\overline{c}_{ab} - 2\, 
                        \frac{h_{\rm SPH} \, \vecvab \cdot \vecrab}{
                                                 |\vecrab|^2} } 
\label{SPH-03} 
\end{equation}
provided $\vecvab \cdot \vecrab < 0$, and $\tpiab = 0$ otherwise. Here
$\vecvab = \vecva - \vecvb$, $\overline{c}_{ab} = \frac{1}{2}(c_a +
c_b)$ is the average sound speed, $\overline{\rho}_{ab} = \frac{1}{2}
(\rho_a + \rho_b)$, and $\alpha \sim 1.0$ is a parameter.

Using the first law of thermodynamics and applying the SPH formalism
one can derive the thermal energy equation in terms of the specific
internal energy $\varepsilon$ (see, \eg \cite{Mo92}). However, when
deriving dissipative terms for SPH guided by the terms arising from
Riemann solutions, there are advantages to use an equation for the
total specific energy $E \equiv \vecv^2/2 + \varepsilon$, which
reads \cite{Mo97}
\begin{equation}
 {d E_a \over d t} = -\sum_b  m_b \rund{ 
                                       \frac{p_a \vecvb}{\rho_a^2}
                                     + \frac{p_b \vecva}{\rho_b^2}
                                     + \tomab } \cdot \nabla_a W_{ab} \; ,
\label{SPH-04}
\end{equation}
where $\tomab$ is the artificial energy dissipation term derived by
Monaghan \cite{Mo97}. For the relativistic case the explicit form of
this term is given in Section\,(\ref{ss:SPH}).

In SPH calculations the density is usually obtained by summing up the
individual particle masses, but a continuity equation may be solved
instead, which is given by
\begin{equation}
 {d \rho_a \over d t} = -\sum_b  m_b (\vecva - \vecvb) \, \nabla_a W_{ab} \; .
\label{SPH-05}
\end{equation}

The capabilities and limits of SPH have been explored, \eg in
\cite{SM93, TT98}. Steinmetz \& M\"uller \cite{SM93} conclude that it
is possible to handle even difficult hydrodynamic test problems
involving interacting strong shocks with SPH provided a sufficiently
large number of particles is used in the simulations.  SPH and finite
volume methods are complementary methods to solve the hydrodynamic
equations each having its own merits and defects.

%\newpage
%\listoffigures
%\newpage
%\listoftables

\end{document}